\documentclass[12pt,letterpaper]{article}
\usepackage{jheppub}
\usepackage{graphicx}
\usepackage[usenames, dvipsnames, table]{xcolor}

\usepackage{amsmath}
\usepackage{amssymb}
\usepackage{amsthm}

\newtheoremstyle{named}{}{}{\itshape}{}{\bfseries}{.}{.5em}{#3}
\theoremstyle{named}
\newtheorem*{namedconjecture}{Conjecture}

\usepackage{mathtools}
\usepackage{mathdots}
\usepackage[mathscr]{euscript}

\usepackage{fnpct}
\usepackage{diagbox}

\newcommand\mathdiagbox[3][]{\hbox{\tabcolsep=\arraycolsep\diagbox[#1]{$#2$}{$#3$}}}

\usepackage{enumerate}
\usepackage[labelformat=simple]{subcaption}

\renewcommand{\arraystretch}{0.95}

\definecolor{KvColor}{rgb}{0.75,0.75,1}
\definecolor{KvpColor}{rgb}{1,0.8,1}
\definecolor{KaColor}{rgb}{1,1,0.92}
\definecolor{KapColor}{rgb}{1,1,0.7}

\usepackage{hhline}

\usepackage{float}
\usepackage{afterpage}

\newcommand{\cF}{\mathcal{F}}
\newcommand{\cG}{\mathcal{G}}
\newcommand{\cH}{\mathcal{H}}

\newcommand{\cQ}{\mathcal{Q}}
\newcommand{\cR}{\mathcal{R}}

\newcommand{\cW}{\mathcal{W}}

\newcommand{\df}{\mathrel{:=}}
\newcommand{\noeq}{\mathrel{\phantom{=}}}

\newcommand{\Mori}{\mathscr{M}}
\newcommand{\Mcap}{\mathscr{M}_{\cap}}
\newcommand{\Eff}{\mathscr{E}}
\newcommand{\Kah}{\mathscr{K}}
\newcommand{\Kv}{\widetilde{\mathscr{K}}}
\newcommand{\Yd}{\tilde{Y}}
\newcommand{\FundK}{\mathscr{F}}

\newcommand{\su}{\mathfrak{su}}

\DeclareMathOperator{\Tr}{Tr}

\newcommand{\pagelabel}[1]{\phantomsection\label{#1}}

\definecolor{cobalt}{RGB}{44, 98, 120}
\definecolor{celadon}{rgb}{0.67, 0.88, 0.69}
\definecolor{dm}{cmyk}{.20, 0, .30, 0}
\definecolor{burgundy}{rgb}{0.5, 0.0, 0.13}
\definecolor{plotBlue}{RGB}{94, 130, 181}

\hypersetup{
  colorlinks,
  citecolor=Violet,
  linkcolor=cobalt,
  urlcolor=Blue}

\begin{document}
\begin{titlepage}

\setcounter{page}{1} \baselineskip=15.5pt \thispagestyle{empty}

{\flushright ACFI-T21-09 \\ }

\bigskip\

\vspace{1.05cm}
\begin{center}
{\fontsize{20}{28} \bfseries The Weak Gravity Conjecture \\ \vspace{0.3cm}and BPS Particles }

 \end{center}

\vspace{1cm}

\begin{center}
\scalebox{0.95}[0.95]{{\fontsize{14}{30}\selectfont Murad Alim,$^{a}$ Ben Heidenreich,$^{b}$ Tom Rudelius$^{c,d}$}}
\end{center}

\begin{center}
\vspace{0.25 cm}
\textsl{$^{a}$Department of Mathematics, University of Hamburg, Bundesstr. 55, 20146, Hamburg, Germany}\\
\textsl{$^{b}$Department of Physics, University of Massachusetts, Amherst, MA 01003 USA}\\
\textsl{$^{c}$Institute for Advanced Study, Princeton, NJ 08540 USA}\\
\textsl{$^{d}$Department of Physics, University of California, Berkeley, CA 94703 USA}\\

\vspace{0.25cm}

\end{center}

\vspace{1cm}
\noindent

Motivated by the Weak Gravity Conjecture, we uncover an intricate interplay between black holes, BPS particle counting, and Calabi-Yau geometry in five dimensions. In particular, we point out that extremal BPS black holes exist only in certain directions in the charge lattice, and we argue that these directions fill out a cone that is dual to the cone of effective divisors of the Calabi-Yau threefold. The tower and sublattice versions of the Weak Gravity Conjecture require an infinite tower of BPS particles in these directions, and therefore imply purely geometric conjectures requiring the existence of infinite towers towers of holomorphic curves in every direction within the dual of the cone of effective divisors. We verify these geometric conjectures in a number of examples by computing Gopakumar-Vafa invariants.

 \vspace{1.1cm}

\bigskip
\noindent\today

\end{titlepage}
\setcounter{tocdepth}{2}
\tableofcontents

\section{Introduction}

Delineating the ``landscape'' of low-energy limits of consistent quantum gravities from the ``swampland''~\cite{Vafa:2005ui} of \emph{seemingly} consistent gravitational effective field theories lacking such an ultraviolet completion is essential to bridging the gap between quantum gravity and experiment. Due to the difficulty of this problem almost nothing is known with certainty, but numerous conjectures have been formulated and tested to probe the boundary between the landscape and the swampland, of which one of the oldest and best-established  is the Weak Gravity Conjecture (WGC)~\cite{ArkaniHamed:2006dz}.
 
 In its minimal form, the WGC states that in any consistent theory of quantum gravity with a massless gauge boson, there exists a charged state with charge-to-mass ratio greater than or equal to that of a large extremal black hole:\footnote{For simplicity, we assume that the cosmological constant vanishes, making the $M \to \infty$ large black hole limit sensible. The limit ensures that the bound is fully determined by the two-derivative low energy effective action for the massless fields.}
\begin{equation}
\text{WGC: }~~~ \frac{q}{m} \geq \lim_{M \to \infty} \frac{Q}{M}\biggr|_{\text{ext BH}}. \label{eqn:WGC}
\end{equation}
We refer to states saturating this bound as \emph{extremal} and those satisfying it as \emph{superextremal}. Note that by this terminology, extremal states are superextremal.

Because the superextremal states required by the WGC could have masses well above the Planck scale, this minimal form of the conjecture places no constraints on the low-energy effective field theory. Nonetheless, there is increasing evidence that quantum gravities contain entire towers of superextremal (possibly unstable) particles of increasing charge~\cite{Heidenreich:2015nta, Heidenreich:2016aqi, Montero:2016tif, Heidenreich:2017sim, Andriolo:2018lvp, Grimm:2018ohb, Heidenreich:2018kpg, Lee:2018urn,Lee:2019tst}, rather than just a few superextremal charged states. This has motivated stronger versions of the WGC, including the tower WGC (TWGC) \cite{Andriolo:2018lvp}---requiring superextremal charged particles of arbitrarily large charge in every direction in the charge lattice---and the yet-stronger sublattice WGC (sLWGC)~\cite{Heidenreich:2016aqi}---requiring a superextremal charged particle for every site on a full-dimensional sublattice of the charge lattice.\footnote{Both are notionally related to the ``magnetic WGC'' of~\cite{ArkaniHamed:2006dz}. Some other strong forms of the WGC, such as the requirement that the lightest charged particle is superextremal, have proven to be false~\cite{Heidenreich:2016aqi}. For a recent, related conjecture about BPS strings, see \cite{Tarazi:2021duw}.} The two conjectures are closely related, so we refer to them collectively as the ``T/sLWGC'' when it is not necessary to distinguish between them.

The original motivation for the TWGC and sLWGC comes from the behavior of the WGC under $S^1$ Kaluza-Klein compactification \cite{Heidenreich:2015nta}. In particular, the WGC can be violated in the $d$-dimensional compactified theory even if it is satisfied in $d+1$ dimensions. If the WGC is indeed a necessary condition on effective field theories coupled to quantum gravity, this implies that the parent $d+1$-dimensional theory must satisfy some stronger condition to ensure consistency of the $d$-dimensional theory. The TWGC and sLWGC both suffice for this purpose and seem to be simplest available options. Thus, we may say that the WGC in $d$ dimensions implies the T/sLWGC (or something similar) in $d+1$ dimensions.

Additional strong evidence for the sLWGC arises in perturbative heterotic string theory. In this context, the sLWGC can be proven to hold at tree-level using worldsheet modular invariance~\cite{Heidenreich:2016aqi, Montero:2016tif} and other worldsheet techniques~\cite{BenMatteoProof}. The TWGC is a consequence of the sLWGC, and therefore holds in these cases as well.

\subsection{Supersymmetry and a conifold conundrum}\label{sec:CC}

While the WGC and its stronger variants constrain the spectrum of massive charged particles in a given quantum gravity, this spectrum is often hard to calculate in concrete examples. 
This makes it difficult to verify or refute these conjectures outside certain contexts (such as tree-level heterotic string theory). For example, in Calabi-Yau compactifications of type II string theory, the charged spectrum depends on the minimum volume in each homology class, which (except for calibrated cycles) cannot be computed without detailed knowledge of the Calabi-Yau metric.

However, it is comparatively straightforward to compute the spectrum of BPS particles, when they exist. Any concrete predictions of the T/sLWGC that depend only on the BPS spectrum can then be tested and potentially falsified. Exactly what the T/sLWGC predict about the BPS spectrum is frequently misunderstood, so we will discuss it at length.

BPS particles preserve some fraction of the supercharges. Equivalently, they saturate one or more BPS bounds, each of the form:
\begin{equation} \label{eqn:BPSbound}
M \ge |Z| \,,
\end{equation}
where the ``central charge'' $Z$ is a (real or complex) moduli-dependent linear combination of the gauge charges.  
Depending on the dimension and number of supercharges, there can be one, several, or even a continuum of BPS bounds.\footnote{For example, there may be real central charges $Z_1, Z_2$ such that $M \ge |Z_1 \cos \theta + Z_2 \sin \theta|$ for all $\theta$. This particular continuum of BPS bounds is equivalent to the bound
$M \ge \sqrt{Z_1^2+Z_2^2}$.
Note that this can be rewritten as $M \ge |Z|$ for complex $Z = Z_1 +i Z_2$; all BPS bounds can be expressed in terms of real central charges, but complex central charges are sometimes more convenient.} 

Whereas the WGC places an upper bound on particle masses, each BPS bound places a lower bound on them, and unlike the WGC bound, BPS bounds are satisfied by \emph{all} states\footnote{For our purposes a ``state'' is a localized excitation above the vacuum with some rest-frame energy (``mass'') $M$ and charge $Q$; this excludes branes extending to infinity, but includes configurations that are not particles (e.g., a pair of mutually BPS particles placed a finite distance apart).} in the theory, not just by a few special states. Thus, a BPS particle of charge $Q$ is always the lightest state in the theory of that charge, and therefore BPS particles are necessarily superextremal. 

It is easy to get carried away by this simple yet powerful observation, and arrive at wrong conclusions by too-hasty reasoning. The naive argument goes as follows. First, in theories with BPS bounds, superextremal particles are necessarily BPS. Next, the tower and sublattice conjectures require not just one but a whole tower of superextremal particles, so in such theories they require BPS particles to come in infinite towers. However, some consistent quantum gravities contain BPS particles that are not part of an infinite tower, and therefore the tower and sublattice conjectures are false.

Before explaining how this argument goes awry---and correcting it to uncover the true predictions of the T/sLWGC for the spectrum of BPS particles---let us see explicitly why the last statement is true: BPS particles do not always come in infinite towers.
Consider, for example, type IIA string theory (or M-theory) compactified to four (or five) dimensions on a Calabi-Yau threefold $X$. In this context, BPS particles arise from D2 (or M2) branes wrapping holomorphic curves with charges dictated by their homology classes (along with their dissolved D0 charge in the D2 case). For special choices of $X$, such as the quintic, there is a unique two-cycle and a holomorphic curve in every multiple of the corresponding homology class~\cite{S.Katz:1986cba}. Thus, by wrapping D2 (M2) branes on the curves, we obtain an infinite tower of BPS particles. But for most $h^{1,1} > 1$ geometries, this is not the case. A generic Calabi-Yau manifold contains holomorphic curves that can shrink to zero size (forming conifold singularities) at special points in the moduli space. These ``flop'' degenerations, which lie at a finite distance in the moduli space, lead to massless BPS particles, but the number of massless particles is finite~\cite{Strominger:1995cz}, not the infinite number that would result if there were an infinite tower of BPS particles of increasing charge.\footnote{While infinite towers of particles can (and do) become massless near a phase transition associated to a strongly interacting conformal field theory, this should never occur at a weakly coupled phase transition per the Emergence Proposal~\cite{Heidenreich:2017sim,Grimm:2018ohb,Heidenreich:2018kpg}, unlike the infinite towers at infinite-distance points in the moduli-space
predicted by the Swampland Distance Conjecture~\cite{Ooguri:2006in}.}

For the Greene-Morrison-Strominger-Vafa (GMSV) conifold \cite{Greene:1995hu, Greene:1996dh}, for instance, 16 BPS particles of charge $(0,1)$ localized at 16 conifold singularities become massless at the flop. A more complete list of BPS particles in this geometry as a function of the charges $(q_1, q_2)$ in the theory (as counted by the genus 0 Gopakumar-Vafa (GV) invariants \cite{Gopakumar:1998ki,Gopakumar:1998jq,Gopakumar:1998ii}) is shown in Table \ref{tab:GMSVinvs0}.\footnote{Some of the GV invariants for this geometry were previously computed in \cite[\S 6.5]{Klemm:2004km} and \cite{Carta:2021sms}.} The table strongly suggests that there are BPS particles of all charges $(q_1,q_2) \in \mathbb{Z}^2$ such that $q_1 \ge q_2/4 \ge 0$, and it is not hard to show that outside this wedge the only BPS particles are the 16 charge $(0,1)$ particles that become massless at the flop.
 
\begin{table}
\begin{center}
$\arraycolsep=5pt
\begin{array}{c|ccccc}
\mathdiagbox[width=1cm,height=0.75cm,innerleftsep=0.1cm,innerrightsep=0cm]{q_2}{q_1}&0&1&2&3&4 \\ \hline
0&-&640&10032&288384&10979984 \\
1&16&2144&231888&23953120&
2388434784 \\
2& 0&120&356368&144785584&36512550816 \\
3& 0&-32&14608&144051072&
115675981232 \\
4& 0&3&-4920&5273880&85456640608\\
5&0&0&1680&-1505472&
3018009984 \\
6& 0&0&-480&512136&-748922304 \\
7&0&0&80&-209856&218062416\\
8& 0&0&-6&75300&-90910176 \\
9&0&0&0&-21600&37721680 \\
10& 0&0&0&4312&-15086208 \\
11&0&0&
0&-512&5300736 
\end{array}
$
\caption{Genus 0 GV invariants of degree $(q_1, q_2)$ for the GMSV conifold. At the flop, 16 hypermultiplets of charge $(0,1)$ become massless; there is no infinite tower of massless hypermultiplets.}
\label{tab:GMSVinvs0}
\end{center}
\end{table}

Naively, this spectrum of BPS particles seems to violate the T/sLWGC in the $(0,1)$ charge direction (more generally, anywhere in the wedge $q_2 > 4 q_1 \ge 0$) due to the absence of an infinite tower of BPS particles. Even worse, violating the T/sLWGC in $d+1$ dimensions typically leads to a violation of the WGC in $d$ dimensions. In particular, if these stronger variants fail in 5d compactifications of M-theory then the WGC itself may fail in the corresponding 4d compactifications of type IIA string theory! We have thus arrived at a ``conifold conundrum'': BPS particles of certain charges do not exist, naively leading to a violation of the WGC.

Fortunately, this is not the case, as the naive reasoning above is flawed.
The misconception arises from conflating the fact that BPS particles are invariably \emph{superextremal} with the misguided notion that they are invariably \emph{extremal}. If a given BPS particle \emph{is} extremal, then since it has the smallest-possible mass-to-charge ratio amongst all particles with parallel charge vectors, any superextremal particle in this direction must also be BPS, and therefore the T/sLWGC require an infinite tower of BPS particles in this direction. 

However, {BPS states are \emph{not always extremal}}.\footnote{All BPS particles are extremal in theories where all gauge bosons reside in the gravity multiplet---such as the quintic theory discussed above---but this \emph{only} occurs in this special class of theories.}\footnote{Although BPS states and extremal black holes both have vanishing long-range self-force~\cite{Heidenreich:2019zkl,Heidenreich:2020upe}, moduli contribute to these forces in addition to gauge fields and gravity;
different couplings to moduli explain how both can have vanishing self-force despite having different charge-to-mass ratios.} This is easy to see in the simple and prototypical~\cite{ArkaniHamed:2006dz} example of heterotic string theory compactified on a torus $T^k$, which has the spectrum of massive charged particles:
\begin{equation}
\frac{\alpha'}{4} m^2 = \frac{1}{2} Q_L^2 + N - 1 = \frac{1}{2} Q_R^2 + \tilde{N} \,,
\end{equation}
where $N, \tilde{N}$ are non-negative integers and the charges $(Q_L, Q_R)$ live on an even unimodular lattice of signature $(16+k,k)$ (so that $Q_L^2 - Q_R^2 \in 2\mathbb{Z}$). The BPS bound
\begin{equation}
\frac{\alpha'}{4} m^2 \ge \frac{1}{2} Q_R^2
\end{equation}
is satisfied by the entire spectrum, whereas the black-hole extremality bound \cite{Sen:1994eb}
\begin{equation}
\frac{\alpha'}{4} m^2 \ge \frac{1}{2} \max(Q_L^2, Q_R^2)
\end{equation}
is not. All sites with $Q_R^2 \ge Q_L^2 - 2$ are populated by BPS particles, but those with $Q_R^2 = Q_L^2 - 2$ are not extremal, as their mass is strictly less than that of an extremal black hole.

Indeed, this example has a number of features in common with the GMSV conifold discussed previously. The BPS particles of charge $(Q_L,Q_R)$ satisfying $Q_R^2 = Q_L^2 - 2$ are massless at special points in the moduli space where $Q_R^2 = 0$, and do not appear in infinite towers, since $(k Q_L, k Q_R)$ satisfies $Q_R^2 < Q_L^2 - 2$ for all $k>1$. By comparison, the charged BPS particles satisfying $Q_R^2 \ge Q_L^2$ appear in infinite towers but are never massless.

Note that the sLWGC and TWGC (and indeed the lattice WGC~\cite{Heidenreich:2015nta}) are satisfied in this example, even though not all BPS particles appear in infinite towers. This is made possible by the fact that the BPS particles that do not appear in infinite towers are not extremal.

With these misconceptions out of the way, it is easy to understand what the T/sLWGC \emph{do} predict about the spectrum of BPS particles. In any charge direction in which the BPS and black hole extremality bounds agree, an infinite tower of BPS particles is required. In any other charge direction, nothing precludes the existence of an infinite tower of superextremal but non-BPS particles, and so the T/sLWGC place no constraints on the spectrum of BPS particles in such directions.\footnote{The WGC places similar constraints on the spectrum of BPS particles, but replacing ``particles'' with ``multiparticle states''. In particular, mutually BPS particles must generate all the charge directions in which the BPS and black hole extremality bounds agree.}

\bigskip

In the remainder of the paper, we show that this is precisely how the conifold conundrum described above is resolved: the BPS particles associated to shrinking curves in the Calabi-Yau are not extremal. Moreover, by computing the GV invariants of several Calabi-Yau threefolds, we find evidence that there are infinite tower of BPS particles in all charge directions in which the BPS and black hole extremality bounds agree. Thus, the constraints placed on the BPS spectrum by the T/sLWGC are satisfied.

Our analysis focuses on 5d theories arising from M-theory compactified on a Calabi-Yau threefold, rather than their 4d relatives arising from type IIA/IIB string theory compactified on a Calabi-Yau threefold. One benefit of these 5d theories is that they avoid a subtlety caused by infrared divergences that we have so far ignored:
in 4d the massless charged hypermultiplets that appear at the flop logarithmically renormalize the corresponding $U(1)$ gauge coupling to zero in the deep infrared. This allows exponentially big black holes to have arbitrarily large charge-to-mass ratios, hence per~\eqref{eqn:WGC} the WGC requires a massless charged particle (which is obviously present)\footnote{The flopping hypermultiplets must be exponentially light to significantly alter the infrared gauge coupling, so they satisfy WGC very near the flop but are parametrically superextremal, not extremal.}
 but then the T/sLWGC naively require an infinite tower of massless charged particles (which, as already noted, is \emph{not} present).

What should we make of this apparent counterexample to the T/sLWGC? Although we cannot apply the $S^1$ compactification argument of~\cite{Heidenreich:2015nta} to relate the 4d T/sLWGC to the ``3d WGC''---there are no asymptotically flat black holes in three dimensions, and thus no 3d WGC bound---many other arguments suggest that these conjectures remain true in four dimensions, e.g.,~\cite{Heidenreich:2016aqi,Heidenreich:2017sim,Andriolo:2018lvp}. How can this be consistent with the behavior near the flop? The obvious resolution is that due to these infrared divergences the 4d T/sLWGC cannot be defined using a strict infrared limit, as in~\eqref{eqn:WGC}, instead requiring renormalization, see, e.g.,~\cite{Heidenreich:2017sim,Klaewer:2020lfg}. (Unfortunately, precise renormalized versions of these conjectures are not yet known.)

These issues (while not insurmountable) are neatly avoided in five dimensions because the 5d gauge coupling does not suffer from infrared divergences. Instead light charged particles only contribute threshold corrections in 5d---indeed, these play a central role in our analysis, see, e.g.,~\eqref{eqn:flop}, \eqref{eqn:shrinkingcurve}---and the gauge coupling remains finite at the flop. Thus, the 5d T/sLWGC can be defined using only infrared quantities and they make the sharp prediction that infinite towers of BPS particles are required in any charge direction in which the BPS and black hole extremality bounds agree, as discussed above. Should this prediction fail, the 4d WGC is almost certainly violated by the arguments of~\cite{Heidenreich:2015nta}, making our work a stringent test of the 4d WGC.

For this reason---in combination with the additional technical difficulties associated to 4d case (some of which are touched on in~\S\ref{sec:4d})---we focus on 5d theories, leaving the four-dimensional case to future work.

\bigskip

As a final comment, note that there is an alternate form of the WGC preferred by some authors, the Repulsive Force Conjecture (RFC)~\cite{Palti:2017elp, Lee:2018spm, Heidenreich:2019zkl}, requiring the existence of self-repulsive rather than superextremal charged particles. Sublattice and tower versions of the RFC can also be formulated. What happens to the conifold conundrum for these alternate conjectures? An infinite tower of BPS particles is clearly sufficient to satisfy the conjectures in directions in which they exist. However, even when the BPS and black hole extremality bounds coincide, there is no reason that these conjectures \emph{must} be satisfied by BPS particles. In particular, heavier non-BPS particles can still be self-repulsive given appropriate moduli couplings, so these conjectures do not place interesting constraints on the spectrum of BPS particles, and we will not discuss them further.

\subsection{Summary of results} \label{subsec:summary}

Much of this paper is devoted to proving theorems and exploring examples, the details of which can get rather technical. We therefore begin by previewing our main results for the reader's benefit.\footnote{For some recent, related results, see \cite{Long:2021lon}.}

In~\S\ref{sec:SUGRA} we review 5d $\mathcal{N}=1$ supergravity and its connection to Calabi-Yau geometry and BPS states, paying close attention to phase transitions (such as flops) and the possibility of wall crossing phenomena. Then in~\S\ref{sec:ExtBound} we discuss how to determine the black hole extremality bound using the attractor mechanism~\cite{Ferrara:1995ih,Cvetic:1995bj,Strominger:1996kf,Ferrara:1996dd,Ferrara:1996um} and the fake-superpotential formalism~\cite{Ceresole:2007wx,Andrianopoli:2007gt,Andrianopoli:2009je,Andrianopoli:2010bj,Trigiante:2012eb}. This brings us to our central question: for which charge directions do the BPS and black hole extremality bounds agree?

By the attractor mechanism, a sufficient condition for the two to agree everywhere in the moduli space is for the associated central charge $Z$ to have (1) a positive local minimum within the moduli space, with (2) a $Z$ gradient flow connecting every other point within the moduli space to this minimum. In the case at hand, the moduli space consists of one or more phases separated by phase transitions (e.g., flops), where each phase corresponds to a different Calabi-Yau manifold. Those connected by flops form 
 the extended K\"ahler cone $\Kah$, and the first condition is satisfied on $\Kah$ when~\cite{Chou:1997ba}
\begin{equation}
\cF_I \df \frac{\partial \cF}{\partial Y^I} \propto +q_I 
\end{equation}
at some point in $\Kah$. Here $\cF[Y]$ is the prepotential associated to the ``very special geometry'' of the Coulomb branch, which is cubic within each phase of $\Kah$
\begin{equation}
\cF := \frac{1}{6} C_{IJK} Y^I Y^J Y^K \,,
\end{equation}
where $Y^I$ are the K\"ahler coordinates, $C_{I J K}$ are the intersection numbers of the Calabi-Yau threefold, and the central charge $Z$ is proportional to $\zeta_q = q_I Y^I$.

As shown in~\S\ref{sec:BPSBHE}, the second condition is also satisfied on $\Kah$ when the first is, as can be seen by rephrasing the problem in ``dual coordinates'' $\Yd_I = \cF_I / \cF$.\footnote{These are the conjugate variables arising from a Legendre transformation of the potential $L[Y] = -\log \cF$, and the mapping between K\"ahler coordinates $Y^I$ and dual coordinates $\Yd_I$ can be shown to be one-to-one using the convexity of $L[Y]$ and of $\Kah$.} We call the dual-coordinate image of $\Kah$ the ``cone of dual coordinates'' $\Kv$ (see Table \ref{glossary} for a glossary of notable cones appearing in our paper). As is well-known to those familiar with the BPS solutions arising in the study of the attractor mechanism, $\zeta_q$ gradient flows are straight lines directed towards $\Yd_I \propto +q_I$ in dual coordinates (see, e.g.,~\cite{Larsen:2006xm}). Thus, a $\zeta_q$ gradient flow connects every point in $\Kah$ to the minimum if $\Kv$ is convex. 

\begin{table}
\begin{center}
\renewcommand{\arraystretch}{1.5}
\begin{tabular}{|c|c|c|}\hline
{\bf Symbol} & { \bf Name} & {\bf Relationships} \\\hline
$\Kah_A$ & K\"ahler cone in phase $A$ & --- \\\hline
$\Kah  $ & \hspace{-0.7em}\rule[-1em]{0pt}{2.7em} {\renewcommand{\arraystretch}{0.8} \begin{tabular}{@{}c@{}}Extended K\"ahler cone \\ (= cone of movable divisors)\end{tabular}}  & $\Kah = \bigcup_A \Kah_A$ \\\hline
$\Eff $ & Effective cone (of divisors)& $\Eff \supseteq \Kah$  \\\hline
$\Mori_A$ & Mori cone in phase $A$ & $\Mori_A = \Kah_A^\ast$  \\\hline
$\Mori_\cap$ & Intersection of the Mori cones & $\Mori_\cap = \bigcap_A \Mori_A = \Kah^\ast$ \\\hline
$\Kv_A$ & \hspace{-0.7em}\rule[-1em]{0pt}{2.7em} {\renewcommand{\arraystretch}{0.8} \begin{tabular}{@{}c@{}}Cone of dual coordinates in phase $A$ \\ (not convex in general)\end{tabular}} & $\Kv_A = \Yd[\Kah_A] $ \\\hline
$\Kv$ & \hspace{-0.7em}\rule[-1em]{0pt}{2.7em} {\renewcommand{\arraystretch}{0.8} \begin{tabular}{@{}c@{}}Cone of dual coordinates \\ (= cone of movable curves [\S\ref{sec:boundaries}])\end{tabular}}  & $\Kv = \Yd[\Kah] \subseteq \Kah^\ast$, $\Kv = \Eff^\ast$ [\S\ref{sec:boundaries}] \\\hline
\end{tabular}
\caption{Glossary of cones. Here, $\mathcal{C}^\ast$ represents the dual cone of $\mathcal{C}$. The relation $\Kv = \Eff^\ast$ (implying that $\Kv$ is the cone of movable curves) is shown in~\S\ref{sec:boundaries}. All cones listed here are convex except $\Kv_A$.}
\label{glossary}
\end{center}
\end{table}

While $\Kah$ is convex, to show that $\Kv$ is also convex is non-trivial because the dual-coordinate map $\Yd_I = \Yd_I[Y]$ is non-linear. Indeed, although the K\"ahler cone $\Kah_A$ associated to a single phase $A$ is convex, its dual-coordinate image $\Kv_A = \Yd[\Kah_A]$ is generally \emph{not} convex, see, e.g., Figure~\ref{sfig:kmvKvPhases}. Nonetheless, as shown in~\S\ref{sec:boundaries}, $\Kv$ is the dual of the effective cone $\Eff$, and since dual cones are convex this implies that $\Kv$ is convex. (The relation $\Kv = \Eff^\ast$---implying $\Kv$ is the cone of movable curves~\cite{Boucksom13}---is interesting in its own right, as it determines the pseudoeffective cone $\overline\Eff = \Kv^\ast$ of a Calabi-Yau manifold $X$ given knowledge of the birationally-equivalent Calabi-Yau manifolds $X'$ filling out $\Kah$ along with their intersection numbers. This is a new result to our knowledge.)

Thus, we conclude that the BPS and black hole extremality bounds agree for all $q_I \in \Kv$. This allows us to formulate novel geometric conjectures that follow from the T/sLWGC in \S\ref{sec:geometric}: for any nontrivial class $q_I \in H_2(X,\mathbb{Z})$ lying within the dual of the effective cone $\Kv = \Eff^\ast$, there is a holomorphic curve in the class $k q_I$ for some positive integer $k$, where the sLWGC further implies that a single $k$ can be chosen independent of $q_I \in H_2(X,\mathbb{Z})$. These geometric conjectures hold (moreover for $k=1$) in all the examples considered in~\S\ref{sec:conifoldexample} up to the highest degree GV invariants that we have been able to compute. For example, they hold for the GV invariants shown in Table~\ref{tab:GMSVinvs0}, as illustrated in Table~\ref{tab:GMSVinvs}.

What happens when $q_I$ lies outside $\Kv$? If $\zeta_q$ is negative anywhere in the moduli space then some $\zeta_q$ gradient flows will pass through zero, and such flows cannot describe black holes (as black holes cannot have negative mass). Thus, the BPS and black hole extremality bounds disagree in at least part of the moduli space when $q_I$ lies outside the cone $\Kah^\ast$ in which $\zeta_q = q_I Y^I$ is non-negative across $\Kah$.
 
While BPS black holes could in principle exist elsewhere in the moduli space, they would have to decay by wall crossing upon entering the part of moduli space where they do not exist. As argued in~\S\ref{subsec:wallcrossing}, wall crossing phenomena should not occur within $\Kah$ in 5d theories, and therefore the BPS and black hole extremality bounds disagree everywhere in the moduli space when $q_I$ lies outside $\Kah^\ast$. Note that $\Kah^\ast = \Mcap$ where $\Mcap = \bigcap_A \Mori_A$ is the intersection of the Mori cones $\Mori_A$ and for all phases $A$ within $\Kah$. If two phases $A$ and $B$ are connected by a flop where hypermultiplets of charge $q_I$ become massless then $q_I \in \Mori_A$ and $(-q_I) \in \Mori_B$ (or vice versa). In particular, $q_I \not\in \Mori_A \cap \Mori_B$, so $q_I$ lies outside $\Mcap = \Kah^\ast$; indeed, $\zeta_q$ changes sign within $\Kah$ (at the flop). Thus, the hypermultiplets that become massless at flops are not extremal, resolving our conifold conundrum. We will see this explicitly in the examples discussed in~\S\ref{sec:conifoldexample}.

Since $\Kv \subseteq \Kah^\ast$ (see Table~\ref{glossary}), the two conclusions reached so far about the relationship between the BPS and black hole extremality bounds are compatible: they agree within $\Kv$, whereas they disagree outside $\Kah^\ast$. When $\Kah$ has no finite-distance boundaries, $\Kv = \Kah^\ast$ (see~\eqref{eqn:KvFormula}) and this is the whole story. Otherwise, $\Kv \ne \Kah^\ast$, and there are some charge directions for which the relationship between the two bounds remains in question. It turns out that for $q_I \in \Kah^\ast \setminus \Kv$, all $\zeta_q$ flows beginning within $\Kah$ reach a finite-distance boundary of $\Kah$ outside the event horizon of the corresponding candidate black hole solution. What happens next depends on the physics at the boundary in question, and since this physics is often strongly-coupled we will not attempt to determine the outcome in general in the present work. However, there are often infinite towers of holomorphic curves throughout $\Kah^\ast$, in which case the T/sLWGC are satisfied regardless of whether the BPS and black hole extremality bounds agree within $\Kah^\ast \setminus \Kv$.\footnote{In the example discussed in~\S\ref{ssec:KMV}, certain charge directions along the boundary of $\Kah^\ast$ lack infinite towers of holomorphic curves. However, the BPS and black hole extremality bounds can be shown to disagree in these directions, see~\S\ref{sssec:GVrevisit}.} 

\begin{figure}
\centering
\begin{subfigure}{0.44\textwidth}
\centering
\includegraphics[height=6.5cm]{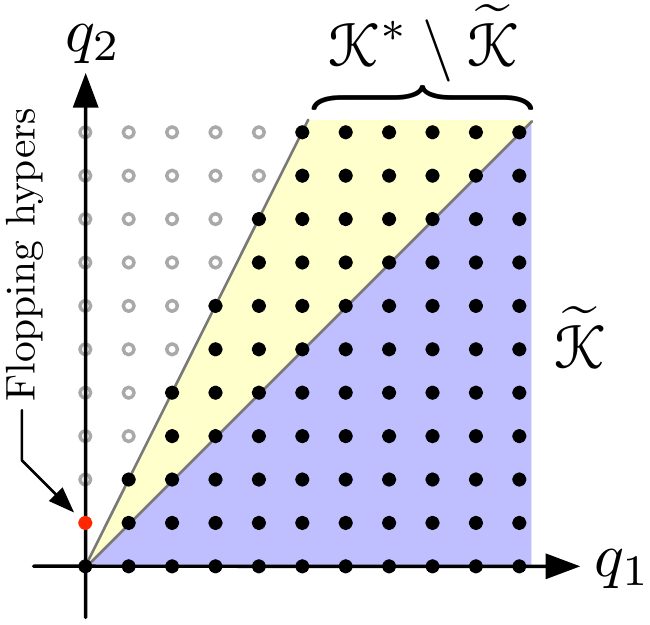}
\caption{Infinite towers of curves} \label{sfig:ConesAndCurves}
\end{subfigure}
\begin{subfigure}{0.54\textwidth}
\centering
\includegraphics[height=6.5cm]{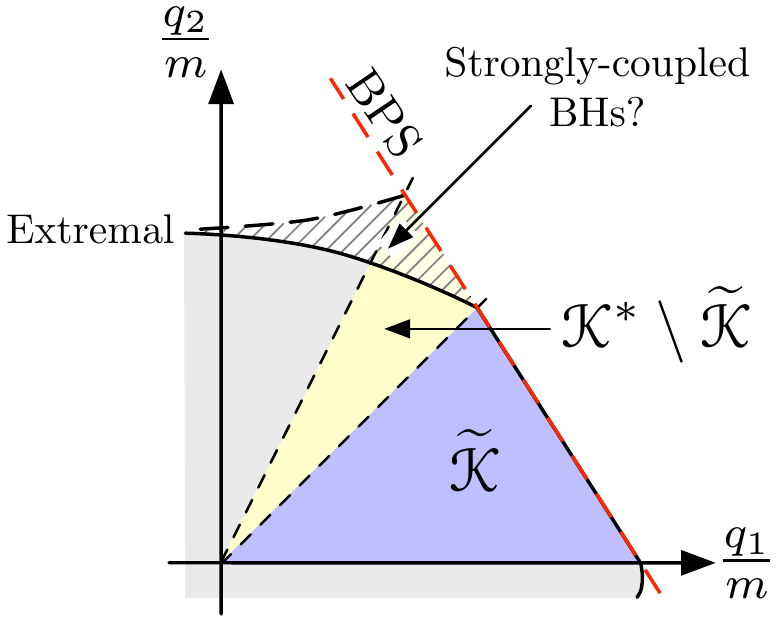}
\caption{BPS versus extremal} \label{sfig:ConesAndExtremality}
\end{subfigure}
\caption{\subref{sfig:ConesAndCurves} The spectrum of holomorphic curves (filled dots) compared with various cones. Infinite towers of holomorphic curves are possible within $\Kah^\ast$ (sometimes occurring in every charge direction therein), whereas they 
are required by the T/sLWGC in every direction within $\Kv \subseteq \Kah^\ast$. \subref{sfig:ConesAndExtremality} The black hole region (shaded) and the BPS bound (dashed red line) compared with various cones. The BPS and black hole extremality bounds coincide within $\Kv$, whereas they differ outside $\Kah^\ast$. The cross-hatched region may contain black holes with strongly-coupled CFT cores; depending on whether this occurs the BPS and black hole extremality bounds may or may not agree within $\Kah^\ast \setminus \Kv$.}
\label{fig:ConesFigs}
\end{figure}

To illustrate the above discussion, we summarize the relationship between $\Kah^\ast$, $\Kv$, the spectrum of holomorphic curves and the black hole extremality bound in Figure~\ref{fig:ConesFigs}.\footnote{A similar interplay between various cones occurs in the analysis of~\cite{Lanza:2021qsu}.}

After considering several examples in detail in~\S\ref{sec:conifoldexample}, we present our conclusions and discuss possible directions for future research in~\S\ref{sec:CONC}. Certain details about computing GV invariants and an analytic determination of the black hole extremality bound in the example of~\S\ref{ssec:symFlop} are included as appendices.

\section{Five-dimensional Supergravity, Geometry, and BPS States}\label{sec:SUGRA}

We begin by reviewing 5d $\mathcal{N}=1$ (ungauged) supergravity and its relationship to M-theory compactified on a Calabi-Yau threefold. At a generic point in the K\"{a}hler moduli space of the threefold, the gauge group is abelian, so we consider supergravity coupled to $n$ abelian vector multiplets. The action for the bosonic fields is\footnote{See, e.g.,~\cite{Lauria:2020rhc}, where $h^I_{\text{(there)}} = \frac{g_5}{\sqrt{3}} Y^I$, $a_{I J}^{\text{(there)}} = \frac{1}{g_5^2} a_{I J}$, $g_{i j}^{\text{(there)}} = \frac{1}{2 \kappa_5^2} \mathfrak{g}_{i j}$, and $\mathcal{C}_{I J K}^{\text{(there)}} = \frac{\sqrt{3}}{2 g_5^3} C_{I J K}$.}
\begin{align}
  S &= \frac{1}{2 \kappa_5^2}  \int d^5 x \sqrt{- g}  \left( R -
  \frac{1}{2} \mathfrak{g}_{i j} (\phi) \partial \phi^i \cdot \partial \phi^j
  \right) - \frac{1}{2 g_5^2} \int a_{I J} (\phi) F^I \wedge
  \star F^J \nonumber \\
  &+ \frac{1}{6(2\pi)^2} \int C_{I J K} A^I \wedge F^J \wedge F^K,
  \label{eqn:5dsugra}
\end{align}
where $I = 0, \ldots, n$, $i = 1, \ldots, n$, and $g_5^2 = (2\pi)^{4/3} (2\kappa_5^2)^{1/3}$. The scalar metric
$\mathfrak{g}_{i j} (\phi)$, the gauge kinetic matrix $a_{I J} (\phi)$, and the
Chern-Simons couplings $C_{I J K}$ are all determined by a prepotential
$\mathcal{F} [Y]$ that is homogenous of degree-three in independent variables $Y^I$ corresponding to the $n+1$ gauge fields. In particular, the $n$-dimensional moduli space is the surface
\begin{equation}
  \mathcal{F} [Y (\phi)] = 1,
\end{equation}
which we parameterize $Y^I = Y^I (\phi)$ in terms of $n$ moduli $\phi^i$. Defining $\mathcal{F}_I \df \partial_I \mathcal{F}$, $\mathcal{F}_{I J}
\df \partial_I \partial_J \mathcal{F}$ and $\mathcal{F}_{I J K} \df
\partial_I \partial_J \partial_K \mathcal{F}$, $C_{I J K} = \mathcal{F}_{I J K} $ fixes the Chern-Simons coupling, whereas
\begin{equation}
  a_{I J} (\phi) =\mathcal{F}_I [Y (\phi)] \mathcal{F}_J [Y (\phi)]
  -\mathcal{F}_{I J} [Y (\phi)], \qquad \mathfrak{g}_{i j} (\phi) = a_{I J}
  (\phi) \partial_i Y^I \partial_j Y^J .
  \label{scalarmetric}
\end{equation}
In a physical region of the moduli space, both $a_{I J}$ and $\mathfrak{g}_{i j}$ must be positive definite, where the two conditions are equivalent due to the identity
\begin{equation}
a^{I J} = \mathfrak{g}^{i j} \partial_i Y^I \partial_j Y^J + \frac{1}{3} Y^I Y^J \,, \label{eqn:inversea}
\end{equation}
where $a^{I J}$ and $\mathfrak{g}^{i j}$ denote the inverses of $a_{I J}$ and $\mathfrak{g}_{i j}$, respectively.

Note that invariance under small and large gauge transformations requires that the Chern-Simons couplings are integers, $C_{I J K} \in \mathbb{Z}$.\footnote{Naively, large gauge transformations impose stronger constraints when $I$, $J$, $K$ are not all distinct. However, these quantization conditions can be shifted by anomalies (e.g.,~\cite{Witten:1996md}). Regardless of the structure of these anomalies, the axions $A_0^I = \oint A_1^I$ resulting from dimensional reduction on a circle must be $2\pi$-periodic. One can check that this imposes the constraints $C_{I J K} \in \mathbb{Z}$ as well as $C_{I I J} + C_{I J J} \in 2 \mathbb{Z}$.} Thus, $\mathcal{F}[Y]$ is a homogeneous cubic polynomial with quantized coefficients and in particular it is not subject to renormalization, except as described in~\S\ref{subsec:PTs}. For M-theory on a Calabi-Yau threefold, $n+1=h^{1,1}$, and the integers $C_{I J K}$ are the intersection numbers of the threefold.

In addition to the supergravity and vector multiplets, there are also in general massless hypermultiplets. However, only charged hypermultiplets couple directly to the vector multiplets. As these are massive at generic points in the moduli space, we will not need the full hypermultiplet action for our analysis.

\subsection{BPS particles and strings}

Consider the above theory coupled to a massive, charged particle.
\begin{equation}
  S_{\mathrm{pp}} = - \int m (\phi) d \tau + q_I \int A^I .
\end{equation}
Alternately, $q_I$ can be measured using the flux integral:
\begin{equation}
  q_I \df \oint_{S^3} \left( \frac{1}{g_5^2} a_{I J}
  (\phi) \,{\star F^J} - \frac{1}{2(2\pi)^2} \mathcal{F}_{I J K} A^J \wedge F^K \right),
\end{equation}
where the quantity in parentheses is closed as a consequence of the $F^I$
equations of motion.\footnote{Despite the explicit appearance of $A^I$, $q_I$
is invariant under both small and large gauge transformations so long as the
electric charge is not superimposed on top of magnetic charge; this is because
$F^I$ is locally exact within an small ball surrounding the charge, so $\oint
A \wedge F$ is invariant upon shifting $A$ by any closed form. Otherwise,
$q_I$ is a Page charge.}

Associated to $q_I$, there is a (rescaled) central charge
\begin{equation}
 \zeta_q (\phi) \df q_I Y^I (\phi) \,, \label{eqn:centralcharge}
\end{equation}
associated to the BPS bound
\begin{equation}
  m (\phi) \geq \frac{g_5}{\sqrt{2} \kappa_5}  | \zeta_q (\phi)
  | = \frac{g_5}{\sqrt{2} \kappa_5}  | q_I Y^I (\phi) | \,.
  \label{eqn:5dBPS}
\end{equation}
Using~\eqref{eqn:inversea}, one can show that (\ref{eqn:5dBPS}) leads to vanishing long-range force between BPS particles with ``aligned'' (same sign) central charges $\zeta_1$, $\zeta_2$.

Similarly, for a string with charge $\tilde{q}^I$ normalized by
\begin{equation}
\tilde{q}^I = \frac{1}{2\pi} \oint_{S^2} F_2^I \,,
\end{equation}
there is a central charge $\tilde{\zeta}_{\tilde q}(\phi)$ and an associated BPS bound on the string tension:
\begin{equation}
\tilde{\zeta}_{\tilde q}(\phi) = \tilde{q}^I \mathcal{F}_I(\phi) \,, \qquad T(\phi) \ge \frac{\tilde{g}_5}{\sqrt{2} \kappa_5} |\tilde{\zeta}_{\tilde q}(\phi)| \,,   \label{eqn:stringBPS}
\end{equation}
where $\tilde{g}_5 = 2\pi/g_5$. As before, one can show that~\eqref{eqn:stringBPS} leads to vanishing long-range force between BPS strings with aligned central charges $\tilde{\zeta}_1$, $\tilde{\zeta}_2$.

\bigskip

Note that the absence of global symmetries requires central charges to be a linear combinations of gauge charges, but what selects these particular linear combinations over any others? Since the BPS bound relates the mass (the ``gravitational charge") of a BPS particle to its central charge, one would expect that the particular combination~\eqref{eqn:centralcharge} should be related to gravity in some way. As all but one of the gauge bosons sit in vector multiplets, with just one graviphoton in the gravity multiplet, the linear combination of charges that defines the central charge should correspond to this graviphoton. We now show that this is indeed the case.

Let $Y^I = Y^I_{(0)}$ be any point of interest in the moduli space. Fixing $Y_{(0)}^I = \delta^I_0$ by a linear transformation on the coordinates,
we write the prepotential near $Y^I_{(0)}$ as
\begin{equation}
  \mathcal{F}= (Y^0)^3 + a_i  (Y^0)^2 Y^i - \frac{1}{2} b_{i j} Y^0 Y^i Y^j +
  \frac{1}{6} C_{i j k} Y^i Y^j Y^k.
\end{equation}
By a further linear transformation $Y^0 \rightarrow Y^0 - \frac{1}{3}
a_i Y^i$ this can be simplified to
\begin{equation}
  \mathcal{F}= (Y^0)^3 - \frac{1}{2} b_{i j} Y^0 Y^i Y^j + \frac{1}{6} C_{i j
  k} Y^i Y^j Y^k. \label{eqn:canonicalF}
\end{equation}
The gauge kinetic matrix now takes the form $a_{I J} = \begin{psmallmatrix} 3 & 0\\ 0 & b_{i j} \end{psmallmatrix}$
at the point of interest, hence there is no kinetic mixing between $A^0$ and the remaining vectors $A^i$.

For small variations of the scalars about this point, we find:
\begin{equation}
  0 = \delta \mathcal{F}= 3 (Y^0)^2 \delta Y^0 - \frac{1}{2} b_{i j} \delta
  Y^0 Y^i Y^j - b_{i j} Y^0 Y^i \delta Y^j + \frac{1}{2} C_{i j k} Y^i Y^j
  \delta Y^k = 3 \delta Y^0
\end{equation}
to linear order about the point of interest. Therefore, for small fluctuations
the $Y^i$ are dynamical whereas $Y^0$ is fixed by the constraint $\mathcal{F}=
1$. Each vector multiplet contains a scalar whereas the gravity multiplet has
no scalars, hence the scalar/vector pairs $(Y^i,A^i)$ sit in vector multiplets whereas the vector $A^0$
belongs to the gravity multiplet.

The central charge that appears in the BPS bound is
\begin{equation}
  \zeta_{q} = q_I Y^I = q_0
\end{equation}
at the point of interest, which is indeed the electric charge associated to the graviphoton $A^0$.
Likewise the string central charge is the graviphoton magnetic charge.\footnote{Since we made an in-general non-integral change of basis to reach~\eqref{eqn:canonicalF}, neither central charge is the \emph{quantized} electric/magnetic charge associated to the graviphoton. Indeed, generally the graviphoton is not by itself a $U(1)$ gauge field, but rather it is a real linear combination of the $U(1)$ gauge fields of the theory.}

\subsection{Massless particles and phase transitions} \label{subsec:PTs}

The low-energy effective action~\eqref{eqn:5dsugra} is valid at generic points in the moduli space, but modifications are required when additional particles become massless at finite-distance points within the moduli space. Where they exist, massless particles necessarily saturate the BPS bound~\eqref{eqn:5dBPS}, thus either a massive, charged BPS state becomes massless, or a long multiplet becomes massless and splits into BPS constituents. The first case can be further subdivided based on whether the BPS state is a hypermultiplet or a vector multiplet, so there are three cases to consider:
\begin{enumerate}
\item
A charged BPS hypermultiplet becomes massless.
\item
A charged BPS vector multiplet becomes massless.
\item
A long multiplet becomes massless and splits into a hypermultiplet and a vector multiplet.
\end{enumerate}
More generally, a phase transition may involve multiple such particles becoming massless simultaneously. When there are finitely many, this is simply an elaboration on the three elementary cases listed above.
However, one additional possibility remains: 
\begin{enumerate}
\setcounter{enumi}{3}
\item
An infinite tower of particles becomes massless.
\end{enumerate}
We will comment on this possibility further after discussing the three elementary cases.

The first case is central to our paper. Let $q_I$ be the charge of the hypermultiplet, which per~\eqref{eqn:5dBPS} is massless along the codimension-one surface $q_I Y^I=0$ in the moduli space. Changing the sign of the hyperino mass induces a shift in the Chern-Simons couplings:
\begin{equation} \label{eqn:flop}
C^+_{I J K} = C^-_{I J K} + q_I q_J q_K \,,
\end{equation}
where $+$ ($-$) denotes the region where $q_I Y^I > 0$ ($q_I Y^I < 0$). Likewise, for $N$ charge $q_I$ hypermultiplets, the formula is the same with an added factor of $N$ multiplying the last term.

This phase transition corresponds to a flop of the Calabi-Yau geometry, with an associated change in the intersection numbers of the Calabi-Yau manifold. Note that the Calabi-Yau manifold develops one or more singular points at the flop. In the simplest case each one is a conifold singularity, with a single massless hypermultiplet associated to each~\cite{Strominger:1995cz}.

At the phase transition $q_I Y^I = 0$, it is also possible to turn on a vacuum expectation value (vev) for a combination of the charged hypermultiplets (provided there are at least two of them), spontaneously breaking the associated gauge symmetry. This generates a long multiplet from the vector and the Higgsed hyper whose mass depends on the remaining hypermultiplet moduli. In fact, this is nothing but the third case in reverse, so these two types of phase transition are the same, viewed from different branches of the moduli space. In geometric language, Higgsing corresponds to a complex structure smoothing of the singular points appearing at the flop, and passing from one branch to the other is a conifold transition \cite{Candelas:1989ug}.

Next, we consider the second case, where a charged BPS vector multiplet becomes massless. The charged vectors are necessarily $W$-bosons, and the gauge symmetry is enhanced to a nonabelian group at the phase transition. For example, suppose the enhanced nonabelian gauge algebra is $\su(2)$. This breaks to $\mathfrak{u}(1)$ along the Coulomb branch, parameterized by $u = \Tr \Phi^2$ where $\Phi$ is the $\su(2)$-adjoint scalar in the vector multiplet. Note that $u \ge 0$, so the $\su(2)$ enhancement occurs at a boundary of the moduli space. Geometrically, this corresponds to divisor shrinking to a curve \cite{Witten:1996qb}, where M5 branes wrapping the divisor describe the associated 't Hooft-Polyakov monopole string. Larger enhanced gauge symmetries occur at higher codimension in the vector multiplet moduli space.

It is interesting to compare this case with the hypermultiplet case discussed above. Consider the double cover of the moduli space, parameterized by $\Phi$, where $\Phi \to - \Phi$ generates the $\su(2)$ Weyl group. Passing through the phase transition at $\Phi = 0$ once again alters the Chern-Simons couplings, this time due to the change in sign of the wino masses:
\begin{equation} \label{eqn:shrinkingcurve}
C^+_{I J K} = C^-_{I J K} - q_I q_J q_K \,,
\end{equation}
where $q_I$ is the W-boson charge and $+$ ($-$) denotes the region where $q_I Y^I > 0$ ($q_I Y^I < 0$).\footnote{Note that $W$-bosons always come in $W^+$, $W^-$ pairs, so there will always be an even multiplicity factor in the formula eventually, e.g., a factor of $2$ in the $\su(2)$ case.} Notice the relative sign difference between~\eqref{eqn:flop} and~\eqref{eqn:shrinkingcurve}. Explicitly, this sign appears because the wino mass depends on the central charge with the opposite sign compared to the hyperino mass, but its presence can also be inferred from the fact that a long multiplet (consisting of both a vector and a hyper) has no effect on the Chern-Simons coupling.\footnote{An $\mathcal{N}_5=1$ long multiplet is the same as an $\mathcal{N}_5 =2$ vector multiplet, so this is required for consistency with enhanced supersymmetry, where the prepotential is not corrected.}

However, since the regions $\Phi > 0$ and $\Phi < 0$ are physically identified, $C^+_{I J K}$ and $C^-_{I J K}$ were already related. For example, consider the $\su(2)$ case and choose a basis where the $W^\pm$ bosons have charges $q_I = (\pm 1,0,\ldots,0)$. 
The $\su(2)$ Weyl group is then generated by
\begin{equation}
Y^0 \to - Y^0 \,, \qquad Y^i \to Y^i + \alpha^i Y^0 \,, \label{eqn:WeylZ2}
\end{equation}
for some $\alpha^i$, relating the equivalent phases $Y^0 > 0$ and $Y^0 < 0$. By a further change of basis $Y^i \to Y^i + (1/2) \alpha^i Y^0$, we set $\alpha^i = 0$.
 In this basis,\footnote{This basis is not integral unless $G=SO(3) \times U(1)^n$. For $G=SU(2)\times U(1)^{n}$ / $G = U(2) \times U(1)^{n-1}$ the $W$-bosons have charges $(\pm 2,0,\ldots)$ / $(\pm 1, \mp 1, 0, \ldots)$ in the naturally associated integral bases, where in the latter case the Weyl group exchanges the first two entries. In either case, changing to a basis where $q^{(W)}_I = (\pm1,0,0,\ldots)$ with Weyl group $Y^0 \to -Y^0$ introduces half-integral charges.} 
  Weyl invariance of the prepotential requires $C^+_{000} = - C^-_{000}$, $C^+_{00i} = C^-_{00i}$, $C^+_{0ij} = -C^-_{0ij}$, and $C^+_{ijk} = C^-_{ijk}$. Per~\eqref{eqn:shrinkingcurve}, $C^+_{000} = C^-_{000} - 2$ with the other components continuous, so we obtain $C^+_{000} = -1$ and $C^+_{0ij} = 0$, constraining the form of the prepotential near the boundary.

\bigskip

Thus, phase transitions in the vector multiplet moduli space are characterized by the appearance of massless charged hyper and/or vector multiplets, with associated changes in the prepotential given by~\eqref{eqn:flop} and/or \eqref{eqn:shrinkingcurve}.\footnote{As we saw above, long multiplets become massless at special points in the \emph{hypermultiplet} moduli space, and so do not generate phase transitions in the vector multiplet moduli space.} Because the third derivative of the prepotential fixes the Chern-Simons couplings, $\mathcal{F}_{I J K} = C_{I J K}$, no other corrections to the prepotential are possible.

In fact, when charged vector multiplets become massless, the moduli space comes to an end: the gauge group is enhanced to a nonabelian group, and the gauge quotient of the Coulomb branch by the associated Weyl group creates an orbifold boundary as we saw above in the case of an $\su(2)$ enhancement. Nonetheless, one can formally remove this boundary by considering the covering space of the Weyl-group orbifold; doing so will turn out to be convenient for determining the black hole extremality bound.

Finally, we consider the fourth possibility, in which an infinite tower of particles becomes massless. This may occur in a Lagrangian theory at infinite distance in moduli space, but it cannot happen at finite distance in the moduli space of any Lagrangian theory, hence such a phase transition is described by a non-trivial conformal field theory (CFT). Geometrically, this corresponds to a divisor shrinking to a point, so that (1) there are infinitely many BPS particles becoming massless at the phase transition, coming from M2 branes wrapped on curves within the divisor (of which there are generally infinitely many), and (2) there are also magnetically-charged BPS strings becoming tensionless at the phase transition, coming from M5 branes wrapping the divisor~\cite{Witten:1996qb}.

An in-depth discussion of such CFTs is beyond the scope of this work, but it will be important to keep this possibility in mind.

\subsection{The extended K\"ahler cone and infinite-distance boundaries} \label{subsec:extendedBoundaries}

Having understood the possible phase transitions in moduli space, we now consider the overall structure of the moduli space. First, consider a single phase, labeled ``A'', corresponding to $Y^I(\phi) \in \cR_A$ where $\cR_A$ is some connected subregion of the moduli space separated from the rest by phase transitions. In addition to its finite-distance boundaries (which are phase transitions by construction) $\cR_A$ can also have boundaries at infinite distance, corresponding geometrically to the entire Calabi-Yau collapsing to a manifold of lower dimension. Such asymptotic boundaries are always accompanied by an infinite tower of BPS particles becoming light~\cite{Grimm:2018ohb,Corvilain:2018lgw}, as required by the Swampland Distance Conjecture~\cite{Ooguri:2006in} (SDC).\footnote{It has been argued that such towers are also required by the T/sLWGC~\cite{Gendler:2020dfp}, see also~\cite{StringWGC}.}

Indeed, all of the possible boundaries of $\cR_A$ involve BPS hypers and/or vectors becoming massless. Each boundary corresponds to one of the geometric possibilities~\cite{Witten:1996qb}:
\begin{enumerate}[(1)] \pagelabel{pp:KAboundarytypes}
\item A complex curve may collapse to a point.
\item A complex divisor may collapse to either (a) a curve or (b) a point.
\item The entire Calabi-Yau may collapse to a manifold of lower dimension.
\end{enumerate}
Cases 1 and 2 were discussed already in~\S\ref{subsec:PTs}. There are massless hypermultiplets in case 1 and massless vector multiplets (and potentially also massless hypermultiplets) in case 2(a), whereas in case 2(b) an infinite tower of BPS particles becomes massless. As just stated, case 3 also results in an infinite tower of BPS particles become massless, albeit at infinite moduli space distance. Thus, there is at least one collapsing curve at every boundary of $\cR_A$. 

An elementary physical consequence of this is as follows. Let $\Kah_A$ be the cone over $\cR_A$, i.e., the set $\{\lambda r \; | \; \lambda >0 , r\in\cR_A\}$, which is the \emph{K\"ahler cone} (or ``ample cone'') associated to this phase. At each point $Y^I$ on the boundary of $\Kah_A$, there is a BPS state of charge $q_I$ associated to a collapsing curve, such that $q_I Y^I = 0$. On the other hand, since $\cR_A$ describes a single phase, all the charged BPS states must be massive throughout the interior of $\cR_A$, hence for any charge $q'_I$ admitting a BPS state, $q'_I Y^I > 0$ for all points $Y^I$ in the interior of $\Kah_A$. Let $\Mori_A$ be the \emph{Mori cone} (or ``cone of curves'') for this phase, i.e., the convex cone generated by the classes of all the holomorphic curves, equivalently by the charges $q_I$ of all the BPS states. Then the above requirements can be summarized as the statement that the K\"ahler cone $\Kah_A$ is the dual cone of the Mori cone $\Mori_A$,
\begin{equation}
 \Kah_A = \Mori_A^\ast \,, \qquad \text{where $\Mori_A^\ast = \{ Y^I \;|\; \forall q_I \in \Mori_A, Y^I q_I \ge 0  \}$.} \label{eqn:KAdualCA}
\end{equation}
(To be precise, $\Mori_A^\ast$ is the \emph{closure} of $\Kah_A$, denoted $\overline \Kah_A$. Henceforward, all cones should be understood as closed unless otherwise specified, so $\Kah_A$ means $\overline \Kah_A$, etc.)
As a corollary, $\Kah_A$ is convex, since dual cones are always convex.

These are well-known geometric facts. We include a physical ``derivation'' of them in anticipation of similar physical arguments that we will make for divisors in~\S\ref{sec:boundaries}.

Next, consider the union of all the phases connected by flops, i.e., by phase transitions involving only a finite number of hypermultiplets becoming massless. The union of the corresponding K\"ahler cones is the \emph{extended K\"ahler cone} $\Kah$:
\begin{equation}
\Kah = \bigcup_A \Kah_A \,. \label{eqn:Kdef}
\end{equation}
Like each of its components, the extended K\"ahler cone is itself convex. To see this, note that $\Kah$ is equal to the cone of movable divisors~\cite{Kawamata88}, i.e., the cone generated by the movable divisors, which is convex by definition. 

The extended K\"ahler cone $\Kah$ describes a larger moduli space consisting of one or more phases connected by flops. What happens when we approach its boundary? If we can reach the boundary inside a single phase $A$, then it must correspond to one of the cases on page~\pageref{pp:KAboundarytypes} other than a flop (case 1). Alternatively, the boundary may lie at an accumulation point of flops, requiring us to traverse infinitely many phases to reach it.\footnote{We thank Callum Brodie for pointing out this possibility to us.} Thus, each boundary of $\Kah$ is one of the following:
\begin{enumerate}[(i)] \pagelabel{pp:Kboundarytypes}
\item A divisor collapses to a curve.
\item A divisor collapses to a point.
\item The entire Calabi-Yau collapses to a manifold of lower dimension.
\item An infinite chain of flops occurs.
\end{enumerate}
Type (i) and (ii) boundaries occur at finite distance, whereas type (iii) boundaries occurs at infinite distance.

To understand type (iv) boundaries better, note that only finitely many distinct Calabi-Yau manifolds are known. Unless there are infinitely many (in particular, infinitely many with the same Hodge numbers), the chain of flops must eventually reach a phase isomorphic to an earlier one along the chain,\footnote{Typically this happens immediately, i.e., each flop in the chain connects two isomorphic phases.} hence there is a discrete symmetry acting on $\Kah$ that maps one isomorphic phase to the other. Moreover, the group $G$ of all such discrete symmetries must have infinite order (since there are infinitely many phases, but only finitely many non-isomorphic ones). Therefore, approaching such a boundary can be accomplished by acting repeatedly with an infinite-order element of $G$. This clearly entails an infinite moduli-space distance, so type (iv) boundaries also lie at infinite distance.\footnote{Note that there is no infinite tower of particles becoming light at such a boundary; rather, infinitely many particles become light at different points in the moduli space, but then become heavy again afterwards. This does \emph{not} violate the SDC because $G$ is gauged, so we are not moving asymptotically far away in the moduli space, but rather traveling many times around a closed loop.}

Finally, we consider whether it is possible to extend beyond $\Kah$ into a yet-larger vector multiplet moduli space. Clearly, we cannot move past infinite-distance boundaries, so we need only consider what happens at boundaries of types (i) and (ii). We have already seen that boundaries of type (i) are orbifold singularies in the moduli space, i.e., we can move past them, but in doing so we end up in a phase that is gauge-equivalent to our starting point, so the moduli space really does end at these boundaries. Boundaries of type (ii) (CFT boundaries) are more subtle due to their strongly-coupled nature. After compactification on a circle, non-geometric phases appear behind such boundaries, but at least some of these disappear in the 5d decompactifcation limit~\cite{Witten:1996qb}. Whether other non-geometric phases persist in 5d in a difficult question that we do not attempt to answer here. If not, then $\Kah$ (or, more precisely, $\Kah/G$) represents the entire vector multiplet moduli space.\footnote{If such non-geometric phases \emph{do} exist, then they could affect our analysis of the extremality bound. Our lack of knowledge about them introduces some uncertainty, but this is the \emph{same} uncertainty that is already introduced by the non-trivial CFT behind which the non-geometric phase would appear.}

\subsection{On the possibility of wall crossing} \label{subsec:wallcrossing}

So far, we have implicitly assumed that the BPS states that are present in one part of moduli space persist elsewhere in the moduli space. However, this is not necessarily the case, as BPS states can appear and disappear upon crossing walls of marginal stability, see, e.g.,~\cite{Denef:2002ru,Denef:2000nb,Denef:2000ar}. To our knowledge, no systematic treatment of wall crossing in 5d theories is available in the literature,\footnote{We thank Edward Witten for discussions on this point.} and we will not attempt one here. However, the simple reasoning discussed below suggests that such phenomena are strongly constrained.

Upon crossing a wall of marginal stability, a BPS particle may disintegrate into several constituent BPS particles. This should occur when the binding energy of the constituents goes to zero, i.e., when the central charges become aligned. However, in 5d the central charge $\zeta_q = q_I Y^I$ is \emph{real}, so alignment between BPS particles of charges $q, q'$ occurs when $\zeta_q$ and $\zeta_{q'}$ have the same sign. In particular, a bound state between $q$ and $q'$ should disintegrate as soon as the binding energy goes to zero, i.e., when either $\zeta_q$ or $\zeta_{q'}$ changes sign and the two constituents come into alignment. This necessarily occurs when either $\zeta_q = 0$ and $\zeta_{q'} = 0$, so this type of wall crossing phenomenon should only occur at a phase transition where massless BPS particles appear.

Alternately, the final state may involve BPS strings. In particular, as the wall of marginal stability is approached a BPS string loop carrying electric charge dissolved in its worldvolume may expand away from the core of the BPS state until it decouples as a long, straight BPS string infinitely far away. To avoid a divergent contribution to the mass of the bound state, this process can only occur if the tension of the BPS string in question vanishes on the wall of marginal stability. Therefore, this second kind of wall crossing phenomenon should only occur at a phase transition where a tensionless BPS string appears.

So far, we have inferred that 5d wall crossing should only occur at phase transitions. This is already dramatically different from 4d wall crossing, which does not require light particles or strings. To go further, we compare the circle compactification these 5d phenomena with what is known about 4d wall crossing. The latter invariably involves constituent BPS particles with a non-trivial Dirac pairing, $q_e q_m' - q_m q_e' \ne 0$. However, BPS particles in 5d reduce to purely electrically charged BPS particles in 4d, so the first type of wall crossing discussed above---involving only BPS particles---evidently does not occur.

Therefore, we conclude that 5d BPS particles should only disappear via wall crossing at a phase transition involving tensionless strings. In particular, no wall crossing should occur in the interior of the extended K\"ahler cone $\Kah$. While the arguments given above are not rigorous, we will see these expectation borne out in examples in~\S\ref{sec:conifoldexample}, see in particular~\S\ref{sssec:GMSVgvInvs}, \S\ref{sssec:GVrevisit}.

\section{The Black Hole Extremality Bound} \label{sec:ExtBound}

To establish the black hole extremality bound, we construct extremal black hole solutions to the action~\eqref{eqn:5dsugra}. This can be done systematically using the attractor mechanism~\cite{Ferrara:1995ih,Cvetic:1995bj,Strominger:1996kf,Ferrara:1996dd,Ferrara:1996um} and the fake superpotential formalism~\cite{Ceresole:2007wx,Andrianopoli:2007gt,Andrianopoli:2009je,Andrianopoli:2010bj,Trigiante:2012eb}. Below we review the results we will need, omitting many details that are not important for our analysis. We refer interested readers to, e.g.,~\cite{Heidenreich:2020upe,BenBH} for a fuller treatment.

For the time being, we assume that the vector multiplet moduli space has no finite-distance boundaries and ignore the interaction between the vector- and hyper-multiplet moduli spaces that occurs during geometric transitions (i.e., at the intersection of Coulomb and Higgs branches). We will discuss those complications below after explaining this simplified scenario.

\subsection{Extremality from fake superpotentials} \label{subsec:fakeW}

With the gauge choice in~\cite{Heidenreich:2020upe}, static spherically symmetric black hole solutions take the form:
\begin{align}
  d s^2 &=  - e^{2 \psi(r)} f (r) d t^2 +
  e^{- \psi(r)}  \left[ \frac{d r^2}{f (r)} + r^2 d
  \Omega^2_3 \right] \,, &  f (r) &= 1 - \frac{r_h^2}{r^2}\,, \nonumber \\ 
  F^I &= \frac{g_5^2}{2 \pi^2} a^{I J}(\phi(r))q_J  \frac{e^{2\psi(r)}}{r^3} d t \wedge d r \,,  
\end{align}
where $r_h \geqslant 0$ is a constant and  $\psi(r)$, $\phi^i(r)$ solve a system of second-order ODEs plus a first-order constraint. 
In terms of $z \df \frac{\sqrt{2} \kappa_5 g_5}{4 \pi^2} \frac{1}{r^2}$ and $\cQ^2(\phi) \df a^{I J}(\phi) q_I q_J$,\footnote{Note that, in comparison to~\cite{Heidenreich:2020upe}, $z^{\text{(here)}} = \frac{\sqrt{2} \kappa_5 g_5}{2 \pi^2} z^{\text{(there)}}$ and $\cQ^2_{\text{(here)}} = \frac{3 (2 \pi^2)^2}{2 \kappa_5^2 g_5^2} \cQ^2_{\text{(there)}}$.}
\begin{subequations} \label{eqn:BHeqns}
\begin{align}
  \frac{d}{dz} [f \dot{\psi}] &= \frac{1}{3} \cQ^2(\phi) e^{2
  \psi}\,,  \label{eqn:psizeqn} \\
  \frac{d}{dz} [f \dot{\phi}^i] + f \Gamma^i_{\; j k} (\phi)  \dot{\phi}^j
  \dot{\phi}^k &= \frac{1}{2} \mathfrak{g}^{i j} (\phi) \cQ^2_{, j} (\phi)
 e^{2 \psi} \,, \label{eqn:phizeqn}  \\
  3 \dot{\psi} (f \dot{\psi} + \dot{f}) + f \mathfrak{g}_{i j} (\phi) 
  \dot{\phi}^i \dot{\phi}^j &= \cQ^2(\phi) e^{2 \psi} \,, \label{eqn:zcons} 
\end{align}
\end{subequations}
where dots denote $z$-derivatives, $\Gamma^i_{\; j k}(\phi)$ is the Levi-Civita connection for the metric on moduli space $\mathfrak{g}_{i j}(\phi)$ and now $f=1-z/z_h$. The ADM mass of the black hole is
\begin{equation}
M_{\text{BH}} =  \frac{3 g_5}{\sqrt{2}\kappa_5} \biggl[ - \dot{\psi}_{\infty} + \frac{1}{2 z_h} \biggr] \,, \label{eqn:MBH}
\end{equation}
where the $\infty$ subscript denotes the quantity evaluated at spatial infinity ($r=\infty$, equivalently $z = 0$).

Solutions to~\eqref{eqn:BHeqns} with a smooth horizon and $r_h >0$ are ``nonextremal'' (finite temperature and entropy) black holes, whereas those with $r_h = 0$ are ``quasiextremal'' (zero temperature or zero entropy) black holes in the language of~\cite{Heidenreich:2020upe}. In particular, a smooth horizon requires $\psi(r)$ to be finite at $r=r_h$ in the nonextremal case, whereas $r^2 e^{-\psi(r)}$ must remain finite as $r\to 0$ in the quasiextremal case. In the latter case, $\psi \sim - \log z$ near the horizon, so that $\dot \psi \sim - \frac{1}{z} < 0$. Likewise, in the former case, \eqref{eqn:psizeqn} implies $-\frac{1}{z_h}\dot{\psi}(z_h) = \frac{1}{3} a^{I J}(\phi_h) q_I q_J e^{2 \psi_h} > 0$, so $\dot{\psi}(z_h) < 0$. An important consequence is as follows: combining the first and third equations of~\eqref{eqn:BHeqns}, we obtain
\begin{equation}
\ddot{\psi} = \dot{\psi}^2 + \frac{1}{3} \mathfrak{g}_{i j} \dot \phi^i \dot \phi^j \ge 0 \,. \label{eqn:ddotpsi}
\end{equation}
Thus, since $\dot{\psi} < 0$ at the horizon we conclude that $\dot{\psi} < 0$ for all $z \le z_h$ for all solutions with a smooth horizon.\footnote{This implies, for instance, that $M_{\rm BH} \ge 0$, see~\eqref{eqn:MBH}.} In fact, under mild assumptions, $\dot{\psi} < 0$ together with regularity outside the horizon is a \emph{sufficient} condition for a nonextremal solution to have a smooth horizon~\cite{BenBH}.

Our primary interest is in ``extremal'' solutions, i.e., those with the least mass for any specified charge. Since by assumption they cannot radiate, extremal solutions are always quasiextremal (though the converse is not always true~\cite{BenBH}). Thus, to construct extremal solutions and determine the extremality bound, we first examine the quasiextremal case more closely.

We use the fake superpotential formalism~\cite{Ceresole:2007wx,Andrianopoli:2007gt,Andrianopoli:2009je,Andrianopoli:2010bj,Trigiante:2012eb}. Suppose we are given a function on the moduli space $W(\phi)$ that satisfies the first-order PDE:
\begin{equation}
\cQ^2(\phi) = \mathfrak{g}^{i j} W_{,i}(\phi) W_{,j}(\phi) + \frac{1}{3} W(\phi)^2 \,. \label{eqn:WPDE}
\end{equation}
Then it is straightforward to check that the first-order gradient-flow ODEs
\begin{equation}
\dot{\psi} = - \frac{1}{3} e^{\psi} W(\phi)\,, \qquad \dot{\phi}^i = - e^{\psi} \mathfrak{g}^{i j} W_{,j}\,,
\label{eq:fogf}
\end{equation}
 imply the black hole equations~\eqref{eqn:BHeqns} with $r_h=0$.
 
 Note that the function $W(\phi)$ need not be globally defined in the moduli space; it is sufficient for $W(\phi)$ to be defined in a small neighborhood of a solution to~\eqref{eq:fogf}. Indeed, due to the non-linearity of the equation, local, real solutions to~\eqref{eqn:WPDE} may hit branch cuts and become complex when extended throughout the rest of the moduli. Nonetheless, global, real solutions to~\eqref{eqn:WPDE} are especially interesting, as we will see.
 
A ``black hole'' solution resulting from~\eqref{eq:fogf} may or may not have a smooth horizon. Suppose first that the gradient flow approaches a local minimum of $W(\phi)$ and that $W=W_0>0$ at the minimum. We obtain $\dot{\psi} \simeq -\frac{1}{3} e^\psi W_0$ near the end of the flow (as $z \to \infty$), hence $e^{-\psi} \sim W_0 z/3$, and the horizon is smooth, with finite area
\begin{equation}
A = 2 \pi^2 \lim_{r \to 0} r^3 e^{-\frac{3}{2}\psi(r)} = \kappa_5^2\, (W_0 / 3)^{3/2} \,.
\end{equation}
If instead the minimum occurs at $W_0 = 0$, then the horizon shrinks to zero area and becomes singular. However, such solutions still satisfy $\dot{\psi} < 0$ outside the horizon, and so can be realized as \emph{limits} of smooth nonextremal solutions. Despite the singular horizon, we nonetheless refer to such solutions as quasiextremal ``black holes,'' since they lie at the boundary of the space of nonextremal black hole solutions with smooth horizons.\footnote{For example, extremal black holes in Einstein-Maxwell-Dilaton theory fall into this class.}

By comparison, if $W_0 < 0$ then $\dot \psi$ becomes positive at finite $z$ and a smooth horizon is impossible. Moreover, such behavior will never arise as a limit of a smooth black hole solution, and solutions of this kind are not black holes in any sense.

\begin{figure}
\centering
\includegraphics[width=7cm]{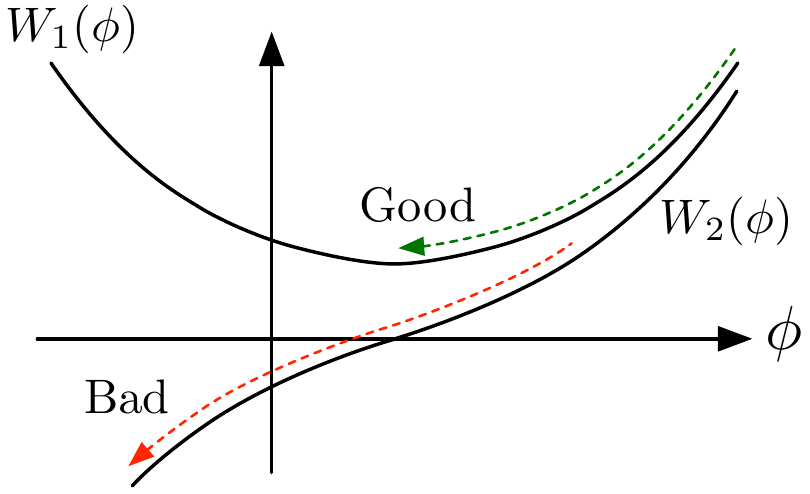}
\caption{Good flows remain entirely in the region $W(\phi) \ge 0$, whereas bad flows cross into the region $W(\phi) < 0$.}
\label{fig:goodbad}
\end{figure}

Note that the gradient flow can end not just at local minima of $W(\phi)$ but also at other critical points (except local maxima). Moreover, the flow could run off to infinity in a noncompact moduli space.\footnote{This occurs, e.g., for extremal black holes in Einstein-Maxwell-Dilaton theory.} So long as $W(\phi) \ge 0$ along the entire flow, we obtain a bona fide quasiextremal black hole solution, either with a smooth horizon or lying at the boundary of the space of nonextremal black holes with smooth horizons. For this reason, we call such flows (with $W(\phi) \ge 0$ everywhere along the flow) ``good'' flows, in contrast with the ``bad'' flows that cross into a region where $W(\phi) < 0$; the difference is illustrated in Figure~\ref{fig:goodbad}.

Obtaining a good flow places requirements on $W(\phi)$ that go beyond~\eqref{eqn:WPDE} and depend on the choice of vacuum $\phi_\infty^i$. In general, no single choice of $W(\phi)$ will satisfy the requirements for all possible $\phi_\infty^i$, but let us suppose to the contrary that a global-defined $W(\phi)$ exists for which all flows are good. We call such a $W(\phi)$ (distinguished amongst all the solutions to~\eqref{eqn:WPDE}) a (global) ``fake superpotential.'' Solutions to~\eqref{eqn:WPDE} are highly nonunique, but if a fake superpotential exists then it is unique, as we now prove.
  
Given a fake superpotential $W(\phi)$, we obtain a quasiextremal black hole solution of mass
\begin{equation}
M_{\rm BH}^{\text{q-ext}} = \frac{g_5}{\sqrt{2}\kappa_5} W(\phi_\infty) \,,
\end{equation}
for any choice of asymptotic vacuum $\phi_\infty^i$. Such solutions are, in fact, extremal, i.e., they saturate the extremality bound, which takes the form:
\begin{equation}
M_{\rm BH} \ge \frac{g_5}{\sqrt{2}\kappa_5} W(\phi_\infty) \,. \label{eqn:Wbound}
\end{equation}
To prove that~\eqref{eqn:Wbound} is the extremality bound, consider the functional
\begin{equation}
  \hat{S}[\psi,\phi] = \frac{g_5}{\sqrt{2} \kappa_5} \int_0^{z_h} \biggl[ \frac{3}{2} (f \dot{\psi} + \dot{f})^2
  + \frac{1}{2} f^2 \mathfrak{g}_{i j}(\phi)  \dot{\phi}^i  \dot{\phi}^j + \frac{1}{2} e^{2
  \psi} \cQ^2(\phi) \biggr] d z \,.
\end{equation}
This reduces to
\begin{equation}
  \hat{S} = \frac{3g_5}{\sqrt{2} \kappa_5} \biggl( \int_0^{z_h} d_z \biggl[ \frac{1 + f}{2} f
  \dot{\psi} \biggr] d z + \frac{1}{2 z_h} \biggr) = \frac{3g_5}{\sqrt{2} \kappa_5} \biggl( -
  \dot{\psi}_{\infty} + \frac{1}{2 z_h} \biggr) = M_{\text{BH}} \,,
\end{equation}
evaluated on a solution to the black hole equations~\eqref{eqn:BHeqns}, where we assume a smooth horizon to justify dropping the boundary term at $z=z_h$.\footnote{Since $\ddot{\psi} \ge 0$, see \eqref{eqn:ddotpsi}, the boundary term vanishes for any solution satisfying $\dot{\psi}(z) < 0$, including the singular quasiextremal solutions that are limits of smooth solutions discussed previously.} 
Now rewrite $\hat{S}$ as 
\begin{multline}
  \hat{S} = \frac{g_5}{\sqrt{2}\kappa_5} \int_0^{z_h} \biggl( \frac{3}{2} \biggl[ f \dot{\psi} +
  \dot{f} + \frac{1}{3} e^{\psi} W(\phi) \biggr]^2 + \frac{1}{2} \mathfrak{g}_{i j} 
  [ f \dot{\phi}^i +e^{\psi} \mathfrak{g}^{i k}  W_{, k} ] [ f
  \dot{\phi}^j + e^{\psi} \mathfrak{g}^{j l}  W_{, l}] \biggr) d z  \\ -
   \frac{g_5}{\sqrt{2}\kappa_5} f e^{\psi} W(\phi) \biggr|_0^{z_h} \,.
\end{multline}
Since the first line is non-negative, we obtain the bound
\begin{equation}
\hat{S} \ge -\frac{g_5}{\sqrt{2}\kappa_5} f e^{\psi} W(\phi) \biggr|_0^{z_h} = \frac{g_5}{\sqrt{2}\kappa_5} W(\phi_\infty) \,,
\end{equation}
again assuming a smooth horizon. Thus, the bound~\eqref{eqn:Wbound} holds for any solution with a smooth horizon, and the quasiextremal solutions arising from $W(\phi)$ gradient flow are extremal.\footnote{Moreover, it is straightforward to show that all extremal solutions to the black hole equations~\eqref{eqn:BHeqns} are $W(\phi)$ gradient flow solutions.} By the same argument, any two fake superpotentials $W_1(\phi)$ and $W_2(\phi)$ must satisfy $W_1(\phi) \ge W_2(\phi)$ and $W_2(\phi) \ge W_1(\phi)$, hence $W_1(\phi) = W_2(\phi)$.

\subsection{Finite distance boundaries in the moduli space} \label{subsec:finitedistance} 

Clearly a fake superpotential, if it exists, must be nonnegative everywhere in the moduli space. What about the converse? Is a global, nowhere-negative solution $W(\phi)$ to~\eqref{eqn:WPDE} necessarily a fake superpotential? If the moduli space is geodesically complete then this is true because there is a gradient flow starting at any point $\phi^i(0) = \phi^i_\infty$ extending all the way to the horizon $z\to\infty$, and obviously $W(\phi) \ge 0$ along this flow. Thus, the fake superpotential, if it exists, is the unique global nowhere-negative solution to the PDE~\eqref{eqn:WPDE}.

\begin{figure}
\centering
\begin{subfigure}{0.48\textwidth}
\centering
\includegraphics[height=4cm]{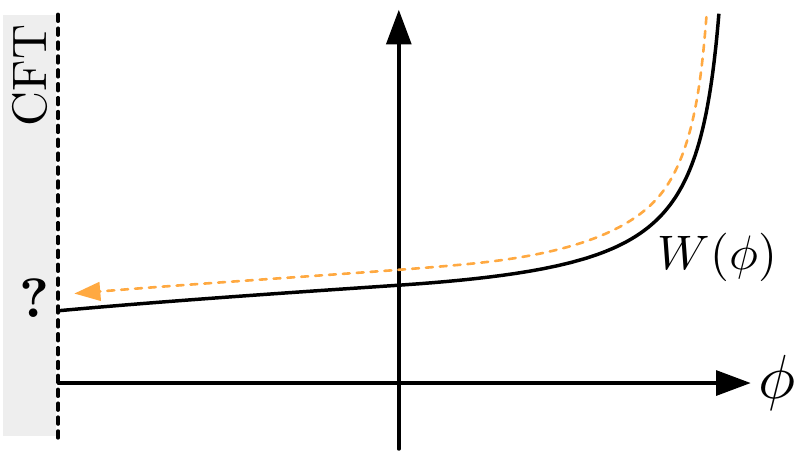}
\caption{An indeterminate flow.} \label{sfig:Indeterminate}
\end{subfigure}
\begin{subfigure}{0.48\textwidth}
\centering
\includegraphics[height=4cm]{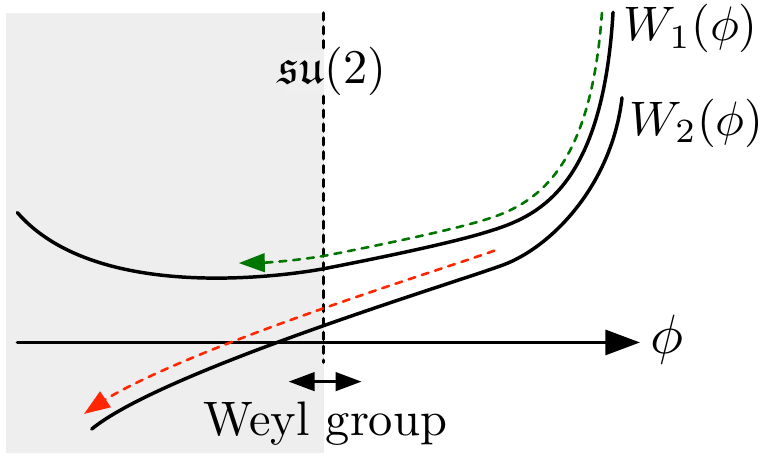}
\caption{Resolving indeterminate flows.} \label{sfig:ResolvedFlow}
\end{subfigure}
\caption{\subref{sfig:Indeterminate} Indeterminate flows reach a finite-distance boundary of the moduli space at finite $z$. \subref{sfig:ResolvedFlow} Indeterminate flows reaching an $\mathfrak{su}(2)$ boundary can be resolved into good and bad flows by passing to the Weyl reflection, unless they subsequently reach yet another finite-distance boundary.}
\end{figure}

However, this need not be true if the moduli space has boundaries at finite distance. In particular, $W(\phi)$ gradient flows can reach a finite-distance boundary in finite $z$. A black hole solution of this kind will enter a regime controlled by the physics associated to the boundary in question a finite distance outside its horizon. Describing such a solution is beyond the scope of the effective action~\eqref{eqn:5dsugra} that we have employed so far; we need further input about the physics near the boundary. For this reason, we will call flows that reach a finite distance boundary in the moduli space at finite $z$ ``indeterminate,'' see Figure~\ref{sfig:Indeterminate}. 

Per the discussion in~\S\ref{subsec:PTs}, finite-distance boundaries in the moduli space are due to either (1) a nonabelian enhancement of the gauge group or (2) a non-trivial CFT. In the former case, we can shift perspective to a covering space of the moduli space where the boundary disappears. Applying the methods of the previous section to this covering space, indeterminate flows can be resolved into good or bad flows, see Figure~\ref{sfig:ResolvedFlow}, unless they reach a different finite-distance boundary that still remains in the covering space. If this further boundary is another nonabelian enhancement, then we take a yet-bigger cover, etc., so all indeterminate flows are resolved into good or bad flows, unless there are CFT boundaries.

Resolving flows that reach a CFT boundary of the moduli space is much more difficult, and we will not attempt it in the present work. This inevitably generates some ambiguity in theories with such CFT boundaries, but does not prevent us from drawing certain unambiguous conclusions as we will see.

\subsection{Interaction with hypermultiplets}

Naively, since the vector and hypermultiplet moduli spaces decouple, and the gauge kinetic matrix $a_{I J}(\phi)$ depends only on the former, charged black hole solutions will not be affected by the presence or interactions of the hypermultiplet moduli. However, this is not quite right: vectors \emph{do} interact directly with \emph{charged} hypermultiplets. Due to the attractor mechanism, vevs for these charged hypers (which partially Higgs the gauge group) may not penetrate all the way to the black hole horizon, i.e., the black hole horizon could sit inside a patch of gauge-symmetric vacuum. If so, then the black hole background would include gradients for both hyper and vector multiplets, with a geometric transition occurring in the internal Calabi-Yau manifold at some finite distance from the horizon.

While this phenomenon merits further study, it cannot occur for BPS black holes, whose mass can only depend on vector multiplet moduli. As such, we defer further consideration of it to a future work.

\section{BPS versus Extremal}\label{sec:BPSBHE}

To resolve the conifold conundrum raised in \S\ref{sec:CC}, we were led to the question of when the BPS and black hole extremality bounds agree, or equivalently when superextremal particles must be BPS. It is clear that the two bounds do not always agree, even in theories with BPS bounds. We already saw one example of this in \S\ref{sec:CC}: namely, there exist non-BPS extremal black holes in heterotic string theory on $T^k$. This raises the question: when exactly do the extremality bounds and BPS bounds coincide?

With the technology of~\S\ref{sec:SUGRA}, \S\ref{sec:ExtBound} in hand, we are now in a position to answer this question in 5d supergravity theories. Crucially, spherically symmetric BPS solutions are solutions to the gradient flow equations~\eqref{eq:fogf} with $W(\phi)$ equal to the central charge $\zeta_q(\phi) = q_I Y^I(\phi)$. Thus, BPS black holes of charge $q_I$ exist (do not exist) at a point $\phi = \phi_0$ in moduli space if the $\zeta_q$ gradient flow starting at $\phi = \phi_0$ is a good (bad) flow.\footnote{Recall that a ``good'' flow is one for which $W(\phi) \geq 0$ everywhere along the flow, whereas a ``bad'' flow has $W(\phi) <0$ somewhere along the flow.} (Likewise, anti-BPS black holes exist when $-\zeta_q(\phi)$ flows are good.) When the flow is indeterminate---i.e., reaching a CFT boundary at finite $z$---then we cannot determine whether BPS black holes of charge $q_I$ exist without better understanding the corresponding strongly-coupled physics. 

Based on this, we will show:
\begin{enumerate}[I.] \pagelabel{pp:zetaCases}
\item If $\zeta_{q}$ has a critical point anywhere in moduli space then either $\zeta_q$ or $-\zeta_q$ is a fake superpotential, and the charge-$q$ BPS and black hole extremality bounds agree \emph{everywhere} in moduli space. 
\item If $\zeta_{q}$ vanishes anywhere in moduli space then all its flows are at best indeterminate, and some are bad, so the BPS and black hole extremality bounds \emph{disagree} in at least part of the moduli space. 
\item If $\zeta_{q}$ neither vanishes nor has a critical point in the moduli space, then all its flows are indeterminate.
\end{enumerate}
To reach these conclusions, we study the flow equations for $\zeta_q$ as well as the structure of its critical points.

\subsection{Homogeneous coordinates and dual coordinates}\label{ssec:coordinates}

The form of the gradient flow equations~\eqref{eqn:WPDE} with $W(\phi) = \zeta_q(\phi)$ will depend on what parameterization $\phi^i$ we choose for the hypersurface $\mathcal{F}[Y] = 1$ within $\mathbb{R}^{n+1}$. We can avoid such a choice by instead describing the moduli space projectively in terms of ``homogeneous coordinates'' $Y^I$ subject to the equivalence $Y^I \cong \lambda Y^I$ for $\lambda > 0$. The prepotential $\mathcal{F}[Y]$ is a weight-3 function on this projective space, meaning that $\mathcal{F} \to \lambda^3 \mathcal{F}$ under $Y^I \to \lambda Y^I$, hence the slice $\mathcal{F}[Y] = 1$ meets each equivalence class at a single point, and there is a one-to-one correspondence between the projective and inhomogeneous ($\mathcal{F}=1$) descriptions.

There is a natural weight-$(-2)$ metric on this projective space:
\begin{equation}
h_{I J} \df \frac{1}{\mathcal{F}^2} \mathcal{F}_I \mathcal{F}_J - \frac{1}{\mathcal{F}} \mathcal{F}_{I J} = - \frac{\partial^2 L}{\partial Y^I \partial Y^J}\,, \qquad L[Y] \df \log \mathcal{F}[Y] \,, \label{eqn:hdef}
\end{equation}
which is positive-definite because $a_{IJ} = \mathcal{F}^{2/3} h_{I J}$ is positive-definite. Therefore, $L[Y]$ is a concave function across the moduli space. It is likewise natural to define the weight-$(-1)$ ``dual coordinates''
\begin{equation}
\Yd_I \df \frac{1}{\mathcal{F}} \mathcal{F}_I = \frac{\partial L}{\partial Y^I} \,. \label{eqn:Ytdef}
\end{equation}

As the name suggests, dual coordinates are equally good projective coordinates on the moduli space. To show this, consider the function
\begin{equation}
\cG[Y, \tilde Y] := Y^I \Yd_I - L[Y],
\end{equation}
with $Y^I$ and $\Yd_I$ viewed as formally independent variables. Holding $\Yd_I$ fixed, the Hessian matrix $\partial_I \partial_J \cG  = - \partial_I \partial_J L = h_{IJ}$ is positive-definite, so $\cG[Y]$ is a convex function of $Y^I$ for fixed $\tilde Y_I$. Moreover, the domain of $\cG[Y]$---the extended K\"ahler cone $\Kah$ of the Calabi-Yau threefold---is convex. As a convex function defined on a convex subset $\Kah \subset \mathbb{R}^{n+1}$, $\cG[Y]$ has a unique minimum. If the minimum occurs in the interior of $\Kah$, then it is located at a critical point of $\cG[Y]$ and
\begin{equation}
\partial_I \cG = \Yd_I - \partial_I L = 0 \,,
\end{equation}
at the minimum, so the minimum is located precisely at the point $Y^I = Y^I[\Yd]$ solving the equation $\Yd_I = \frac{\partial L}{\partial Y^I}$, and the solution to this equation is unique as a consequence of the uniqueness of the minimum. Thus, there is a one-to-one mapping between the interior of $\Kah$ and the interior of $\Kv$, where $\Kv$ is the cone of dual coordinates, defined as the image of $\Kah$ under the map $\Yd = \Yd[Y]$.\footnote{The map between the boundaries of $\Kah$ and $\Kv$ is more complicated, as we will see in~\S\ref{sec:boundaries}.} (Likewise, $\Kv_A$ denotes the image of $\Kah_A$ under the map $\Yd = \Yd[Y]$, so that $\Kv = \bigcup_A \Kv_A$.)

Note that
\begin{equation}
\tilde{L}[\Yd] \df \min_Y \cG[Y, \Yd] = Y^I \Yd_I - L[Y]\biggr|_{Y=Y[\Yd]} \,,
\end{equation}
is precisely the Legendre transform of the concave function $L[Y]$ with $Y^I$ and $\Yd_I$ the corresponding conjugate variables. Some further, easily-derived identities are
\begin{subequations}
\begin{align}
h_{I J} Y^J &= \Yd_I, & \Yd_I Y^I &= h_{I J} Y^I Y^J = 3 \,, &
\frac{\partial \Yd_I}{\partial Y^J} &= -h_{IJ} \,, \label{eqn:partialJtYI} \\
\frac{\partial(\log \mathcal{F})}{\partial \Yd^I} &= -Y^I \,, & \frac{\partial^2(\log \mathcal{F})}{\partial \Yd^I \partial \Yd^J} &= - \frac{\partial Y^I}{\partial \Yd_J} = h^{IJ} \,, \label{eqn:dualYhdef}
\end{align}
\end{subequations}
where $Y_I \Yd^I = 3$ implies $L[\Yd] = 3 - \log \mathcal{F}$ and~\eqref{eqn:dualYhdef} is analogous to~\eqref{eqn:hdef}, \eqref{eqn:Ytdef}.

\subsection{BPS flows are straight lines in dual coordinates} \label{subsec:straightline}

Functions on the vector multiplet moduli space lift to weight-zero functions of the homogeneous coordinates $Y^I$. In particular, expressed in homogeneous coordinates, the central charge becomes
\begin{equation}
  \zeta_q[Y] = \frac{q_I Y^I}{\mathcal{F}^{1 / 3}} \,.
  \label{centralchargehom}
\end{equation}
Therefore
\begin{equation}
  \frac{\partial \zeta_q}{\partial Y^I} = \frac{1}{\mathcal{F}^{1 / 3}} q_I -
  \frac{\zeta_q}{3}  \Yd_I , \label{eqn:dzeta}
\end{equation}
and the critical points of $\zeta_q$ are the rays $\Yd_I \propto \pm q_I$ (whenever these lie within $\Kv$).\footnote{The constant of proportionality drops out, as required by the projective description; in particular $\Yd_I \propto \pm q_I$ implies $q_I \Yd_J = q_J \Yd_I$, so that $\frac{\zeta_q}{3}\Yd_I = \frac{1}{3 \mathcal{F}^{1/3}} q_J Y^J \Yd_I = \frac{1}{3 \mathcal{F}^{1/3}} q_I Y^J \Yd_J = \frac{1}{\mathcal{F}^{1/3}} q_I$.} Note that $\zeta_q > 0$ ($\zeta_q < 0$) at the critical point $\Yd_I \propto + q_I$ ($\Yd_I \propto - q_I$). Moreover, at the critical point 
\begin{equation}
\partial_I \partial_J \zeta_q = -
  \frac{1}{3\mathcal{F}^{1 / 3}}  (q_I  \Yd_J + q_J  \Yd_I) +
  \frac{\zeta_q}{9}  \Yd_I \Yd_J + \frac{\zeta_q}{3} h_{I J} =  \frac{\zeta_q}{3}  \biggl( h_{I J} -
  \frac{1}{3}  \Yd_I  \Yd_J \biggr) \,,
\end{equation}
where we use $\frac{1}{\cF^{1/3}} q_I = \frac{\zeta_q}{3} \Yd_I$ in the second equality. The matrix $h_{I J} - \frac{1}{3}  \Yd_I  \Yd_J$ is positive semi-definite with a single zero eigenvector $Y^I$ corresponding to the projective rescaling $Y^I \to \lambda Y^I$ under which $\zeta_q$ is unchanged. Thus, the critical point is a local minimum (maximum) for $\zeta_q > 0$ ($\zeta_q < 0$).\footnote{This was derived in~\cite{Chou:1997ba}. However, it does not immediately follow as suggested there that the local minimum or maximum is unique without additional information about the behavior of $\zeta_q$ at the boundaries of moduli space. We thank Edward Witten for pointing this out to us. In our treatment, the uniqueness of the minimum or maximum is a consequence of the invertibility of the dual coordinate map $\Yd_I=\Yd_I[Y]$ (shown above).} There are no critical points at $\zeta_q = 0$ except in the trivial, $q_I = 0$ case in which $\zeta_q$ vanishes everywhere.

Suppose now that we have some extremal, BPS black hole solution. Such a solution solves the first-order gradient-flow ODEs \eqref{eq:fogf} with a fake superpotential $W(\phi) = \zeta_q(\phi)$:
\begin{equation}
\dot{\psi} = - \frac{1}{3} e^{\psi} \zeta \,, \qquad \dot{\phi}^i = - e^{\psi} \mathfrak{g}^{i j} \zeta_{,j}\,.
\end{equation}
Thus,
\begin{equation}
\frac{d}{d z} Y^I(\phi) = \dot{\phi}^i \partial_i Y^I = - e^{\psi} \mathfrak{g}^{i j} \partial_i Y^I \partial_j Y^J \frac{\partial \zeta}{\partial Y^J} = - e^{\psi} a^{I J} \frac{\partial \zeta}{\partial Y^J} \,, \label{eqn:dYdz}
\end{equation}
using~\eqref{eqn:inversea} as well as $Y^I \frac{\partial \zeta_q}{\partial Y^I} = 0$ (since $\zeta_q$ has weight 0). By matching weights, it is straightforward to lift \eqref{eqn:dYdz} to homogeneous coordinates, giving
\begin{equation}
\dot{\psi} = - \frac{1}{3} e^{\psi} \zeta_q \,, \qquad \dot{Y}^I = - e^{\psi} h^{IJ} \zeta_{q,J}\,.
\label{bechom}
\end{equation}
In fact, there is an ambiguity at this stage, because numerous paths through the space of homogeneous coordinates project to the same path through the moduli space; corresponding to this, an arbitrary multiple of $Y^I$ could be added to the second equation in~\eqref{bechom}. To remove this ambiguity, we demand that $\mathcal{F}$ is constant along the flow; then, since~\eqref{bechom} already implies $\dot{\cF}= \cF_{I} \dot{Y}^I = -e^{\psi} \cF \Yd_I h^{I J} \zeta_{q,J}=-e^{\psi} \cF Y^J \zeta_{q,J} = 0$, there is no extra term.

Applying~\eqref{eqn:partialJtYI} to~\eqref{bechom}, the corresponding flow equation in dual coordinates is
\begin{equation}
\dot{\tilde Y}_I =  e^\psi \zeta_{q,I} = \frac{e^\psi}{\cF^{1/3}} q_I - \frac{e^\psi}{3} \tilde Y_I \zeta_q = \frac{e^\psi}{\cF^{1/3}} q_I + \dot{\psi} \tilde Y_I .
\end{equation}
Therefore,\footnote{This solution and its analogs in other dimensions are well known in the literature on the attractor mechanism~\cite{Ferrara:1995ih,Cvetic:1995bj,Strominger:1996kf,Ferrara:1996dd,Ferrara:1996um}, see, e.g.,~\cite{Larsen:2006xm}.}
\begin{equation}
\frac{d}{dz} [e^{- \psi} {\tilde Y}_I ] =  \frac{q_I}{\cF^{1/3}}  \qquad \Longrightarrow \qquad \tilde Y_I = e^{\psi} \biggl[\tilde Y_I^{\infty} + z \frac{q_I}{\cF^{1/3}} \biggr], \label{eqn:zetaFlows}
\end{equation}
where we exploit $\dot{\cF} = 0$ to integrate the flow, and $\Yd_I^{\infty}$ is an integration constant specifying the starting point of the flow at $z=0$ (infinitely far from the black hole).\footnote{Note that $\psi(z)$ is fixed implicitly by the constraints $\psi(z=0) = 0$ and $\mathcal{F}(z) = \mathcal{F}(z=0)$; there is no need for further integration to determine it.}

\begin{figure}
\centering
\begin{subfigure}{0.5\textwidth}
\centering
\includegraphics[height=38mm]{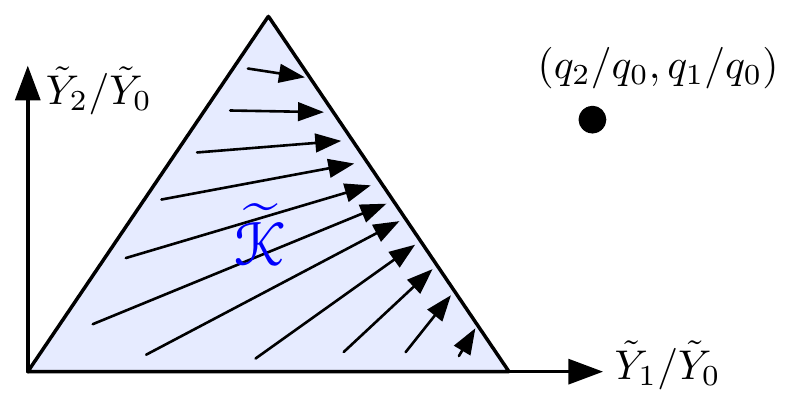}
\caption{Flows are straight lines in $\Kv$.} \label{sfig:straightflow}
\end{subfigure}
\begin{subfigure}{0.47\textwidth}
\centering
\includegraphics[height=38mm]{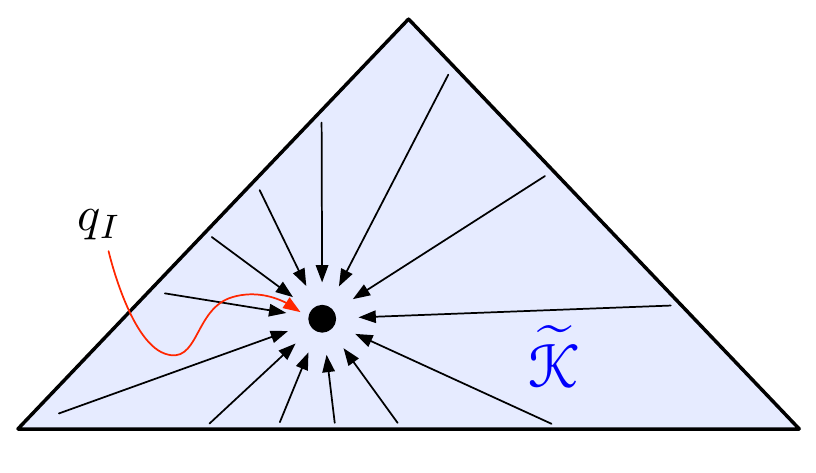}
\caption{All flows are good when $q_I \in \Kv$.} \label{sfig:insideflow}
\end{subfigure}
\caption{\subref{sfig:straightflow} Gradient flows of $\zeta_q$ are straight lines in the cone of dual coordinates $\Kv$ (up to projective rescaling), directed towards the ray $\Yd_I \propto + q_I$. \subref{sfig:insideflow} If $q_I$ lies inside $\Kv$, which is convex, then any gradient flow beginning inside $\Kv$ will end at this attractor point, and a BPS black hole solution is guaranteed.}
\label{fig:flow}
\end{figure}

Thus, when plotted in the cone of dual coordinates $\Kv$, the gradient flow is simply a straight line (up to projective rescaling) directed towards the ray $\Yd_I \propto +q_I$. This simple result is illustrated in Figure~\ref{sfig:straightflow}.

\subsection{Consequences for BPS black holes} \label{subsec:BPSconsequences}

We now revisit the three cases discussed on page~\pageref{pp:zetaCases}. First, suppose that $\zeta_q$ has a critical point inside the moduli space (case I).
 Applying charge conjugation ($q_I \to - q_I$) as needed, we can assume that $\zeta_q > 0$ at the critical point without loss of generality. Then, from any starting point $\Yd_I^\infty$ inside $\Kv$, there is a flow to the critical point $\Yd_I^h \propto +q_I$ given explicitly by~\eqref{eqn:zetaFlows}, unless the flow passes through the boundary of the moduli space along the way. In fact, $\Kv$ is convex (as shown in~\S\ref{sec:boundaries}), and a straight line between any two points inside a convex set lies inside the set, hence the flow lies entirely inside the moduli space, as illustrated in Figure~\ref{sfig:insideflow}. By construction, this is a good flow, so there are extremal charge-$q_I$ BPS black holes everywhere in the moduli space.

Note that, since $\zeta_q$ decreases along the gradient flow, it follows that a $\zeta_q > 0$ critical point is a \emph{global minimum} of $\zeta_q$. Likewise, a $\zeta_q < 0$ critical point is a global maximum of $\zeta_q$. Thus, $\zeta_q$ has at most one critical point in the moduli space. Another corollary is that for any two points in the moduli space, $Y^I_{(1)} \Yd_I^{(2)} \ge 0$ where the inequality can only be saturated if both points lie on the boundary.\footnote{To show this, choose $q_I = \Yd_I^{(2)}$ and evaluate $\zeta_{q}[Y] \ge \zeta_{q}[Y_{(2)}] \ge 0$ at $Y^I = Y_{(1)}^I$.} Thus, $\Kv \subseteq \Kah^\ast$ or equivalently $\Kah \subseteq \Kv^\ast$, where $\Kah^\ast$ ($\Kv^\ast$) is the dual cone of $\Kah$ ($\Kv$).

If $q_I$ lies \emph{on the boundary} of $\Kv$ then the above reasoning does not directly apply. However, $\zeta_q \ge 0$ is non-negative across $\Kah$ as a consequence of $\Kah \subseteq \Kv^\ast$, and moreover flows originating in the interior of $\Kah$ remain in the interior for all finite $z$ per~\eqref{eqn:zetaFlows}, hence all of them remain good flows.

Next, suppose that $\zeta_q$ vanishes somewhere in the interior of the moduli space (case II). This cannot occur at a critical point, since all critical points have $\zeta_q \ne 0$, so $\zeta_q$ must change sign in the neighborhood of this point. By the above reasoning, this implies that there is no critical point anywhere else in the moduli space either, so case I and case II are mutually exclusive. Thus, $\zeta_q$ flows in case II can be divided into three classes: (1) those that cross into the $\zeta_q < 0$ region (bad flows), (2) those that reach a finite-distance boundary within the $\zeta_q > 0$ region (indeterminate flows), and (3) those that approach an infinite-distance boundary within the $\zeta_q > 0$ region (good flows). However, as shown in~\S\ref{sec:boundaries}, any flow approaching an infinite-distance boundary is a bad flow unless $q_I$ lies on the boundary itself (which is case I), so there are no flows in the third class. Thus, all flows in case II are either (1) bad or (2) indeterminate, and at least some flows are bad (those passing through the $\zeta_q = 0$ locus).

Finally, suppose that $\zeta_q$ neither vanishes nor has a critical point in the moduli space (case III). This is much like the previous case, but there are no bad flows, so all flows are indeterminate. In particular, this can only happen when there are finite-distance boundaries.

\bigskip

Recalling that $\zeta_q$ has a critical point if either $q_I$ or $-q_I$ is in $\Kv$ and noting that $\zeta_q$ changes sign in the moduli space if and only if neither $q_I$ nor $-q_I$ is in $\Kah^\ast$, we can summarize these results as follows:
\begin{enumerate}[I.] \pagelabel{pp:coneCases}
\item If $q_I$ or $-q_I$ lies within or on the boundary of $\Kv$, then the charge-$q_I$ BPS and black hole extremality bounds agree everywhere in moduli space.
\item If neither $q_I$ nor $-q_I$ is in $\Kah^\ast$,\footnote{Per~\eqref{eqn:KAdualCA} and~\eqref{eqn:Kdef}, we can also write $\Kah^\ast = \Mori_{\cap}$, where $\Mori_{\cap}=\bigcap_A \Mori_A$ is the intersection of the Mori cones for all the phases.} then the BPS and black hole extremality bounds \emph{disagree} in at least part of the moduli space; 
elsewhere, the answer depends on physics at the finite-distance boundaries of moduli space (if present).
\item If $q_I$ or $-q_I$ is in $\Kah^\ast$ but not in $\Kv$, then the answer depends on physics at the finite-distance boundaries of moduli space.
\end{enumerate} 
Note that indeterminate flows that reach an $\su(2)$-enhancement boundary can be further resolved by moving to a cover of the moduli space, as discussed in~\S\ref{subsec:PTs}, \S\ref{subsec:finitedistance}. The above statements can therefore be refined to resolve some of the ambiguous outcomes in such cases, but we will not attempt a general treatment here. (For an example and further discussion, see~\S\ref{ssec:KMV} and \cite{CornellGV}.)

Finally, note that if some of the indeterminate $\zeta_q$ flows appearing in case II resolve to good flows then the resulting BPS black holes must disappear via wall crossing somewhere in the interior of $\Kah$, since some of the $\zeta_q$ flows are bad. Per the discussion in~\S\ref{subsec:wallcrossing}, we do not expect this to occur, hence we expect the BPS and extremality bounds to disagree everywhere outside $\Kah^\ast$, regardless of where we are in the moduli space.

\section{Moduli Space Boundaries and the Effective Cone}\label{sec:boundaries}

In the previous section, we saw that the cone of dual coordinates $\Kv$ plays an important role in delineating which charges $q_I$ do or do not admit BPS black hole solutions. What, then, is the geometric interpretation of $\Kv$? Let $\Eff$ be the cone generated by the effective divisors, which is unchanged by flops and thus constant across the extended K\"ahler cone. In this section, we argue that $\Kv$ is the dual of the effective cone,
$\Kv = \Eff^\ast$, where $\Eff^\ast$ is also the cone of movable curves~\cite{Boucksom13}.\footnote{The connection between $\Kv$ and $\Eff^\ast$ was noticed independently in \cite{Lanza:2021qsu}.}
 Since dual cones are always convex, $\Kv$ is convex, a fact already used in~\S\ref{subsec:BPSconsequences}. (However, the subcone $\Kv_A \subseteq \Kv$ corresponding to a single phase $A$ is not in general convex.)

In particular, for any effective divisor $\tilde{q}^I \in \Eff$ there is a BPS string of charge $\tilde{q}^I$ coming from an M5 brane wrapping the corresponding divisor. Per~\eqref{eqn:stringBPS}, the tension of this string is
\begin{equation}
T(\phi) = \frac{\tilde{g}_5}{\sqrt{2} \kappa_5} |\tilde{\zeta}_{\tilde q}(\phi)| , \qquad \tilde{\zeta}_{\tilde q}(\phi) = \mathcal{F}^{1/3} \tilde{q}^I \Yd_I(\phi) \,, \label{eqn:stringBPS2}
\end{equation}
rewritten in homogeneous coordinates. Following the discussion in~\S\ref{subsec:PTs}, tensionless strings can only appear at the boundaries of the moduli space, so $\tilde{q}^I \Yd_I > 0$ for any $\tilde{q}^I \in \Eff$ and any $\Yd_I$ in the \emph{interior} of $\Kv$. Thus, $\Kv \subseteq \Eff^\ast$, and to show that $\Kv = \Eff^\ast$ we need only show that for any $\Yd_I$ on the \emph{boundary} of $\Kv$ there is some $\tilde{q}^I$ in $\Eff$ such that $\Yd_I \tilde{q}^I = 0$.\footnote{To be precise, there is a divisor $\tilde{q}^I$ in the \emph{closure} of $\Eff$, also known as a \emph{pseudo}effective divisor. We defer further discussion of whether the divisor is effective to~\cite{StringWGC}.}

To do so, we recall from page~\pageref{pp:Kboundarytypes} that $\Kah$ (and by extension $\Kv$) has four types of boundaries
\begin{enumerate}[(i)] \pagelabel{pp:Kboundarytypes2}
\item A divisor collapses to a curve (an $\su(2)$ boundary)
\item A divisor collapses to a point (a CFT boundary)
\item The entire Calabi-Yau collapses to a manifold of lower dimension (an asymptotic boundary)\footnote{Occasionally such boundaries are actually periodic boundaries, i.e., when there is an infinite-order automorphism mapping one phase to itself (e.g., for the Schoen threefold~\cite{Grassi93}). However, this will not matter for our present analysis.}
\item An infinite chain of flops occurs (a periodic boundary).
\end{enumerate}
Types (i) and (ii) lie at finite distance, and each is accompanied by a tensionless string (coming from an M5 brane wrapping the divisor in question), so $\tilde{q}^I \in \Eff$ satisfying $\Yd_I \tilde{q}^I = 0$ certainly exists.

Thus, only boundaries of types (iii) and (iv) (lying at infinite distance) remain to be considered. To show that a divisor $\tilde{q}^I \in \Eff$ satisfying $\Yd_I \tilde{q}^I = 0$ exists at each such boundary, note that $\Kah \subseteq \Eff$.\footnote{In particular, the K\"ahler cone (ample cone) $\Kah_A$ for each phase  lies within the effective cone, $\Kah_A \subseteq \Eff$. Since the effective cone is constant across $\Kah = \bigcup_A \Kah_A$, this implies that $\Kah \subseteq \Eff$.} Now consider an infinite-distance point $Y^I$ on the boundary of $\Kah$ and let $\Yd_I$ be the corresponding infinite-distance point on the boundary of $\Kv$. We will show that $\Yd_I Y^I = 0$, hence we can pick $\tilde{q}^I \propto +Y^I$ as our divisor, so that $\tilde{q}^I \tilde Y_I = 0$ by construction and $\tilde{q}^I \in \Kah \subseteq \Eff$. 

How can this be consistent with $\Yd_I Y^I = 3$, as claimed in~\eqref{eqn:partialJtYI}? In approaching an infinite-distance boundary, we take a limit; some components of $Y^I$ and/or $\Yd_I$ may blow up in this limit, in which case we use a projective rescaling to keep the largest component of each finite and nonzero in the limit, preserving information about the directions of both $Y^I$ and $\Yd_I$. Let us denote such rescaled boundary coordinates as $Y^I_\ast$ and/or $\Yd_{\ast I}$. The required rescalings may be different for $Y^I$ and $\Yd_I$, hence $\Yd_{\ast I} Y^I_\ast \ne 3$. We will see that in fact $\Yd_{\ast I} Y^I_\ast = 0$ at all infinite-distance boundaries. (Conversely, $\Yd_{\ast I} Y^I_\ast > 0$ at all finite distance boundaries.)

\bigskip

Before showing this, we note a useful corollary: if $\Yd_I^{\text{in}}$ and $\Yd_I^{\text{out}}$ are points inside and outside $\Kv$, respectively, and the straight line between them crosses the boundary of $\Kv$ at an infinite-distance point $\Yd_{\ast I}$ corresponding to $Y_\ast^I \in \Kah$, then $Y^I_\ast \Yd_I^{\text{out}} < 0$.\footnote{Writing $\Yd_{\ast I} = \alpha \Yd_I^{\text{in}} + \beta \Yd_I^{\text{out}}$ for $\alpha, \beta > 0$, apply $Y_\ast^I \Yd_{\ast I} = 0$ along with $Y_\ast^I \Yd^{\text{in}}_{I} > 0$ (since $\Kv \subseteq \Kah^\ast$).} In particular, this implies (setting $\Yd_I^{\text{out}} = q_I$) that for $q_I$ outside $\Kv$, $\zeta_q$ flows approaching an asymptotic boundary are bad flows, another fact used in~\S\ref{subsec:BPSconsequences}. 

Another corollary is that every boundary of $\Kv$ must be either a finite-distance boundary or a boundary of $\Kah^\ast$. Each finite-distance boundary is associated to a string $s$ of charge $\tilde{q}^I_{(s)}$ becoming tensionless, and therefore lies in the plane $\tilde{q}^I_{(s)} \Yd_I = 0$ with $\tilde{q}^I_{(s)} \Yd_I > 0$ in the interior of $\Kv$. Thus, each boundary of $\Kv$ is either a boundary of $\mathcal{H}_s = \{\Yd | \tilde{q}^I_{(s)} \Yd_I \ge 0 \}$ for some such $s$ or a boundary of $\Kah^\ast$. Since moreover $\Kv \subseteq \Kah^\ast$ and $\Kv \subseteq \mathcal{H}_s$, we find
\begin{equation}
\Kv = \Kah^\ast \cap \bigcap_{s \in \mathfrak{B}} \mathcal{H}_s\,, \label{eqn:KvFormula}
\end{equation}
where $\mathfrak{B}$ is the set of strings that become tensionless at finite-distance boundaries. In particular, $\Kv = \Kah^\ast$ in the absence of finite-distance boundaries.

\subsection{Asymptotic boundaries} \label{subsec:asymptotic}

Boundaries of type (iii) are \emph{asymptotic} boundaries of the extended K\"ahler cone: the geodesic distance from any point inside moduli space to such a boundary is infinite.
To generate an infinite distance, the
prepotential must vanish at such a boundary, see, e.g,~\cite{Heidenreich:2020ptx}. In particular, expressed in homogeneous coordinates the metric on moduli space is
\begin{equation}
d s^2 = g_{I J} d Y^I d Y^J, \qquad g_{I J} = \frac{2}{3} \frac{\cF_I \cF_J}{\cF^2} - \frac{\cF_{I J}}{\cF} = h_{I J} - \frac{1}{3} \Yd_I \Yd_J ,
\end{equation}
where $g_{I J}$ has one zero eigenvalue corresponding to the fact that projectively equivalent points have vanishing distance between them. Since the boundaries of $\Kah$ lie at finite $Y^I$ (up to projective equivalence), $Y^I$ has bounded variation along a path approaching such a boundary, and the moduli space distance can only diverge if components of $g_{I J}$ diverge. However, for any phase $A$ adjoining the asymptotic boundary, $\cF$ is a cubic polynomial in $Y^I$ within $A$,\footnote{Note that such a phase must exist per the discussion in~\S\ref{subsec:extendedBoundaries}.} so $\cF_I$ and $\cF_{I J}$ remain finite for finite $Y^I$, and we must have $\cF \to 0$ at the boundary for components of $g_{I J}$ to diverge and generate an infinite distance there.

Thus, $\cF = 0$ at an asymptotic boundary. Since $\cF_I Y^I = 3 \cF \to 0$ as $Y^I$ approaches a finite point $Y_\ast^I$ on the asymptotic boundary and $\Yd_I \propto \cF_I$, we naively conclude that $\Yd_{\ast I} Y_\ast^I = 0$ at such a boundary, as promised above. This argument is too quick, however, because in some cases $\cF_I$ also vanishes at the boundary, so a further rescaling is required to extract its direction.

For example, consider a theory with one vector multiplet ($n = h^{1,1} - 1 = 1$). By a change of basis, we take the asymptotic boundary to lie at $Y^I \propto Y^I_\ast = \bigl(\begin{smallmatrix}1\\0\end{smallmatrix}\bigr)$, after which the
prepotential in the phase adjoining the asymptotic boundary takes the general form:
\begin{equation}
  \mathcal{F}= \beta (Y^0)^2 Y^1 + \gamma Y^0 (Y^1)^2 + \lambda (Y^1)^3 .
\end{equation}
There are two cases to consider. If $\beta \ne 0$ then we set $\beta = 1$ by rescaling $Y^0$, after which we set $\gamma = 0$ by shifting $Y^0 \rightarrow Y^0 - \frac{1}{2} \gamma Y^1$. The prepotential is now $\cF= (Y^0)^2 Y^1 + \lambda (Y^1)^3$, so that
\begin{equation}
  \mathcal{F}_I = \begin{pmatrix}
    2 Y^0 Y^1 \\
   (Y^0)^2+ 3 \lambda (Y^1)^2
  \end{pmatrix} \rightarrow \begin{pmatrix}
    0\\
    1
  \end{pmatrix} \quad \text{as} \quad Y^I \rightarrow Y^I_\ast =
  \begin{pmatrix}
    1\\
    0
  \end{pmatrix} .
\end{equation}
Thus, the corresponding asymptotic boundary of $\Kv$ lies at $\Yd_{\ast I} = \bigl(\begin{smallmatrix}0\\1\end{smallmatrix}\bigr)$, and $\Yd_{\ast I} Y^I_\ast = 0$ at the boundary, in agreement with the naive argument above.

On the other hand, if $\beta = 0$, then we need $\gamma \ne 0$ to obtain a positive-definite gauge-kinetic matrix $a_{I J}$, so we set $\gamma = 1$ by rescaling $Y^1$, after which we set $\lambda = 0$ by shifting $Y^0 \rightarrow Y^0 -
\lambda Y^1$, leaving $\cF= Y^0 (Y^1)^2$, so that
\begin{equation}
\mathcal{F}_I =
  \begin{pmatrix}
    (Y^1)^2\\
    2 Y^0 Y^1
  \end{pmatrix} \rightarrow 2 Y^0 Y^1 \begin{pmatrix}
    0\\
    1
  \end{pmatrix} \quad \text{as} \quad Y^I \rightarrow Y^I_\ast =
  \begin{pmatrix}
    1\\
    0
  \end{pmatrix} .
\end{equation}
Scaling away the overall prefactor of $\cF_I$, we obtain the same result as before: the asymptotic boundary of $\Kv$ lies at $\Yd_{\ast I} = \bigl(\begin{smallmatrix}0\\1\end{smallmatrix}\bigr)$, so that $\Yd_{\ast I} Y^I_\ast = 0$ at the boundary. However, this no longer directly follows from $\cF_I Y^I = 3 \cF = 0$, which is a trivial consequence of $\cF_I=0$ at the boundary.

To formulate a general argument accounting for such cases, note that along any path from $Y^I = Y_1^I$ to $Y^I = Y_2^I$ through $\Kah$,
\begin{equation}
\Delta(\cF[Y]) = \log \cF[Y_2] - \log \cF[Y_1] = \int_{Y_1^I}^{Y_2^I} (\partial_I \log \cF) d Y^I = \int_{Y_1^I}^{Y_2^I} \Yd_I d Y^I \,. \label{eqn:genYdYarg}
\end{equation}
We choose a path of finite coordinate length in homogeneous coordinates $Y^I$, with $Y_1^I=Y^I_\ast$ lying on an asymptotic boundary and the rest of the path lying in the interior of $\Kah$. Then $\Delta(\log \cF) = \log \cF[Y_2] - \log 0 = \infty$, and the integral must diverge. Since $Y^I$ has bounded variation, this requires $\Yd_I$ to diverge somewhere along the path, which can only happen at the boundary of $\Kah$. Thus, $\Yd_I$ must diverge at the asymptotic boundary. We conclude that $\Yd_{\ast I} = \lambda \Yd_I$ with $\lambda \to 0$ as $Y^I \to Y^I_\ast$, and so $\Yd_{\ast I} Y^I = 3 \lambda \to 0$ at the boundary, as promised.

\bigskip

We illustrate this general result by considering some examples with $n>1$ vector multiplets. The general form of the prepotential in a phase adjoining the asymptotic boundary is
\begin{equation}
  \mathcal{F}= \alpha (Y^0)^3 + \beta_i  (Y^0)^2 Y^i + \gamma_{i j} Y^0 Y^i
  Y^j + \lambda_{i j k} Y^i Y^j Y^k \,,
\end{equation}
for $i=1,\ldots,n$. Choosing $Y^I_\ast = \delta^{I, 0}$ to be a point on the asymptotic boundary, we must have $\alpha = 0$. We then obtain
\begin{equation}
  \mathcal{F}_I = \begin{pmatrix}
    2 \beta_i Y^0 Y^i + \gamma_{i j} Y^i Y^j\\
    \beta_i  (Y^0)^2 + 2 \gamma_{i j} Y^0 Y^j + 3 \lambda_{i j k} Y^j Y^k
  \end{pmatrix} \rightarrow \begin{pmatrix}
    0 \\
    \beta_i
  \end{pmatrix} \quad \text{as} \quad Y^I \rightarrow Y^I_\ast = \begin{pmatrix}1\\0\end{pmatrix} .
\end{equation}
Thus, provided that $\beta_i \ne 0$, $\Yd_{\ast I} = \bigl(\begin{smallmatrix}0\\\beta_i \end{smallmatrix}\bigr)$, and $\Yd_{\ast I} Y^I_\ast = 0$ as expected.

If, however, $\beta_i = 0$, then a subtlety arises: what \emph{point} we reach on the boundary of $\Kv$ depends on what \emph{direction} we approach $Y_\ast^I \in \Kah$ from. In particular, consider the linear path $Y^I(\varepsilon) = \bigl(\begin{smallmatrix}1\\ \varepsilon \hat{Y}^i \end{smallmatrix}\bigr)$ approaching $Y^I_\ast$ as $\varepsilon \to 0$. Then,
\begin{equation}
  \mathcal{F}_I = \begin{pmatrix}
    \gamma_{i j} Y^i Y^j\\
    2 \gamma_{i j} Y^0 Y^j + 3 \lambda_{i j k} Y^j Y^k
  \end{pmatrix} \rightarrow \varepsilon \begin{pmatrix}
    0\\
    2 \gamma_{i j}  \hat{Y}^j
  \end{pmatrix} \quad \text{as} \quad Y^I \rightarrow Y^I_\ast = \begin{pmatrix}1\\0\end{pmatrix}.
\end{equation}
Note that $\gamma_{i j} \hat{Y}^i \hat{Y}^j > 0$ is required to ensure $\cF>0$ and $h_{00} > 0$ along the path away from $Y_\ast^I$, so in particular $\gamma_{i j} \hat{Y}^j  \ne 0$, and we end up at $\Yd_{\ast I} = \bigl(\begin{smallmatrix}0\\ \gamma_{i j}  \hat{Y}^j \end{smallmatrix}\bigr)$, which depends not only on $Y^I_\ast$ but also on the direction $\hat{Y}^i$ from which we approach it. Regardless of where we end up, however, we recover $\Yd_{\ast I} Y^I_\ast = 0$ on the asymptotic boundary, as implied by the general argument given above.

The above example implies that the mapping between the boundaries of $\Kah$ and $\Kv$ is not one-to-one. This is true in both directions: just as a single point on an asymptotic boundary of $\Kah$ can correspond to a higher-dimensional region on the boundary of $\Kv$, likewise an entire region within an asymptotic boundary of $\Kah$ can correspond to a single point on the boundary of $\Kv$. For example, suppose a facet of $\Kah$ forms an asymptotic boundary. By a change of basis, we take the facet to be $Y^0 = 0$, so that
\begin{equation}
  \mathcal{F}= \alpha (Y^0)^3 + \beta_i (Y^0)^2  Y^i + \gamma_{i j} Y^0 Y^i Y^j,
\end{equation}
within a phase adjoining the facet. Approaching the boundary, we obtain
\begin{equation}
  \mathcal{F}_I = \begin{pmatrix}
    3 \alpha (Y^0)^2 + 2 \beta_j Y^0 Y^j + \gamma_{j k} Y^j Y^k\\
    \beta_i  (Y^0)^2 + 2 \gamma_{i j} Y^0 Y^j
  \end{pmatrix} \rightarrow \gamma_{j k} Y^j Y^k \begin{pmatrix}
    1\\
    0
  \end{pmatrix} \quad \text{as} \quad Y^I \rightarrow Y^I_\ast = \begin{pmatrix}0\\Y^i\end{pmatrix} .
\end{equation}
For $n>1$, we need $\gamma_{i j} \ne 0$ to obtain a positive-definite gauge kinetic matrix $a_{I J}$. This implies that $\gamma_{j k} Y^j Y^k \ne 0$ for generic $Y^i$, so we reach the point $\Yd_{\ast I} = \bigl(\begin{smallmatrix}1\\ 0 \end{smallmatrix}\bigr)$ on the boundary of $\Kv$, regardless of which part of the boundary facet of $\Kah$ we are approaching. Thus, in this example many different points $Y_\ast^I = \bigl(\begin{smallmatrix}0\\ Y^i \end{smallmatrix}\bigr)$ on the boundary of $\Kah$ correspond to the same point $\Yd_{\ast I} = \bigl(\begin{smallmatrix}1\\ 0 \end{smallmatrix}\bigr)$ on the boundary of $\Kv$. Regardless, we always obtain $\Yd_{\ast I} Y^I_\ast = 0$, in agreement with the general argument given previously.

In fact, in the above examples, this complicated mapping between the boundaries of $\Kah$ and $\Kv$ is a direct consequence of $\Yd_{\ast I} Y^I_\ast = 0$ and $\Kv \subseteq \Kah^\ast$. For instance, if the boundary of $\Kah$ contains a facet $q_I Y^I = 0$, then $q_I$ lies at a corner of $\Kah^\ast$, i.e., for any $Y_\ast^I$ in the interior of the facet, $q_I' \in \Kah^\ast$ satisfying $q_I' Y_\ast^I = 0$ must be parallel to $q_I$. In particular, if this facet is an asymptotic boundary, then $\Yd_{\ast I} \in \Kv \subseteq \Kah^\ast$ satisfies $\Yd_{\ast I} Y_\ast^I = 0$, implying $\Yd_{\ast I} \propto q_I$, so that the interior of the facet corresponds to a single point $\Yd_{\ast I}$ in dual coordinates (up to projective equivalence). The same argument implies that if a facet of $\Kv$ forms an asymptotic boundary then its interior corresponds to a single point $Y^I_\ast$ in homogeneous coordinates (again up to projective equivalence).

\subsection{Periodic boundaries} \label{subsec:periodic}

Boundaries of type (iv) are \emph{periodic boundaries} of the extended K\"ahler cone: as argued in \S\ref{subsec:extendedBoundaries}, such boundaries---naively involving infinitely many phases and flops between them---are necessarily accompanied by an infinite-order discrete gauge symmetry $G$ that periodically identifies different phases in $\Kah$, leaving only finitely-many distinct phases up to the periodic identification. Since the spectrum of particle masses and string tensions measured in 5d Planck units must be periodic, such boundaries are fundamentally different from asymptotic boundaries. For example, no tower of light states appears ``near'' the boundary; in fact, being ``near'' to the boundary is not a gauge-invariant notion, since the discrete gauge symmetry maps points with homogeneous coordinates arbitrarily close the boundary of $\Kah$ to others well inside $\Kah$.

Despite these differences, certain properties of these boundaries are closely analogous to those of asymptotic boundaries. For example, $\cF \to 0$ as we approach a periodic boundary of $\Kah$ while keeping the homogeneous coordinates finite, $Y^I \to Y^I_\ast$, and therefore $\Yd_{\ast I} Y^I_\ast = 0$ at the boundary by the same general argument as was given in the previous section. Likewise, the map between $\Kah$ and $\Kv$ is not one-to-one at periodic boundaries, much like at asymptotic boundaries.

To see that $\cF \to 0$ at a periodic boundary, we make use of the infinite order discrete symmetry $G$ that necessarily accompanies such a boundary. Any quantum symmetry of the supergravity action~\eqref{eqn:5dsugra} must map the charge lattice into itself, i.e.,
\begin{equation}
A^I \to (A^{I})' = \Lambda^I_{\; J} A^J , \quad \Lambda^I_{\; J} \in GL(n+1,\mathbb{Z}) , \qquad \text{so that} \quad q_I \to q_I' = (\Lambda^{-1})^J_{\; I} q_J \,.
\end{equation}
The associated central charges~\eqref{eqn:centralcharge} must also be mapped to each other, implying that\footnote{One could also consider $Y^I \to (Y^I)' = -\Lambda^I_{\; J} Y^J$, which must be combined with parity to leave the Chern-Simons term invariant. For example, the combination of charge conjugation ($A^I \to - A^I$) with parity is such a symmetry, regardless of what $\cF[Y]$ is. By composing with this always-present CP symmetry, we can fix $Y^I \to (Y^I)' = +\Lambda^I_{\; J} Y^J$ for all elements of $G$ without loss of generality.}
\begin{equation}
Y^I \to (Y^I)' = \Lambda^I_{\; J} Y^J ,
\end{equation}
where the $\cF[Y] = 1$ slice must be mapped to itself, or equivalently $\cF[Y'] = \cF[Y]$ when written in homogeneous coordinates. These necessary conditions are also sufficient to ensure the invariance of the action, hence we conclude that each element of $G$ must take the form
\begin{equation}
Y^I \to (Y^I)' = \Lambda^I_{\; J} Y^J , \quad \text{for} \quad \Lambda^I_{\; J} \in GL(n+1,\mathbb{Z}) , \qquad \text{where} \quad \cF[Y'] = \cF[Y] . \label{eqn:generalG}
\end{equation}

Now fix a fundamental domain $\FundK_{G}$ for $G$ that lies inside $\Kah$, so that any point inside $\Kah$ can be mapped uniquely to a point inside $\FundK_{G}$ by some element of $G$, and
choose some integral point $Y^I_{(0)} \in \mathbb{Z}^{n+1}$ within $\FundK_{G}$. The elements of $g\in G$ map $Y^I_{(0)}$ to an infinite set of integral points $Y^I_{(g)}$ within $\Kah$, where in particular these points accumulate near the periodic boundaries of $\Kah$ due to the periodic identifications there. Since there are only finitely many integral points with components bounded by a given integer, $|Y^I| \le k$, this implies that the components of $Y^I_{(g)}$ grow without bound as we approach the periodic boundaries of $\Kah$, hence to keep them finite we need a projective rescaling $Y^I_\ast = \lambda Y^I_{(g)}$ with $\lambda \to 0$ as we approach the boundary. Per~\eqref{eqn:generalG}, $\cF[Y^I_{(g)}] = \cF[Y^I_{(0)}]$, so $\cF[Y^I_\ast] =  \lambda^3 \cF[\lambda Y^I_{(0)}] \to 0$ as we approach the boundary, hence $\cF[Y] \to 0$ as $Y^I$ approaches a finite point $Y^I_\ast$ on the periodic boundary, as promised.

With this information in hand, the general argument following~\eqref{eqn:genYdYarg} can now be repeated without modification to show that $\Yd_{\ast I} Y^I_\ast = 0$ at a periodic boundary.

\bigskip

To illustrate this general and rather abstract reasoning, we now discuss a concrete example with a single vector multiplet ($n = h^{1,1} - 1 = 1$). Let the K\"ahler cone $\Kah_{\rm I}$ for ``phase I'' of this theory be $X, Y \ge 0$, with a prepotential
\begin{equation}
\cF^{\rm (I)} = \frac{a}{6} X^3 + \frac{a k}{4} X^2 Y + \frac{a k}{4} X Y^2 + \frac{a}{6} Y^3 \,, \label{eqn:symmF}
\end{equation}
in this phase, where $a, k$ are positive integers (at least one of them even) with $k>2$.\footnote{For example, M-theory compactified on the Calabi-Yau threefold given by the complete intersection of five bidegree $(1,1)$ hypersurfaces in $\mathbb{P}^4 \times \mathbb{P}^4$ has this form for $a=5$ and $k=4$. We thank Callum Brodie for bringing this example to our attention.} At both $X=0$ and $Y=0$ there are flops, each with $N = \frac{(k+1)(k-2)^2}{2} a$ hypermultiplets becoming massless. Passing through the flop at $Y=0$, we reach ``phase II'', with prepotential:
\begin{equation}
\cF^{\rm (II)} =  \frac{a}{6} X^3 + \frac{a k}{4} X^2 Y + \frac{a k}{4} X Y^2 - \frac{a}{6} \biggl(1+\frac{k^2(k-3)}{2}\biggr) Y^3 \,.
\end{equation}
Changing variables to $X' = X + k Y$, $Y' = - Y$, this becomes:
\begin{equation}
\cF^{\rm (II)} =  \frac{a}{6} (X')^3 + \frac{a k}{4} (X')^2 Y' + \frac{a k}{4} X' (Y')^2 + \frac{a}{6} (Y')^3 \,,
\end{equation}
which matches~\eqref{eqn:symmF} upon exchanging $X\leftrightarrow X'$ and $Y \leftrightarrow Y'$. Thus, the low-energy effective field theory admits a symmetry
\begin{equation}
\begin{pmatrix}X \\ Y\end{pmatrix} \longrightarrow \begin{pmatrix}X' \\ Y' \end{pmatrix} = \begin{pmatrix}1 & k \\ 0 & -1 \end{pmatrix} \begin{pmatrix}X \\ Y\end{pmatrix} \,,
\end{equation}
exchanging phases I and II. Moreover, phase I admits a symmetry $X\leftrightarrow Y$ exchanging its two boundaries. Let us assume that both symmetries are gauged. Then we obtain a gauged discrete symmetry group $G$ generated by
\begin{equation}
\begin{pmatrix}0 & 1 \\ 1 & 0 \end{pmatrix}\,, \qquad \begin{pmatrix}1 & k \\ 0 & -1 \end{pmatrix} \,.
\end{equation}
Both generators have order two, but they do not commute with each other. In particular,
\begin{equation}
\Lambda = \begin{pmatrix}1 & k \\ 0 & -1 \end{pmatrix} \begin{pmatrix}0 & 1 \\ 1 & 0 \end{pmatrix} = \begin{pmatrix}k & 1 \\ -1 & 0 \end{pmatrix} \,,
\end{equation}
has infinite order, and maps phase I to phase II, phase II to phase III (obtained by passing through the flop at $X' = 0$ in phase II), and so on.

\begin{figure}
\centering
\begin{subfigure}{0.48\textwidth}
\centering
\includegraphics[width=5cm]{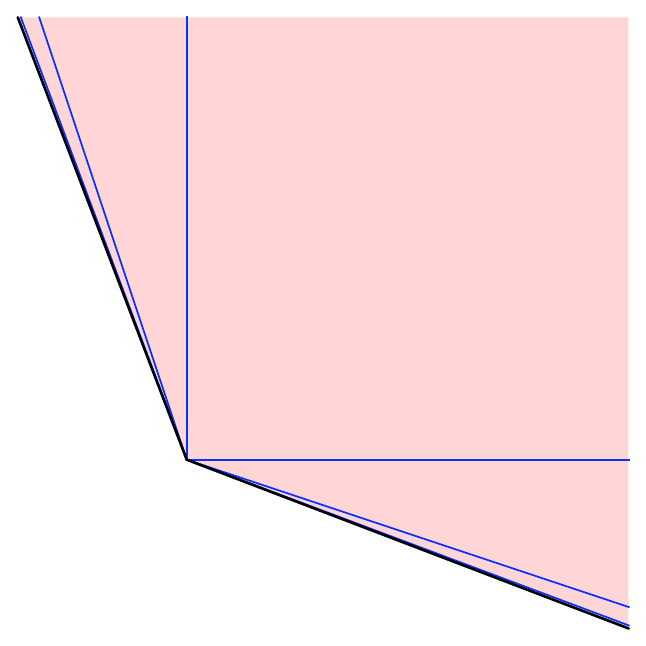}
\caption{The extended K\"ahler cone $\Kah$.} \label{sfig:PeriodicK}
\end{subfigure}
\begin{subfigure}{0.48\textwidth}
\centering
\includegraphics[width=5cm]{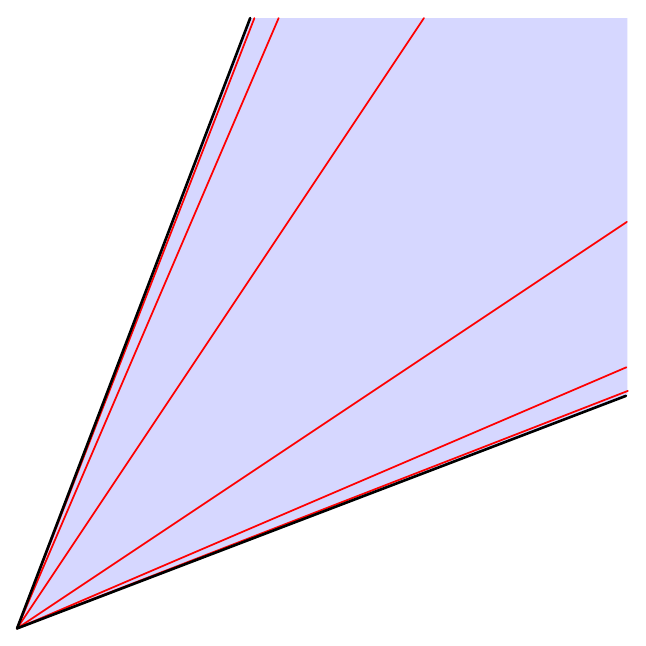}
\caption{The cone of dual coordinates $\Kv$.} \label{sfig:PeriodicKv}
\end{subfigure}
\caption{\subref{sfig:PeriodicK} The extended K\"ahler cone $\Kah$ and \subref{sfig:PeriodicKv} the cone of dual coordinates $\Kv$ for the periodic example with $n=h^{1,1} - 1 = 1$ discussed in the text. (Both figures are drawn using $k=3$, but other values give qualitatively similar results.)}
\label{fig:periodic}
\end{figure}

To find the extended K\"ahler cone $\Kah$ resulting from this infinite sequence of flops, we consider the right eigenbasis for $\Lambda$: 
\begin{equation}
Y^I_+ = \begin{pmatrix} \lambda \\ -1 \end{pmatrix}, \quad Y^I_- = \begin{pmatrix} -1 \\ \lambda \end{pmatrix}, \qquad \Lambda^I_{\; J} Y^J_{\pm} = \lambda^{\pm 1} Y^I_{\pm} \,,
\end{equation}
where $\lambda = \frac{k + \sqrt{k^2-4}}{2} > 1$ solves $\lambda + \lambda^{-1} = k$. We decompose $Y^I = y_+ Y^I_+ + y_- Y^I_-$ where $y_{\pm} > 0$ for $Y^I \in \Kah_{\rm (I)}$. Applying $\Lambda$ to map to phase II takes $y_+ \to \lambda y_+$ and $y_- \to \lambda^{-1} y_-$, so after applying $\Lambda$ many times in sequence we have $y_+ \gg y_-$, and we asymptotically approach the ray $Y^I \to Y_\ast^I = Y^I_+$. Likewise, proceeding through the flop at $X=0$ to ``phase 0'' and onwards corresponds to repeatedly applying $\Lambda^{-1}$, which has the opposite effect, i.e., $y_- \gg y_+$ after repeatedly applying $\Lambda^{-1}$, and we asymptotically approach the ray $Y^I \to Y_\ast^I = Y^I_-$. Thus, $\Kah$ is the cone generated by $Y^I_\pm$, as shown in Figure~\ref{sfig:PeriodicK}.

We can find $\Kv$ by similar reasoning. First, note that $\Kv_{\rm I}$ is the cone generated by the rays $\Yd_I = \bigl(\begin{smallmatrix} 1 \\ k/2 \end{smallmatrix}\bigr)$ and $\Yd_I = \bigl(\begin{smallmatrix} k/2 \\ 1 \end{smallmatrix}\bigr)$, corresponding to the rays $Y^I = 3 \bigl(\begin{smallmatrix} 1 \\ 0 \end{smallmatrix}\bigr)$ and $Y^I = 3 \bigl(\begin{smallmatrix} 0 \\ 1 \end{smallmatrix}\bigr)$, respectively, generating $\Kah_{\rm I}$. Next, note that the left action $Y \to \Lambda Y$ by $\Lambda$ on the homogenous coordinates corresponds to the right action $\Yd \to \Yd \Lambda^{-1}$ by $\Lambda^{-1}$ on the dual coordinates. Thus, we construct a left eigenbasis for $\Lambda^{-1} = \bigl(\begin{smallmatrix} 0 & -1 \\ 1 & k\end{smallmatrix}\bigr)$:
\begin{equation}
\Yd_I^+ = \begin{pmatrix} 1 & \lambda \end{pmatrix}, \quad \Yd_I^- = \begin{pmatrix} \lambda & 1 \end{pmatrix}, \qquad \Yd_I^{\pm} (\Lambda^{-1})^I_{\; J}  = \lambda^{\pm 1} \Yd_I^{\pm} \,,
\end{equation}
with signs chosen such that $\Yd_I = \tilde{y}_+ \Yd_I^+ + \tilde{y}_- \Yd_I^-$ with $\tilde{y}_\pm > 0$ within $\Kv_{\rm I}$, as shown in Figure~\ref{sfig:PeriodicKv}. By the same reasoning as above, we conclude that $\Kv$ is generated by the rays $\Yd_I^\pm$. Note that $\Yd_I^+ Y_+^I = 0$ and $\Yd_I^- Y_-^I = 0$, implying that $\Yd_{\ast I} Y^I_\ast = 0$ at both boundaries of $\Kah$, as expected for periodic boundaries, and moreover that $\Kv = \Kah^\ast$, as expected in the absence of finite-distance boundaries.

Finally, let us examine how $\cF[Y]$ behaves near the boundary of $\Kah$. To do so, note that the quadratic polynomial $\cH[Y] =X^2 +k X Y +Y^2$ is invariant under both $X \leftrightarrow Y$ and $X\to X+kY, Y \to -Y$, so it is $G$-invariant. It is straightforward to check that, within $\Kah_{\rm I}$ ($X,Y\ge 0$),
\begin{equation}
\frac{3 k+2}{(k+2)^{3/2}} \le \frac{\cF_{\rm I}[Y]}{\frac{a}{6} \cH[Y]^{3/2}} \le 1 \,,
\end{equation}
hence $\cF_{\rm I}[Y] \sim \frac{a}{6} \cH[Y]^{3/2}$ throughout $\Kah_{\rm I}$, up to a bounded factor. Since $\cF[Y]$ is also $G$-invariant, this implies that $\cF[Y] \sim \frac{a}{6} \cH[Y]^{3/2}$ throughout $\Kah$, up to the same bounded factor. In particular, $\cH[Y] = (X+\lambda Y)(X+Y/\lambda)$ vanishes at both boundaries (located at $Y = - X/\lambda$ and $X=-Y/\lambda$), and therefore so does $\cF[Y] \sim (X+\lambda Y)^{3/2} (X+Y/\lambda)^{3/2}$, in agreement with the general argument for periodic boundaries given previously.

\bigskip

Similar methods can be applied to examples with $n = h^{1,1} - 1 > 1$ vector multiplets. We will not discuss many details here, but only note that, as in the previous subsection, the mapping between the boundaries of $\Kah$ and $\Kv$ is no longer one-to-one. For example, consider a periodic boundary associated to some infinite-order element $\Lambda^I_{\; J} \in G$, and suppose that $\Lambda^I_{\; J}$ is real diagonalizable. As above, $Y_\ast^I$ at the boundary lies within the right-eigenspace with the largest eigenvalue, whereas $\Yd_{\ast I}$ lies within the left-eigenspace with the smallest eigenvalue.\footnote{The largest and smallest eigenvalues must be distinct, since $\Lambda$ is not the identity.} The dimensions of these eigenspaces (and by extension, the dimension of the corresponding periodic boundaries of $\Kah$ and $\Kv$) need not be the same; indeed, the two eigenspaces are orthogonal to each other, so the sum of their dimensions is at most $n+1$. In particular, if $Y_\ast^I$ lies in an $n$ dimensional right-eigenspaces (a facet of $\Kah$) then $\Yd_{\ast I}$ must lie in a $1$-dimensional left-eigenspace (a single point on the boundary of $\Kv$, up to projective equivalence) and vice versa.

\section{The Geometric T/sLWGC}\label{sec:geometric}

Let $\Gamma_q = \mathbb{Z}^{n+1}$ be the lattice of particle charges allowed by Dirac quantization in a given theory. As discussed in~\S\ref{sec:CC}, the tower WGC and sublattice WGC make the following predictions about the spectrum of BPS particles:
\begin{namedconjecture}[The BPS tower WGC]
For any $q_I \in \Gamma_q$ in a direction in which the BPS and black hole extremality bounds coincide, there exists an integer $k \ge 1$ such that there is a BPS particle of charge $k q_I$.
\end{namedconjecture}
\begin{namedconjecture}[The BPS sublattice WGC]
There exists an integer $k \ge 1$ such that for any $q_I \in \Gamma_q$ in a direction in which the BPS and black hole extremality bounds coincide, there is a BPS particle of charge $k q_I$.
\end{namedconjecture}
\noindent Note that the sublattice WGC is stronger, as it requires the integer $k$ to be independent of the choice of $q_I \in \Gamma_q$, but both conjectures require an infinite tower of BPS particles in any direction in which the BPS and black hole extremality bounds coincide.

With the results of~\S\ref{sec:BPSBHE} and~\S\ref{sec:boundaries} in hand, we have acquired a detailed, geometric understanding of when the BPS and black hole extremality bounds coincide in the particular case of a 5d $\mathcal{N}=1$ theory arising from M-theory compactified on a Calabi-Yau threefold $X$. We can thereby reduce the tower and sublattice WGC to geometric conjectures:
\begin{namedconjecture}[The tower WGC for holomorphic curves]
For any Calabi-Yau threefold $X$ and nontrivial class $q_I \in H_2(X,\mathbb{Z})$ in the dual to the effective cone $\Eff^\ast = \Kv$, there exists an integer $k \ge 1$ such that there is a holomorphic curve in the class $k q_I$.
\end{namedconjecture}
\begin{namedconjecture}[The sublattice WGC for holomorphic curves]
For any Calabi-Yau threefold $X$ there exists an integer $k \ge 1$ such that for any nontrivial class $q_I \in H_2(X,\mathbb{Z})$ in the dual to the effective cone $\Eff^\ast = \Kv$, there is a holomorphic curve in the class $k q_I$.
\end{namedconjecture}
\noindent Once again the sublattice version is stronger because the integer $k$ must be independent of the choice of $q_I \in H_2(X,\mathbb{Z})$, though it can depend on $X$.

Note that the simplest possibility one could contemplate (producing the strongest conjecture) is that $k=1$ in all cases, i.e., there is a holomorphic curve in every nontrivial integral class within $\Eff^\ast = \Kv$. This would follow from the lattice WGC~\cite{Heidenreich:2015nta}, however the lattice WGC is false in general~\cite{Heidenreich:2016aqi}, though it does hold in many examples. Thus, this stronger conjecture could only hold for some subclass of Calabi-Yau threefolds $X$; for instance, we are not aware of any counterexamples to it for simply-connected $X$.\footnote{We thank M.~Demirtas, N.~Gendler, M.~Kim, L.~McAllister, J.~Moritz, and A.~Rios Tascon for discussions on this point.} We defer further consideration of this interesting possibility to future work.

The above geometrized versions of the BPS tower and sublattice WGC are stated conservatively, since we have been able to show that the BPS and black hole extremality bounds coincide within $\Kv$, but not that they differ everywhere else. Indeed, in examples with $\mathfrak{su}(2)$ boundaries, the cone where they coincide will generally be larger than $\Kv$ (including at minimum its Weyl-group images), as discussed in~\S\ref{ssec:KMV} and more systematically in~\cite{CornellGV}. When this occurs, the geometrized conjectures should be strengthened to apply in this larger region, requiring holomorphic curves of arbitrarily high degree in every direction within it.

What happens outside the cone where the BPS and black hole extremality bounds coincide? Only finitely many holomorphic curves are possible along any ray outside $\Kah^\ast = \Mori_\cap$, since any curve in this region will flop somewhere inside $\Kah$ and only finitely many curves can flop at one time. Corresponding to this, we saw in \S\ref{subsec:BPSconsequences} that the BPS and black hole extremality bounds disagree in this region. 

By comparison, infinitely many holomorphic curves can occur along rays that are inside $\Kah^\ast = \Mori_\cap$, but this is not mandated by the tower or sublattice WGC if the ray lies outside the cone where the BPS and black hole extremality bounds coincide. In the examples presented in~\S\ref{sec:conifoldexample}, such infinite towers seem to be present everywhere in the interior of $\Kah^\ast$, but not necessarily on its boundaries. In more general examples, certain directions in the interior of $\Kah^\ast$ also lack infinite towers, but never those within $\Kv$~\cite{CornellGV}.

\section{Examples}\label{sec:conifoldexample}

We now consider several examples, both to illustrate our general results as well as to check the geometric conjectures formulated in the previous section, which we will do by computing the genus 0 Gopakumar-Vafa (GV) invariants. These invariants count the BPS particles of a given charge with signs, so cancellations may occur between hypermultiplets (which contribute positively to the GV invariants) and vector multiplets (which  contribute negatively). However, a nonvanishing GV invariant guarantees the existence of at least one BPS particle of a given charge, and correspondingly at least one holomorphic curve in a given homology class,
so by examining the nonvanishing GV invariants for each geometry under consideration we can verify that there are infinite towers of holomorphic curves in every direction within $\Kv$. In particular, all of the examples we consider appear to satisfy the sublattice WGC with $k=1$, as every non-trivial class within $\Kv$ has a non-zero GV invariant, up to the maximum degree that we have computed.

In each example, we pick a Calabi-Yau manifold $X_{\text{I}}$, which defines one phase $\Kah_{\text{I}}$ in the extended K\"ahler cone $\Kah$. We then determine all the phases by passing through each boundary of $\Kah_{\text{I}}$ corresponding to a flop and repeating as necessary in each new phase (all of our examples turn out to have exactly two phases, so that $\Kah = \Kah_{\text{I}} \cup \Kah_{\text{II}}$). Finding the prepotential in each phase using the intersection numbers $C_{I J K}$ of the Calabi-Yau manifold, we thereby obtain $\Kv$ (hence also the (pseudo)effective cone $\overline\Eff = \Kv^\ast$). We then compute the genus 0 GV invariants and check the geometrized tower and sublattice conjectures. Finally, we discuss some features of the black hole extremality bound for both BPS and non-BPS black holes, emphasizing those which are novel in each successive example.

\subsection{Example 1: a flop-symmetric conifold}\label{ssec:symFlop}

Our first example is the $(h^{1, 1},h^{2, 1}) = (2,86)$ Calabi-Yau manifold given by the intersection of two bidegree $(4,1)$ hypersurfaces inside the weighted projective space $\mathbb{P}^4[4,1,1,1,1] \times \mathbb{P}^1$. This seemingly obscure choice will lead to several simplifications, as we will see.

\subsubsection{Geometry}

Computing the intersection numbers, one finds the prepotential for this phase
\begin{equation}
\cF^{\text{(I)}} = \frac{1}{3} X^3 + 2 X^2 Y \qquad \text{with K\"ahler cone} \qquad \Kah_{\text{I}} = \{X,Y\ge 0\} \,,
\end{equation}
where $X$ and $Y$ correspond to curves in $\mathbb{P}^4[4,1,1,1,1]$ and $\mathbb{P}^1$, respectively. Since $\cF\to0$ as $X\to 0$, $X=0$ is an infinite-distance, asymptotic boundary. At the other boundary of $\Kah_{\text{I}}$ ($Y=0$) the geometry develops 64 isolated conifold singularities. Passing through the flop, one reaches an isomorphic Calabi-Yau, with prepotential
\begin{equation}
\cF^{\text{(II)}} = \frac{1}{3} X'^3 + 2 X'^2 Y' \qquad \text{with} \qquad \Kah_{\text{II}} = \{X', Y' \ge 0\} \,,
\end{equation}
where the two phases are related by $X' = X + 4 Y$, $Y' = - Y$. In the original basis,
\begin{equation}
\cF^{\text{(II)}} = \frac{1}{3} X^3 + 2 X^2 Y - \frac{64}{6} Y^3 \qquad \text{with} \qquad \Kah_{\text{II}} = \{X \ge -4Y \ge 0\} \,,
\end{equation}
where the last term is in agreement with the general result~\eqref{eqn:flop}, since there are 64 flopping curves of charge $q_I = \bigl(\begin{smallmatrix}0\\1\end{smallmatrix}\bigr)$.

The boundary $X'=0$ ($X = -4Y$) is again an asymptotic boundary, so the extended K\"ahler cone is
\begin{equation}
\Kah = \Kah_{\text{I}} \cup \Kah_{\text{II}} = \{ X \ge 0; X \ge -4Y \} \,, \label{eqn:ex1Kah}
\end{equation}
both of whose boundaries are asymptotic. Note that the absence of finite-distance boundaries is a major simplification, as this implies that there are no indeterminate flows and moreover that $\Kv = \Kah^\ast$ with no BPS black holes outside this cone.

\begin{figure}
\centering
\begin{subfigure}{0.48\textwidth}
\centering
\includegraphics[width=5cm]{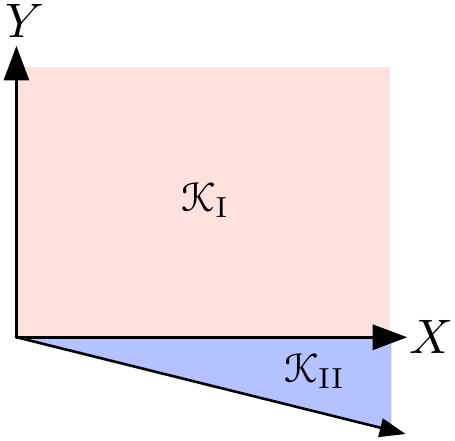}
\caption{The extended K\"ahler cone $\Kah$.} \label{sfig:SymFlopK}
\end{subfigure}
\begin{subfigure}{0.48\textwidth}
\centering
\includegraphics[width=5cm]{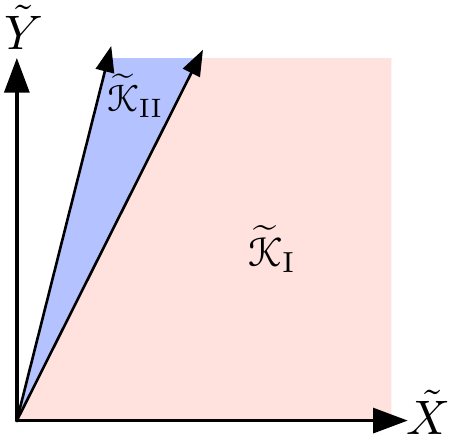}
\caption{The cone of dual coordinates $\Kv$.} \label{sfig:SymFlopKv}
\end{subfigure}
\caption{\subref{sfig:SymFlopK} The extended K\"ahler cone $\Kah$ and \subref{sfig:SymFlopKv} the cone of dual coordinates $\Kv$ for the symmetric flop example.}
\label{fig:SymFlopCones}
\end{figure}

For completeness, let us also calculate $\Kv$ directly. In phase I, we have
\begin{equation}
\Yd_I \propto \cF_I = \begin{pmatrix} X (X+4Y) \\ 2 X^2 \end{pmatrix} \,.
\end{equation}
This approaches the ray $\Yd_{\ast I} = \bigl(\begin{smallmatrix}1\\0\end{smallmatrix}\bigr)$ as $X \to 0$ and the ray $\Yd_{\ast I} = \bigl(\begin{smallmatrix}1\\2\end{smallmatrix}\bigr)$ as $Y \to 0$, so $\Kv_{\text{I}} = \{\tilde{X} \ge \tilde{Y}/2 \ge 0 \}$ is generated by $\bigl(\begin{smallmatrix}1\\0\end{smallmatrix}\bigr)$ and $\bigl(\begin{smallmatrix}1\\2\end{smallmatrix}\bigr)$. By a similar calculation in phase II, $X \to -4 Y$ corresponds to the ray $\Yd_{\ast I} = \bigl(\begin{smallmatrix}1\\4\end{smallmatrix}\bigr)$, so $\Kv_{\text{II}} = \{\tilde{Y}/2 \ge \tilde{X} \ge \tilde{Y}/4\}$ is generated by $\bigl(\begin{smallmatrix}1\\2\end{smallmatrix}\bigr)$ and $\bigl(\begin{smallmatrix}1\\4\end{smallmatrix}\bigr)$. Thus,
\begin{equation}
\Kv = \Kv_{\text{I}} \cup \Kv_{\text{II}}= \{\tilde{X} \ge \tilde{Y}/4 \ge 0 \}\,,
\end{equation}
which is generated by $\bigl(\begin{smallmatrix}1\\0\end{smallmatrix}\bigr)$ and $\bigl(\begin{smallmatrix}1\\4\end{smallmatrix}\bigr)$. Comparing with~\eqref{eqn:ex1Kah}, it is easily seen that $\Kv = \Kah^\ast$ as expected. Both cones are shown in Figure~\ref{fig:SymFlopCones}.

Finally, note that a divisor shrinks at each boundary of $\Kah$, in agreement with $\Kv = \Eff^\ast$. At the $X=0$ ($X'=0$) boundary of phase I (phase II) this is the toric divisor corresponding to a point on $\mathbb{P}^1$ (a point on the flopped $\mathbb{P}^1$).

\subsubsection{GV invariants and the T/sLWGC} Having determined $\Kv$, we now compute the genus 0 GV invariants and compare with the predictions of the geometrized tower and sublattice WGC. Details of the calculation are presented in Appendix~\ref{sec:GVinvariants}; here we simply quote the result, shown in Table~\ref{tab:SymFlopGV}. In particular, up to the degree computed, every GV invariant within $\Kv$ is non-zero, hence there is a holomorphic curve in every non-trivial class within $\Kv$, in agreement with both the tower and sublattice WGC with $k=1$.

\begin{table}
\begin{center}
$\arraycolsep=5pt
\begin{array}{c|cccc}
\mathdiagbox[width=1cm,height=0.75cm,innerleftsep=0.1cm,innerrightsep=0cm]{q_2}{q_1}&0&1&2&3 \\ \hline
0&\cellcolor{KvColor}-&\cellcolor{KvColor}640&\cellcolor{KvColor}10032&\cellcolor{KvColor}288384 \\ \hhline{~|-|>{\arrayrulecolor{KvColor}}--->{\arrayrulecolor{black}}}
1& \multicolumn{1}{c|}{64} &\cellcolor{KvColor}6912&\cellcolor{KvColor}742784&\cellcolor{KvColor}75933184 \\
2& \multicolumn{1}{c|}{0} &\cellcolor{KvColor}14400&\cellcolor{KvColor}8271360&\cellcolor{KvColor}2445747712 \\
3& \multicolumn{1}{c|}{0} &\cellcolor{KvColor}6912&\cellcolor{KvColor}31344000&\cellcolor{KvColor}26556152064 \\
4& \multicolumn{1}{c|}{0} &\cellcolor{KvColor}640&\cellcolor{KvColor}48098560&\cellcolor{KvColor}130867460608 \\ \hhline{~|~|-|>{\arrayrulecolor{KvColor}}-->{\arrayrulecolor{black}}}
5&0&\multicolumn{1}{c|}{0}&\cellcolor{KvColor}31344000&\cellcolor{KvColor}329212616704 \\
6&0&\multicolumn{1}{c|}{0}&\cellcolor{KvColor}8271360&\cellcolor{KvColor}445404149568 \\
7&0&\multicolumn{1}{c|}{0}&\cellcolor{KvColor}742784&\cellcolor{KvColor}329212616704 \\
8&0&\multicolumn{1}{c|}{0}&\cellcolor{KvColor}10032&\cellcolor{KvColor}130867460608 \\ \hhline{~|~~|-|>{\arrayrulecolor{KvColor}}->{\arrayrulecolor{black}}}
9&0&0&\multicolumn{1}{c|}{0}&\cellcolor{KvColor}26556152064 \\
10&0&0&\multicolumn{1}{c|}{0}&\cellcolor{KvColor}2445747712 \\
11&0&0&\multicolumn{1}{c|}{0}&\cellcolor{KvColor}75933184 \\
12&0&0&\multicolumn{1}{c|}{0}&\cellcolor{KvColor}288384 \\ \hhline{~|~~~|-}
13&0&0&0&0 
\end{array}
$
\caption{Genus 0 GV invariants of degree $(q_1, q_2)$ for the symmetric flop example. The shaded region $0 \leq q_2 \leq 4 q_1$ is the cone $\Kv$ where the BPS and black hole extremality bounds agree. There is a holomorphic curve in every non-trivial class within this cone to the degree calculated, in agreement with the $k=1$ sublattice WGC.}
\label{tab:SymFlopGV}
\end{center}
\end{table}

Note that the only non-vanishing GV invariant outside $\Kv = \Kah^\ast$ is $64$ at degree $(0,1)$, corresponding to the flopping curves at the 64 conifold singularities that appear at $Y=0$. This is expected, since any curves outside $\Kah^\ast$ will flop somewhere within $\Kah$.

Moreover, note that the infinite towers of BPS particles at the \emph{boundaries} of $\Kv$ play a dual role: they satisfy the SDC at the two asymptotic boundaries of the moduli space, exemplifying a more general connection between the T/sLWGC and the SDC~\cite{Gendler:2020dfp}. However, the infinite towers of BPS particles in the \emph{interior} of $\Kv$ do not become light anywhere in the moduli space, hence their existence is not mandated by the SDC, only by the T/sLWGC.

\subsubsection{Extremal versus BPS}

According to the general results of~\S\ref{sec:BPSBHE}, since $\Kah$ has no finite-distance boundaries, the BPS and black hole extremality bounds \emph{disagree} everywhere outside $\Kv$, including in the direction of the flopping curves, $q_I = \bigl(\begin{smallmatrix}0\\1\end{smallmatrix}\bigr)$. We can make this more explicit by determining the extremality bound, i.e., the region within the $(q_1/m, q_2/m)$-plane where black holes exist.

To do so, it is sufficient to find a fake superpotential, i.e., a global, positive solution to~\eqref{eqn:WPDE}, as explained in~\S\ref{sec:ExtBound}. First, define $Z = X+2Y = X'+2Y'$, so that
\begin{equation}
\cF = \frac{1}{3}  (Z - 2 | Y |)^2 (Z + 4 | Y |) \label{eqn:symFlopF}
\end{equation}
encapsulates the prepotential in both phases. Parameterizing the $\cF=1$ hypersurface as
\begin{equation}
  Z (\phi) = \frac{1}{3^{2 / 3}}  (e^{2 | \phi |} + 2 e^{- | \phi |}), \qquad
  Y (\phi) = \frac{\mathrm{sgn} (\phi)}{2 \cdot 3^{2 / 3}}  (e^{2 | \phi |} -
  e^{- | \phi |}),
  \label{eqn:ZYparam}
\end{equation}
a straightforward calculation gives $\mathfrak{g}_{\phi \phi} = 6$ as well as
\begin{equation}
\cQ^2(\phi) \df a^{I J} q_I q_J = \begin{cases} P_+^2 e^{4 \phi} + 2 Q_+^2 e^{-2 \phi}\,, & \phi \ge 0\,, \\ 
                                                    P_-^2 e^{-4 \phi} + 2 Q_-^2 e^{2 \phi}\,, & \phi \le 0\,, \end{cases} 
\end{equation}
where
\begin{equation}
P_{\pm} \df \frac{1}{3^{2/3}} \Bigl[q_Z \pm \frac{q_Y}{2}\Bigr] \,, \quad Q_{\pm} \df \frac{1}{3^{2/3}} \Bigl[q_Z \mp \frac{q_Y}{4}\Bigr] \,,
\end{equation}
encode the charges $q_I$, with $q_Z = q_1$ and $q_Y = q_2 - 2 q_1$ in terms of the charges $(q_1, q_2)$ in the original $X, Y$ basis.

Thus, \eqref{eqn:WPDE} becomes
\begin{equation}
[W(\phi)]^2 + \frac{1}{2} [W'(\phi)]^2 = 3\cQ^2(\phi) \,, \qquad \cQ^2(\phi) \df \begin{cases}P_+^2 e^{4 \phi} + 2Q_+^2 e^{-2 \phi}\,, & \phi \ge 0\,, \\ 
                                                                                  P_-^2 e^{-4 \phi} + 2Q_-^2 e^{2 \phi}\,, & \phi \le 0\,. \end{cases} \label{eqn:symWPDE}
\end{equation}
This equation can be solved either analytically (using an implicit function) or by numerical integration. The latter approach generalizes more readily to other examples, so we explain it here, deferring the analytic approach (which of course yields the same results) to Appendix~\ref{app:analytic}.

\begin{figure}
\centering
\includegraphics[width=6.5cm]{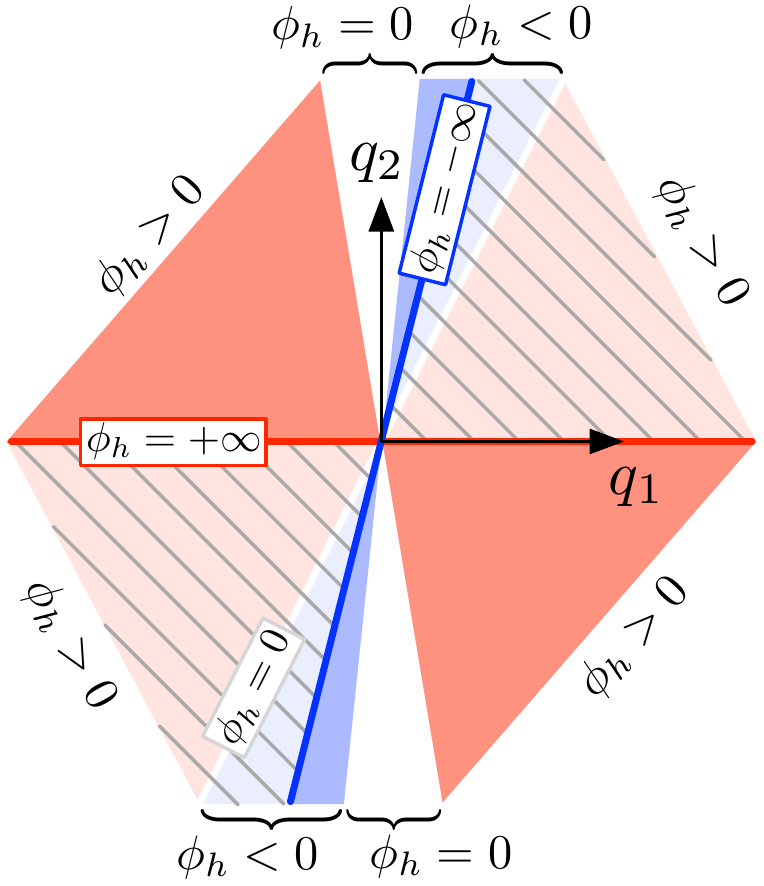}
\caption{The location of the attractor point $\phi_h$ depends on $(q_1,q_2)$. In the crosshatched region $\Kv \cup (-\Kv)$ the attractor is (anti-)BPS ($\sqrt{3}\cQ(\phi_h) = |\zeta_q(\phi_h)|$), whereas elsewhere it is non-BPS ($\sqrt{3}\cQ(\phi_h) > |\zeta_q(\phi_h)|$).}
\label{fig:SymFlopAttractor}
\end{figure}

\begin{figure}
\centering
\begin{subfigure}{0.48\textwidth}
\centering
\includegraphics[width=7.25cm]{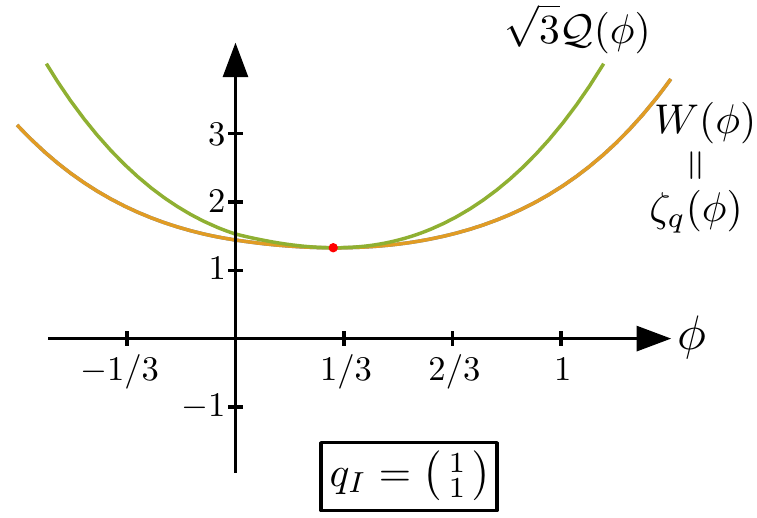}
\caption{A BPS attractor.} \label{sfig:BPSattractor}
\end{subfigure}
\begin{subfigure}{0.48\textwidth}
\centering
\includegraphics[width=7.25cm]{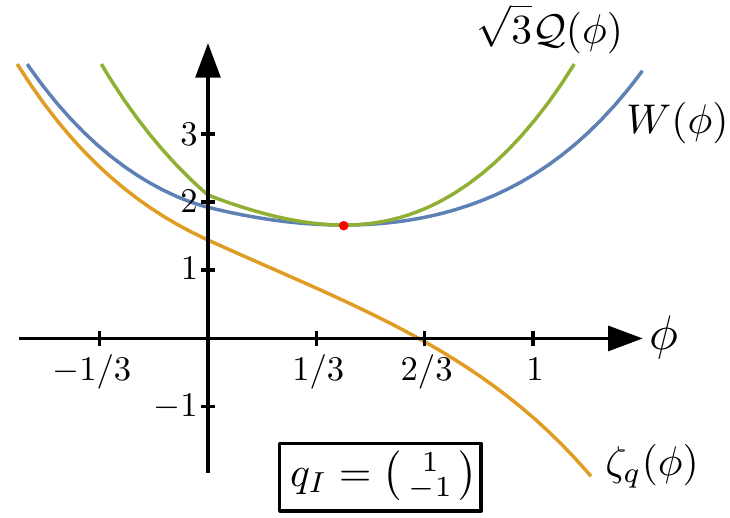}
\caption{A non-BPS attractor.} \label{sfig:nonBPSattractor}
\end{subfigure}
\caption{The fake superpotential $W(\phi)$ can be determined numerically by integrating~\eqref{eqn:WIntegrate}, regardless of whether the attractor in question is \subref{sfig:BPSattractor} BPS, or \subref{sfig:nonBPSattractor} not BPS.}
\label{fig:FakeWPlot}
\end{figure}

\begin{figure}
\centering
\includegraphics[width=8cm]{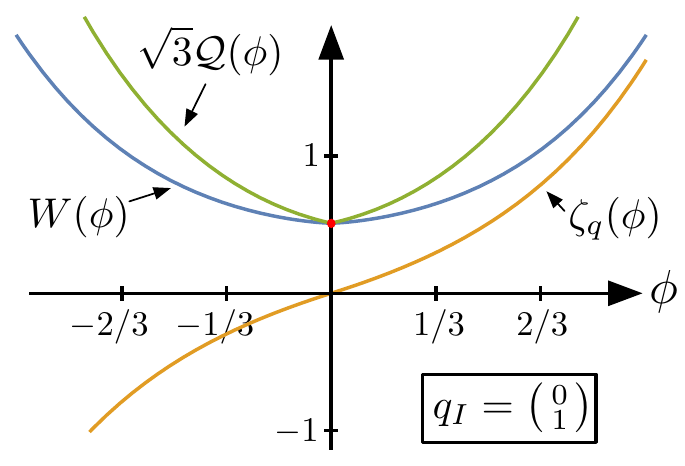}
\caption{The hypermultiplets that become massless at the $\phi = 0$ flop are strictly superextremal (i.e., $W(\phi) > |\zeta_q(\phi)|$) everywhere in the moduli space, although they are nearly extremal (i.e., $W(\phi) \simeq |\zeta_q(\phi)|$) far from the flop.} 
\label{fig:ConifoldAttractor}
\end{figure}

For any choice of charge $q_I$, the function $\cQ^2(\phi)$ either has a single critical point (a minimum) at finite $\phi = \phi_{\text{min}}$ or it is monotonically decreasing (increasing) with $\cQ^2(\phi) \to 0$ as $\phi \to +\infty$ ($\phi \to -\infty$). Thus, $\cQ^2(\phi)$ has a unique attractor point $\phi_h = \phi_{\text{min}}$, whose location depends on $q_I$, as shown in Figure~\ref{fig:SymFlopAttractor}. At this point, $W(\phi_h) = \sqrt{3} \cQ(\phi_h)$, and $W(\phi)$ can be determined elsewhere by integrating
\begin{equation}
W'(\phi) = \begin{cases} +\sqrt{2} \sqrt{3\cQ^2(\phi) - [W(\phi)]^2}\,, & \phi > \phi_h\,, \\ -\sqrt{2} \sqrt{3\cQ^2(\phi) - [W(\phi)]^2}\,, & \phi < \phi_h\,, \end{cases} \label{eqn:WIntegrate}
\end{equation}
from the attractor point outwards to obtain a global solution to~\eqref{eqn:symWPDE} with a single minimum at $\phi = \phi_h$.\footnote{Here we assume for simplicity that $\phi_h$ is finite, since if $\phi_h = \pm \infty$ then the solution is BPS and $W(\phi) = \pm \zeta_q(\phi)$ is already known.} Some examples are shown in Figures~\ref{fig:FakeWPlot} and~\ref{fig:ConifoldAttractor}.

\begin{figure}
\centering
\begin{subfigure}{0.32\textwidth}
\centering
\includegraphics[width=4.5cm]{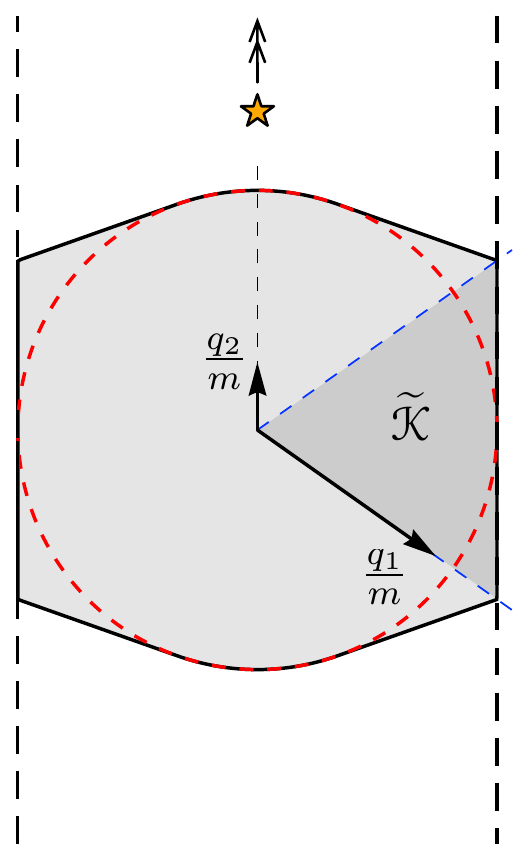}
\caption{$\phi_\infty = 0$.} \label{sfig:ExtBoundFlop}
\end{subfigure}
\begin{subfigure}{0.32\textwidth}
\centering
\includegraphics[width=4.5cm]{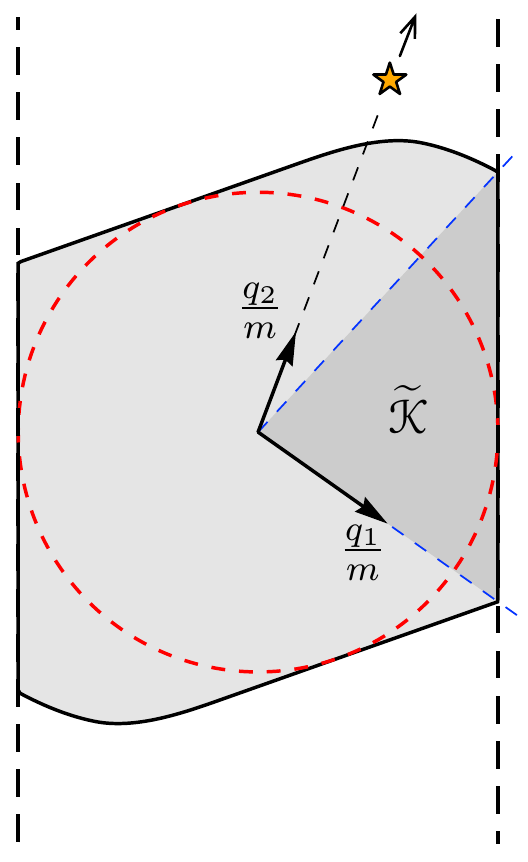}
\caption{$\phi_\infty = 1/3$.} \label{sfig:ExtBoundClose}
\end{subfigure}
\begin{subfigure}{0.32\textwidth}
\centering
\includegraphics[width=4.5cm]{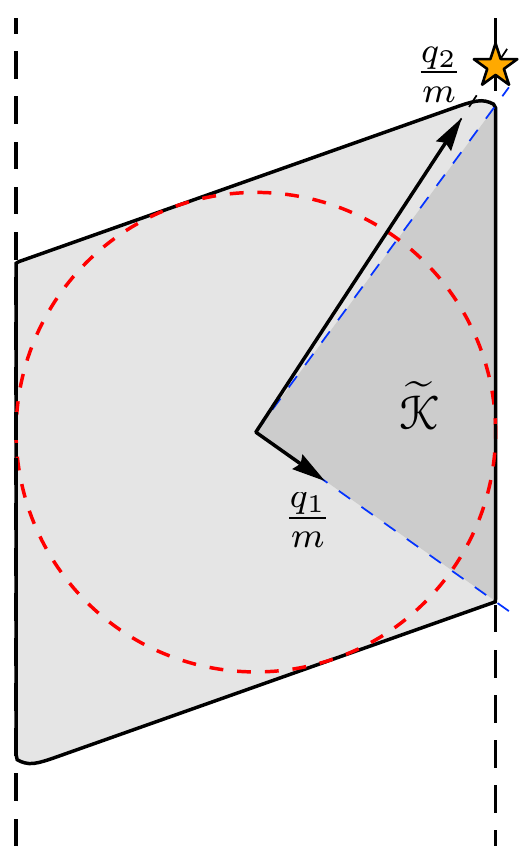}
\caption{$\phi_\infty = 1$.} \label{sfig:ExtBoundFar}
\end{subfigure}
\caption{The black hole region in charge-to-mass space for three different choices of vacuum $\phi_\infty$. The phase~I lattice basis $(q_1, q_2)$ is indicated by the axes, scaled appropriately to make the Reissner-Nordstr\"om black hole region (whose perimeter is indicated by the dashed circle) into the unit disk. The dashed line to the right (left) indicates the BPS (anti-BPS) bound, and the cone $\Kv$ where the BPS and black hole extremality bounds agree is shaded. The flopping hypers are marked by a star in~\subref{sfig:ExtBoundFar}, whereas they lie outside the field of view in \subref{sfig:ExtBoundClose} and at infinity (since $m=0$) in \subref{sfig:ExtBoundFlop}.}
\label{fig:ExtBound}
\end{figure}

Having determined the fake superpotential $W_q(\phi)$ for all choices of $q_I$, we read off the black hole region by fixing a choice of vacuum $\phi_\infty$ and scanning over all charges, where $M_{\text{ext}} = \frac{g_5}{\sqrt{2}\kappa_5} W_q(\phi_\infty)$. The result, compared to the corresponding Reissner-Nordstr\"om (RN) extremality bound as well as to the BPS bound, is shown for three representative vacua in Figure~\ref{fig:ExtBound}. Regardless of which vacuum we choose, the 64 hypermultiplets that become massless at the flop are strictly superextremal (i.e., strictly outside the black hole region). This can also be seen by examining the fake superpotential for $q_I = \bigl( \begin{smallmatrix} 0 \\ 1 \end{smallmatrix} \bigr)$, shown in Figure~\ref{fig:ConifoldAttractor}.

Finally, we note that far out in the moduli space the black hole region becomes approximately a parallelogram, as can be seen in Figure~\ref{sfig:ExtBoundFar}.
Similar asymptotic simplifications are frequently used to analyze the WGC, e.g., in~\cite{Lee:2018spm,Lee:2019tst,Gendler:2020dfp}. However, such an asymptotic analysis misses some important physics only visible in the interior of the moduli space. For example, the 64 flopping hypermultiplets become asymptotically extremal (see Figures~\ref{sfig:ExtBoundFar} and~\ref{fig:ConifoldAttractor}), even though they are strictly superextremal at any finite distance point within the moduli space. (More generally, all BPS states are asymptotically extremal~\cite{Gendler:2020dfp}.) An analysis of the extremality bound based on these asymptotic simplifications would erroneously conclude that flopping hypermultiplets present a counterexample to the T/sLWGC for BPS particles. As we have seen, this is not the case.

\subsection{Example 2: the GMSV conifold, revisited}\label{ssec:GMSV}

Our second example is the $(h^{1,1},h^{2,1}) = (2,86)$ Calabi-Yau given by the intersection of bidegree $(4,1)$ and bidegree $(1,1)$ hypersurfaces inside $\mathbb{P}^4 \times \mathbb{P}^1$, as studied in~\cite{Greene:1995hu, Greene:1996dh} and discussed previously in \S\ref{sec:CC}. This is similar to the previous example in many ways, so we will be brief, highlighting the interesting differences.

\subsubsection{Geometry} From the intersection numbers, one finds
\begin{equation}
\cF^{\text{(I)}} = \frac{5}{6} X^3 + 2 X^2 Y\qquad \text{with K\"ahler cone} \qquad \Kah_{\text{I}} = \{X,Y\ge 0\} \,,
\end{equation}
where $X$ and $Y$ correspond to curves in $\mathbb{P}^4$ and $\mathbb{P}^1$, respectively. $X=0$ is again an asymptotic boundary and $Y=0$ a flop, this time with 16 flopping curves. Passing through the flop, we reach the phase
\begin{equation}
\cF^{\text{(II)}} = \frac{5}{6} X'^3 + 8 X'^2 Y' + 24 X' Y'^2 + 24 Y'^3\qquad \text{with} \qquad \Kah_{\text{II}} = \{X',Y'\ge 0\} \,, 
\end{equation}
where $X' = X+4Y$ and $Y'=-Y$. In the original basis
\begin{equation}
\cF^{\text{(II)}} = \frac{5}{6} X^3 + 2 X^2 Y - \frac{16}{6} Y^3 \qquad \text{with} \qquad \Kah_{\text{II}} = \{X\ge-4Y\ge 0\} \,.
\end{equation}
Unlike the previous example, phase II and phase I are not isomorphic. Moreover, the boundary $X'=0$ is not an asymptotic boundary, but rather a strongly-coupled, CFT boundary, where a $\mathbb{P}^2$ divisor within the Calabi-Yau shrinks to a point.

\begin{figure}
\centering
\begin{subfigure}{0.48\textwidth}
\centering
\includegraphics[width=5cm]{SymFlopK.pdf}
\caption{The extended K\"ahler cone $\Kah$.} \label{sfig:GMSVK}
\end{subfigure}
\begin{subfigure}{0.48\textwidth}
\centering
\includegraphics[width=5cm]{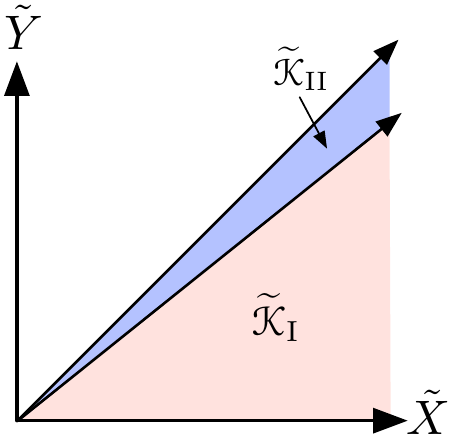}
\caption{The cone of dual coordinates $\Kv$.} \label{sfig:GMSVKv}
\end{subfigure}
\caption{\subref{sfig:GMSVK} The extended K\"ahler cone $\Kah$ and \subref{sfig:GMSVKv} the cone of dual coordinates $\Kv$ for the GMSV geometry.}
\label{fig:GMSVcones}
\end{figure}

The extended K\"ahler cone $\Kah = \{X \ge 0; X \ge -4Y\}$ is unchanged, but the cone of dual coordinates $\Kv = \{ \tilde{X} \ge \tilde{Y} \ge 0\}$ generated by $\bigl(\begin{smallmatrix}1\\0\end{smallmatrix}\bigr)$ and $\bigl(\begin{smallmatrix}1\\1\end{smallmatrix}\bigr)$ is different, see Figure~\ref{fig:GMSVcones}. Now $\Kv$ is a proper subset of $\Kah^\ast = \{\tilde{X} \ge \tilde{Y}/4 \ge 0\}$, as expected in the presence of finite-distance boundaries.

As before, a divisor shrinks at each boundary of $\Kah$, confirming $\Kv = \Eff^\ast$. At the $X=0$ phase I boundary this is the toric divisor corresponding to a point on $\mathbb{P}^1$, whereas at the $X'=0$ phase II boundary it is the aforementioned $\mathbb{P}^2$ exceptional divisor.

\subsubsection{GV invariants and the T/sLWGC} \label{sssec:GMSVgvInvs}

 The genus 0 GV invariants for phase I of this geometry were already shown in Table~\ref{tab:GMSVinvs0}. For easy reference, we repeat them in Table~\ref{tab:GMSVinvs} with the cone $\Kv$ where the BPS and black hole extremality bounds agree highlighted. As in the previous example, there is a holomorphic curve in every non-trivial class within $\Kv$ to the degree calculated, in agreement with the $k=1$ T/sLWGC.

\begin{table}
\begin{center}
$\arraycolsep=5pt
\begin{array}{c|ccccc}
\mathdiagbox[width=1cm,height=0.75cm,innerleftsep=0.1cm,innerrightsep=0cm]{q_2}{q_1}
&0&1&2&3&4 \\ \hline
0&\cellcolor{KvColor}-&\cellcolor{KvColor}640&\cellcolor{KvColor}10032&\cellcolor{KvColor}288384&\cellcolor{KvColor}10979984 \\ \hhline{~|->{\arrayrulecolor{KvColor}}---->{\arrayrulecolor{black}}}
1&\multicolumn{1}{c|}{16}&\cellcolor{KvColor}2144&\cellcolor{KvColor}231888&\cellcolor{KvColor}23953120&\cellcolor{KvColor}2388434784 \\  \hhline{~|~|->{\arrayrulecolor{KvColor}}--->{\arrayrulecolor{black}}}
2& \multicolumn{1}{c|}{0}&\multicolumn{1}{c|}{\cellcolor{KaColor}120}&\cellcolor{KvColor}356368&\cellcolor{KvColor}144785584&\cellcolor{KvColor}36512550816 \\  \hhline{~|~|>{\arrayrulecolor{KaColor}}->{\arrayrulecolor{black}}|-|>{\arrayrulecolor{KvColor}}-->{\arrayrulecolor{black}}}
3& \multicolumn{1}{c|}{0}&\cellcolor{KaColor}-32&\multicolumn{1}{c|}{\cellcolor{KaColor}14608}&\cellcolor{KvColor}144051072& \cellcolor{KvColor}115675981232 \\   \hhline{~|~|>{\arrayrulecolor{KaColor}}-->{\arrayrulecolor{black}}|-|>{\arrayrulecolor{KvColor}}->{\arrayrulecolor{black}}}
4& \multicolumn{1}{c|}{0}&\cellcolor{KaColor}3&\cellcolor{KaColor}-4920&\multicolumn{1}{c|}{\cellcolor{KaColor}5273880}&\cellcolor{KvColor}85456640608\\  \hhline{~|~|-|>{\arrayrulecolor{KaColor}}-->{\arrayrulecolor{black}}|-}
5&0&\multicolumn{1}{c|}{0}&\cellcolor{KaColor}1680&\cellcolor{KaColor}-1505472&\cellcolor{KaColor}3018009984 \\
6& 0&\multicolumn{1}{c|}{0}&\cellcolor{KaColor}-480&\cellcolor{KaColor}512136&\cellcolor{KaColor}-748922304 \\
7&0&\multicolumn{1}{c|}{0}&\cellcolor{KaColor}80&\cellcolor{KaColor}-209856&\cellcolor{KaColor}218062416\\
8& 0&\multicolumn{1}{c|}{0}&\cellcolor{KaColor}-6&\cellcolor{KaColor}75300&\cellcolor{KaColor}-90910176 \\ \hhline{~|~~|-|>{\arrayrulecolor{KaColor}}-->{\arrayrulecolor{black}}}
9&0&0&\multicolumn{1}{c|}{0}&\cellcolor{KaColor}-21600&\cellcolor{KaColor}37721680 \\
10& 0&0&\multicolumn{1}{c|}{0}&\cellcolor{KaColor}4312&\cellcolor{KaColor}-15086208 \\
11&0&0&\multicolumn{1}{c|}{0}&\cellcolor{KaColor}-512&\cellcolor{KaColor}5300736 
\end{array}
$
\caption{Genus 0 GV invariants of degree $(q_1, q_2)$ for the GMSV geometry. The dark blue region $0 \leq q_2 \leq q_1$ is the cone $\Kv$ where the BPS and black hole extremality bounds agree. There is a holomorphic curve in every non-trivial class within this cone to the degree calculated, in agreement with the $k=1$ sublattice WGC. The light yellow region $q_1 < q_2 \leq 4q_1$ is $\Kah^\ast \setminus \Kv$, in which infinite towers of BPS particles are possible but not necessarily required by the T/sLWGC.}
\label{tab:GMSVinvs}
\end{center}
\end{table}

Unlike in the previous example, however, there are parts of $\Kah^\ast$ that are outside $\Kv$. Due to indeterminate flows ending on the CFT boundary (see further discussion below), we cannot say with certainty whether the BPS and black hole extremality bounds agree in this region. If they do then the T/sLWGC predict infinite towers of BPS particles in these directions as well. We observe that such infinite towers do appear to be present, whether related to extremal black holes or not. Thus, the predictions of the T/sLWGC are fulfilled either way.

The GV invariants in phase II, displayed in Table~\ref{tab:GMSVflopped}, have similar properties. In fact, noting that $q_1' = q_1$ and $q_2' = 4 q_1 - q_2$, we see that they are \emph{unchanged} within $\Kah^\ast = \Mori_{\text{I}} \cap \Mori_{\text{II}}$. The only difference, then, is that the 16 flopping hypers appear with opposite charges: $q_I = \bigl(\begin{smallmatrix}0\\1\end{smallmatrix} \bigr)$ in phase I and $q_I' = \bigl(\begin{smallmatrix}0\\1\end{smallmatrix} \bigr)$ hence $q_I = \bigl(\begin{smallmatrix}0\\-1\end{smallmatrix} \bigr)$ in phase II. Note, however, that a BPS ($\zeta_q > 0$) hypermultiplet of charge $q_I$ is equivalent to an anti-BPS ($\zeta_q < 0$) hypermultiplet of charge $-q_I$. After passing through the flop, the 16 charge $q_I = \bigl(\begin{smallmatrix}0\\1\end{smallmatrix} \bigr)$ hypermultiplets become anti-BPS, and can be re-expressed as 16 charge $q_I = \bigl(\begin{smallmatrix}0\\-1\end{smallmatrix} \bigr)$ BPS hypermultiplets. Thus, the GV invariants in Table~\ref{tab:GMSVinvs} and Table~\ref{tab:GMSVflopped} show no sign of wall crossing phenomena at the flop, in agreement with the general expectations laid out in~\S\ref{subsec:wallcrossing}. Similarly, the reflection symmetry in the columns of Table~\ref{tab:SymFlopGV} is consistent with the absence of wall crossing at the flop in the previous example.

\begin{table}
\centering
$\arraycolsep=5pt
\begin{array}{c|cccc}
 \mathdiagbox[width=1cm,height=0.8cm,innerleftsep=0.08cm,innerrightsep=0cm]{q_2'}{q_1'} &0&1&2&3\\
 \hline
 0&\multicolumn{1}{c|}{\cellcolor{KvColor}-} & \cellcolor{KaColor}3 & \cellcolor{KaColor}-6 & \cellcolor{KaColor}27 \\ \hhline{~|-|>{\arrayrulecolor{KaColor}}--->{\arrayrulecolor{black}}}
 1&\multicolumn{1}{c|}{16} & \cellcolor{KaColor}-32 & \cellcolor{KaColor}80 & \cellcolor{KaColor}-512 \\
 2&\multicolumn{1}{c|}{0} & \cellcolor{KaColor}120 & \cellcolor{KaColor}-480 & \cellcolor{KaColor}4312 \\  \hhline{~|~|-|>{\arrayrulecolor{KaColor}}-->{\arrayrulecolor{black}}}
 3&\multicolumn{1}{c|}{0} & \multicolumn{1}{c|}{\cellcolor{KvColor}2144} & \cellcolor{KaColor}1680 & \cellcolor{KaColor}-21600 \\
 4&\multicolumn{1}{c|}{0} & \multicolumn{1}{c|}{\cellcolor{KvColor}640} & \cellcolor{KaColor}-4920 & \cellcolor{KaColor}75300 \\ \hhline{~|~|-|>{\arrayrulecolor{KaColor}}-->{\arrayrulecolor{black}}}
 5&0 & \multicolumn{1}{c|}{0} & \cellcolor{KaColor}14608 & \cellcolor{KaColor}-209856 \\ \hhline{~|~~|-|>{\arrayrulecolor{KaColor}}->{\arrayrulecolor{black}}}
 6&0 & \multicolumn{1}{c|}{0} & \multicolumn{1}{c|}{\cellcolor{KvColor}356368} & \cellcolor{KaColor}512136 \\
 7&0 & \multicolumn{1}{c|}{0} &\multicolumn{1}{c|}{\cellcolor{KvColor} 231888} & \cellcolor{KaColor}-1505472 \\
 8&0 & \multicolumn{1}{c|}{0} & \multicolumn{1}{c|}{\cellcolor{KvColor}10032} & \cellcolor{KaColor}5273880 \\ \hhline{~|~~|-|-}
 9&0 & 0 & \multicolumn{1}{c|}{0} & \cellcolor{KvColor}144051072 \\
 10&0 & 0 & \multicolumn{1}{c|}{0} & \cellcolor{KvColor}144785584 \\
 11&0 & 0 & \multicolumn{1}{c|}{0} & \cellcolor{KvColor}23953120 \\
 12&0 & 0 & \multicolumn{1}{c|}{0} & \cellcolor{KvColor}288384 \\ \hhline{~|~~~|-}
 13&0 & 0 & 0 & 0 \\
 \end{array}
 $
 \caption{Genus 0 GV invariants of degree $(q_1', q_2')$ for phase II of the GMSV geometry. These are related to the phase I GV invariants (Table~\ref{tab:GMSVinvs}) as described in the main text.} 
 \label{tab:GMSVflopped}
 \end{table}

\subsubsection{Extremal versus BPS} 

Due to the presence of a finite-distance CFT boundary, 
there are indeterminate flows ending on the boundary that prevent us from completely fixing the black hole extremality bound outside $\Kv$. 
In principle, this issue should be resolved by a more careful treatment of the strongly-coupled boundary physics, but we will not attempt such an analysis in this paper. Nonetheless, a lot can be said explicitly about the extremality bound, as we now illustrate. In particular, the flopping hypermultiplets are strictly superextremal completely independent of these issues (all $\zeta_q(\phi)$ flows are bad when $q_I$ lies outside $\Kah^\ast$), a fact that we illustrate further below.

\begin{figure}
\centering
\includegraphics[width=6.5cm]{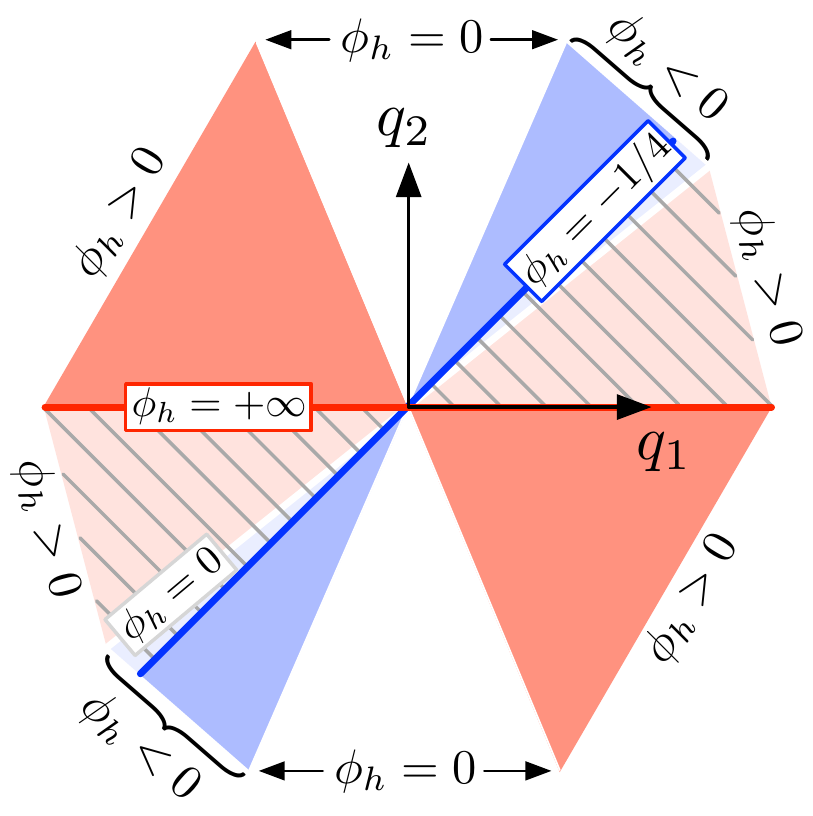}
\caption{The location of the attractor point $\phi_h = Y_h/X_h$ for the GSMV geometry as a function of $(q_1,q_2)$. In the crosshatched region $\Kv \cup (-\Kv)$ the attractor is (anti-)BPS.}
\label{fig:GMSVAttractor}
\end{figure}

We begin by ignoring the indeterminate flows. We choose the parameterization $\phi = Y/X$, so that $-1/4 \le \phi < \infty$, with the flop at $\phi = 0$, the CFT boundary at $\phi = -1/4$, and the asymptotic boundary at $\phi = \infty$. As before, for any charge $q_I$, $\cQ^2(\phi) \df a^{I J} q_I q_J$ either has a single critical point (a minimum) at $\phi = \phi_{\rm min}$ for $-1/4 < \phi_{\rm min} < \infty$, or it decreases monotonically upon approaching one of the boundaries. Thus, there is a unique attractor point $\phi_h = \phi_{\rm min}$ whose location depends on $q_I$, as shown in Figure~\ref{fig:GMSVAttractor}.

\begin{figure}
\centering
\begin{subfigure}{0.32\textwidth}
\centering
\includegraphics[height=8.5cm]{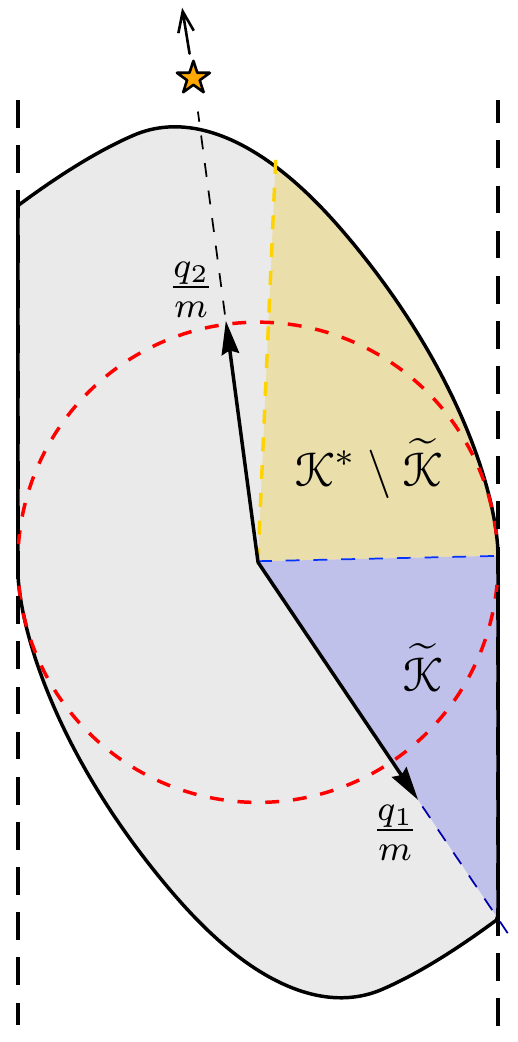}
\caption{$\phi_\infty = -1/5$.} \label{sfig:ExtBoundCFT}
\end{subfigure}
\begin{subfigure}{0.32\textwidth}
\centering
\includegraphics[height=8.5cm]{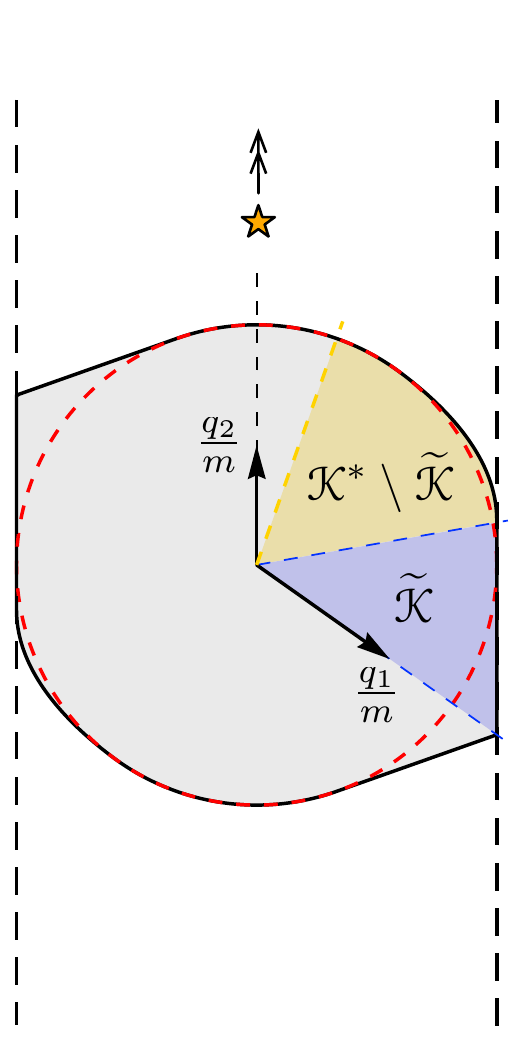}
\caption{$\phi_\infty = 0$.} \label{sfig:ExtBoundGMSVFlop}
\end{subfigure}
\begin{subfigure}{0.32\textwidth}
\centering
\includegraphics[height=8.5cm]{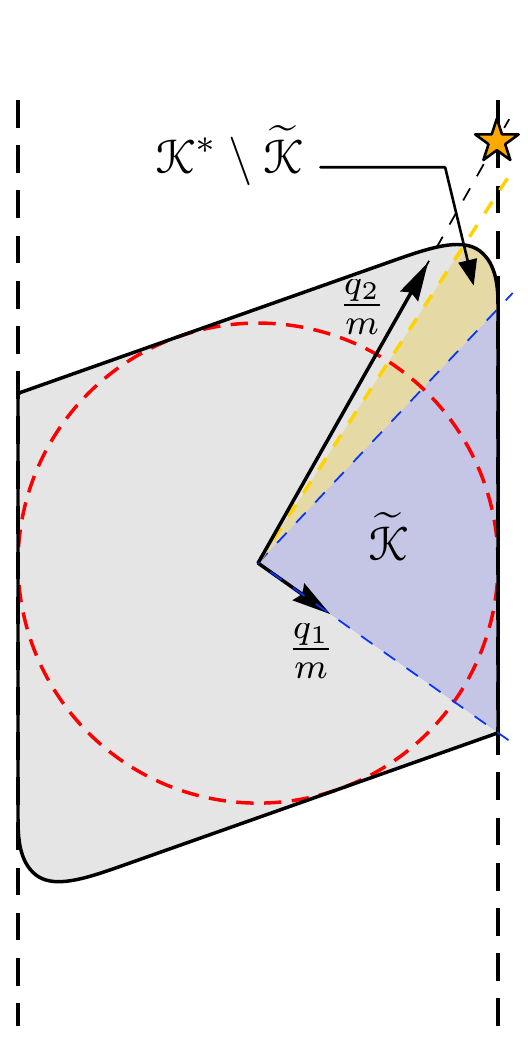}
\caption{$\phi_\infty = 5/2$.} \label{sfig:ExtBoundGMSVFar}
\end{subfigure}
\caption{The apparent extremality bound for the GMSV geometry in three representative vacua, obtained by assuming all indeterminate flows are bad flows. The phase~I lattice basis $(q_1, q_2)$ indicated by the axes is scaled to make the RN black hole region into a disk (outlined by the dashed circle). The dashed line to the right (left) is the BPS (anti-BPS) bound, the cone $\Kv$ where extremal black holes are BPS is shaded blue and the rest of $\Kah^\ast$ is shaded yellow. The flopping hypers are marked by a star in~\subref{sfig:ExtBoundGMSVFar}, whereas they lie outside the field of view in \subref{sfig:ExtBoundCFT} and at infinity in \subref{sfig:ExtBoundGMSVFlop}.}
\label{fig:GMSVExtBound}
\end{figure}

As before, numerically integrating~\eqref{eqn:WPDE} away from the minimum with the initial condition $W(\phi_h) = \sqrt{3} \cQ(\phi_h)$ yields an (apparent) fake superpotential, which determines an (apparent) extremality bound, shown in Figure~\ref{fig:GMSVExtBound}.

This is not the end of the story, however, because although we obtained a fake superpotential satisfying the criteria of~\S\ref{subsec:fakeW} (i.e., globally defined, with all good flows), the bound~\eqref{eqn:Wbound} need not apply to a black hole solution that reaches the CFT boundary a finite distance outside its horizon (should any such solution exist), because the low-energy effective action we used to derive~\eqref{eqn:Wbound} ceases to valid at the boundary. This type of solution is exactly what we would obtain if an indeterminate flow resolves to a good flow upon properly understanding its strongly-coupled portion.

\begin{figure}
\centering
\begin{subfigure}{0.58\textwidth}
\centering
\vspace{1cm}
\includegraphics[height=6cm]{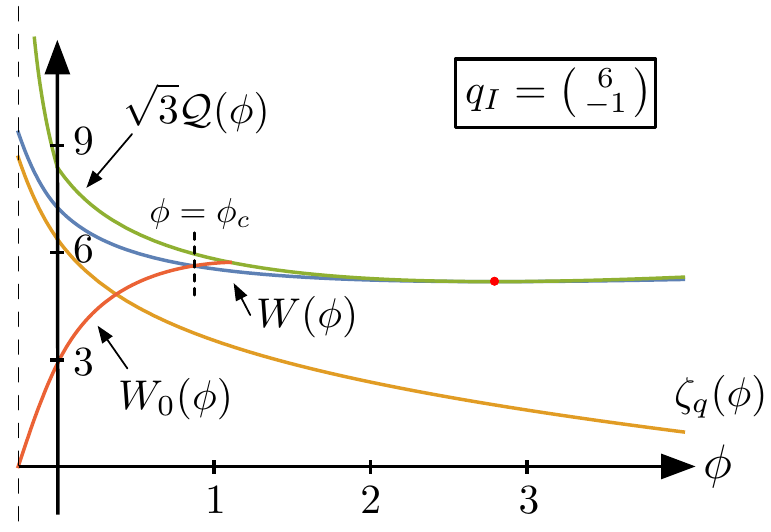}
\vspace{1cm}
\caption{Indeterminate flows are bounded by $W_0(\phi)$\\ \ } \label{sfig:fakeWindet}
\end{subfigure}
\begin{subfigure}{0.4\textwidth}
\centering
\includegraphics[height=8cm]{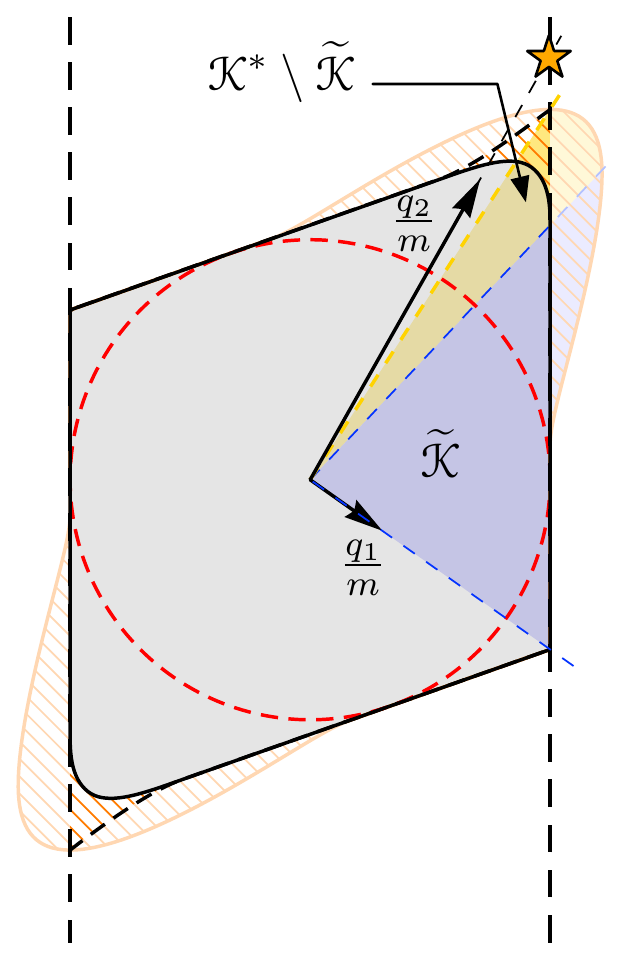}
\caption{Possible modifications to the extremality bound for $\phi_\infty = 5/2$} \label{sfig:extBoundIndet}
\end{subfigure}
\caption{\subref{sfig:fakeWindet}~Indeterminate flows are bounded by $W_0(\phi)$. If $W_0(\phi)$ crosses the apparent fake superpotential $W(\phi)$ at $\phi = \phi_c$ then indeterminate flows are heavier than the attractor flow for $\phi_\infty > \phi_c$, and have no affect on the extremality bound. \subref{sfig:extBoundIndet} If $\phi_c > \phi_\infty$ or if $W_0(\phi)$ and $W(\phi)$ do not cross then some indeterminate flows are lighter than the attractor flow. If all indeterminate flows were good flows, the black hole region would expand to fill the crosshatched area. However, this is not compatible with the BPS bound $W(\phi) \ge |\zeta_q(\phi)|$. Eliminating flows that violate this bound restricts the possible expansion of the black hole region to lie within the dashed lines.}
\label{fig:extBoundIndet}
\end{figure}

To constrain the effect that such indeterminate flows could have, note that they must satisfy $W(\phi) \ge 0$ and $W'(\phi) > 0$. In particular, this implies $W(\phi) \ge W_0(\phi)$ where $W_0(\phi)$ is the $W'(\phi)>0$ solution to~\eqref{eqn:WPDE} satisfying the boundary condition ${W_0(-1/4) = 0}$. Thus, numerically integrating to find this ``lightest possible indeterminate flow'' as in, e.g., Figure~\ref{sfig:fakeWindet}, we obtain a conservative lower bound on the black hole mass that accounts for the possibility of indeterminate flows resolving to good flows. The result is illustrated for a representative choice of vacuum in Figure~\ref{sfig:extBoundIndet}, where the crosshatched areas indicate a conservative outer bound on the black hole region.

Several comments are in order. First, as seen in Figure~\ref{sfig:extBoundIndet},  allowing indeterminate flows has no effect on the extremality bound in certain charge directions. This occurs when $W_0(\phi)$ crosses over the previously-obtained fake superpotential $W(\phi)$ before reaching the chosen vacuum $\phi = \phi_\infty$ (as in, e.g., Figure~\ref{sfig:fakeWindet}), implying that indeterminate flows are heavier than the previously-obtained attractor flow in this vacuum.

Second, the crosshatched region in Figure~\ref{sfig:extBoundIndet} does not respect the BPS bound, so it is not possible for \emph{all} of the indeterminate flows to be good. Indeed, we have not yet leveraged supersymmetry in constraining these flows. The BPS bound implies $W(\phi) \ge |\zeta_q(\phi)|$ at every point along a good flow, hence $W(\phi) \ge W_1(\phi)$ where $W_1(\phi)$ is the $W'(\phi)>0$ solution to~\eqref{eqn:WPDE} satisfying the boundary condition $W_1(-1/4) = |\zeta_q(-1/4)|$. Proceeding analogously to before, one finds that indeterminate flows satisfying the BPS bound are only possible within the dashed wedges in Figure~\ref{sfig:extBoundIndet}.

The possible effect on the extremality bound is therefore rather modest in the $\phi_\infty = +5/2$ vacuum shown in Figure~\ref{sfig:extBoundIndet}. Essentially, this is because the chosen vacuum is relatively ``far'' from the $\phi = -1/4$ CFT boundary. Moving yet farther from the CFT boundary (and towards the asymptotic boundary), the possible effects of indeterminate flows become yet smaller, and the black hole region asymptotes to a parallelogram infinitely far away regardless of the status of these flows, just as in~\S\ref{ssec:symFlop}. By the same token, the possible effects on the extremality bound increase as we approach the CFT boundary.

Regardless of where we are in moduli space, however, certain conclusions remain unchanged: BPS black hole solutions always exist within $\Kv$, and none exist outside $\Kah^\ast$. In particular, the flopping hypermultiplets of charge $q_I = \bigl(\begin{smallmatrix}0\\1\end{smallmatrix} \bigr)$ are always strictly superextremal. For example, the star in Figure~\ref{sfig:extBoundIndet} lies outside the black hole region, regardless of the status of indeterminate flows.

Finally, we emphasize that only the region shown in Figure~\ref{fig:GMSVExtBound} is known to be filled in by black hole solutions. Whether the black hole region is further expanded as in Figure~\ref{sfig:extBoundIndet} depends on the status of various indeterminate flows, which we leave as an interesting question for future research.

\subsection{Example 3: the $h^{1,1} = 3$ KMV conifold}\label{ssec:KMV}

Our final example is the $(h^{1,1},h^{2,1}) = (3,243)$ Calabi-Yau given by the degree $(6,0,0)$ hypersurface in the ambient toric fourfold described by the gauged linear sigma model (GLSM):
\begin{equation}
\begin{array}{c|ccccccc}
& x_1 & x_2 & x_3 & x_4 & x_5 & x_6 & x_7 \\ \hline
U(1)_X & 0 & 0 & 0 & 0 & 2 & 3 & 1 \\ 
U(1)_Y & 0 & 0 & 1 & 1 & 0 & 0 & -2 \\
U(1)_Z & 1 & 1 & 0 & -1 & 0 & 0 & -1 
\end{array} \;. \label{eqn:KMVglsm}
\end{equation}
This geometry, previously studied by Klemm, Mayr, and Vafa \cite{Klemm:1996hh}, is an elliptic fibration over $\mathbb{F}_1 = dP_1$ (with base coordinates $x_1, \ldots, x_4$), and also a K3 fibration over $\mathbb{P}^1$ (with base coordinates $x_1, x_2$).

\subsubsection{Geometry} 
\begin{figure}
\centering
\begin{subfigure}{0.52\textwidth}
\centering
\includegraphics[height=6.5cm]{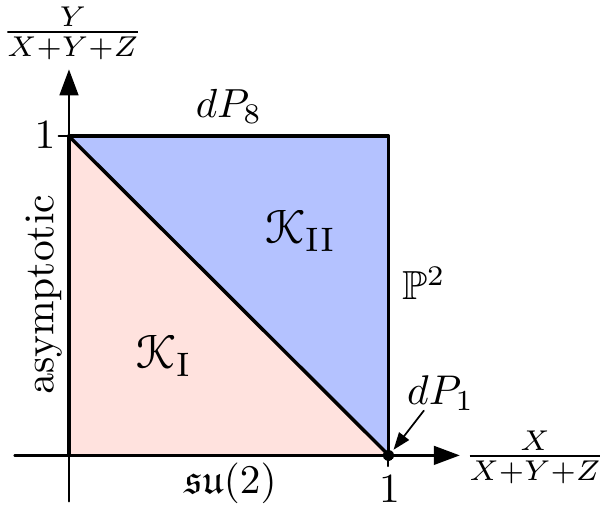}
\caption{The extended K\"ahler cone $\Kah$.} \label{sfig:kmvK}
\end{subfigure}
\begin{subfigure}{0.47\textwidth}
\centering
\includegraphics[height=6.5cm]{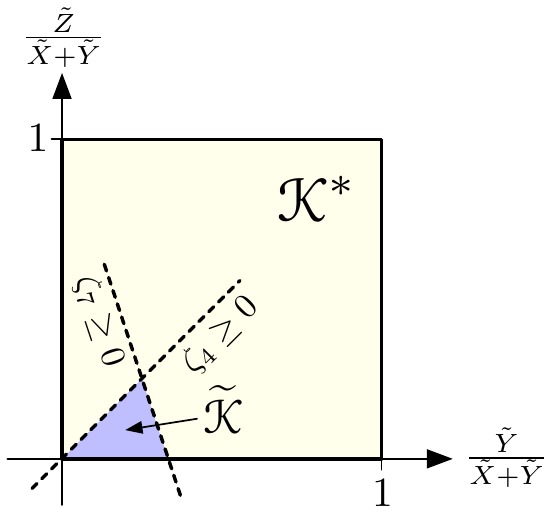}
\caption{Determining $\Kv$ using~\eqref{eqn:KvFormula}.} \label{sfig:kmvKvInt}
\end{subfigure}
\caption{\subref{sfig:kmvK} The extended K\"ahler cone $\Kah$ for the KMV geometry sliced through the $X+Y+Z=1$ plane, with labeled boundaries. \subref{sfig:kmvKvInt} The cone of dual coordinates $\Kv$ can be determined from $\Kah^\ast$ and the charges of the tensionless strings at the finite-distance boundaries using~\eqref{eqn:KvFormula}.}
\label{fig:KMVcones1}
\end{figure}

There are two phases, with associated Fayet-Iliopoulos parameters $\xi_X, \xi_Y, \xi_Z \ge 0$ in phase I and  $\xi_X, \xi_Y \ge - \xi_Z \ge 0$ in phase II. The phase I prepotential is
\begin{equation}
\cF^{\text{(I)}} = \frac{4}{3} X^3 + \frac{3}{2} X^2 Y + \frac{1}{2} X Y^2+ X^2 Z + X Y Z \,, \qquad \text{with} \qquad \Kah_{\text{I}} = \{ X,Y,Z \ge 0 \}\,.
\end{equation}
There is an asymptotic boundary at $X=0$, whereas the $dP_1$ toric divisor $x_7 = 0$ shrinks to a $\mathbb{P}^1$ at $Y=0$ and a single curve inside $dP_1$ flops at $Z=0$. Passing through the flop we reach phase II, with prepotential
\begin{multline}
\cF^{\text{(II)}} =  \frac{4}{3}  \check{X}^3 + \frac{3}{2} \check{X}^2 \check{Y} + \frac{1}{2}   \check{X} \check{Y}^2 + \frac{9}{2} \check{X}^2 \check{Z} + \frac{9}{2} \check{X} \check{Z}^2 + \frac{1}{2} \check{Y}^2 \check{Z}  \\
+ \frac{3}{2} \check{Y}  \check{Z}^2+ \frac{3}{2} \check{Z}^3 +3 \check{X} \check{Y} \check{Z} \,  \qquad \text{with} \qquad \Kah_{\text{II}} = \{ \check{X}, \check{Y}, \check{Z} \ge 0 \}\,,
\end{multline}
where $\check{X} = X+Z$, $\check{Y}=Y+Z$, $\check{Z} = -Z$, and $\cF^{\text{(II)}}  = \cF^{\text{(I)}}  - \frac{1}{6} Z^3$ in agreement with~\eqref{eqn:flop}. In phase II, the $dP_8$ toric divisor $x_4 = 0$ shrinks to a point at $\check{X}=0$, whereas the $\mathbb{P}^2$ toric divisor $x_7 = 0$ shrinks to a point at $\check{Y}=0$; the latter is the same divisor that shrinks to a $\mathbb{P}^1$ at $Y=0$ in phase I, but with a changed topology owing to the flop. Thus, the extended K\"ahler cone $\Kah = \{X,Y\ge 0; X,Y \ge -Z\}$ pictured in Figure~\ref{sfig:kmvK} has an asymptotic boundary, an $\mathfrak{su}(2)$ boundary, and two CFT boundaries. Note that the $dP_1$ toric divisor $x_7 = 0$ shrinks to a point at the intersection of the $Y=0$ boundary with the flop, hence the $dP_1$ CFT sits at the intersection of the $\mathfrak{su}(2)$ boundary with the $\mathbb{P}^2$ CFT boundary, straddling the two phases.

The cone of dual coordinates $\Kv$ can be determined in multiple ways. The quickest is to apply~\eqref{eqn:KvFormula}, noting that the divisors shrinking at finite-distance boundaries are the toric divisors $x_4 = 0$ and $x_7=0$ of charges $(0,1,-1)$ and $(1,-2,-1)$ respectively.\footnote{The charge $\tilde{q}^I$ of the toric divisor $x_i =0$ is given by the $i$th column of the corresponding GLSM, \eqref{eqn:KMVglsm}.} Thus, since $\Kah^\ast = \{ \tilde{X},\tilde{Y},\tilde{Z} \ge 0; \tilde{X}+\tilde{Y} \ge \tilde{Z}\}$,
\begin{equation}
\Kv = \Kah^\ast \cap \{\tilde{Y}\ge\tilde{Z} \} \cap \{\tilde{X} \ge 2 \tilde{Y} + \tilde{Z}\} = \{\tilde{X} \ge 2 \tilde{Y} + \tilde{Z}; \tilde{Y} \ge \tilde{Z} \ge 0\} \,,
\end{equation}
as illustrated in Figure~\ref{sfig:kmvKvInt}.
 
More explicitly, we can find $\Kv$ by taking the image of $\Kah$ under the map $\tilde{Y} = \tilde{Y}[Y]$ determined by the prepotential.  
For instance, in phase I,
\begin{equation}
\cF_I = \begin{pmatrix} 4 X^2 + 3 X Y + \frac{1}{2}Y^2 + 2 X Z+Y Z \\ \frac{3}{2} X^2 + X Y + X Z \\ X^2 + X Y \end{pmatrix} \,.
\end{equation}
Thus, $\cF_I = {\scriptsize \biggl(\begin{smallmatrix} 4 X^2 + 2 X Z \\ \frac{3}{2} X^2 +X Z \\ X^2  \end{smallmatrix}\biggr)}$ on the $Y=0$ boundary, varying between $\Yd_{\ast I} = \Bigl(\begin{smallmatrix}2\\1\\0\end{smallmatrix}\Bigr)$ and $\Yd_{\ast I} = \Bigl(\begin{smallmatrix}8\\3\\2\end{smallmatrix}\Bigr)$ along the plane $\tilde{X} - 2 \tilde{Y} - \tilde{Z} = 0$ as we go from $X=0$ to $Z=0$. 
 Taking $X\to 0$ with fixed $Y \ne 0$, we obtain $\Yd_{\ast I} = \Bigl(\begin{smallmatrix}1\\0\\0\end{smallmatrix}\Bigr)$ independent of $Y$ and $Z$. On the other hand, taking $X, Y \to 0$ with fixed $Z \ne 0$, we obtain some point in the cone generated by $\Yd_{\ast I} = \Bigl(\begin{smallmatrix}1\\0\\0\end{smallmatrix}\Bigr)$ and $\Yd_{\ast I} = \Bigl(\begin{smallmatrix}2\\1\\0\end{smallmatrix}\Bigr)$, depending on $Y/Z$. Thus, the map between the asymptotic boundaries of $\Kah$ and $\Kv$ is not one-to-one, as anticipated in~\S\ref{subsec:asymptotic}.
 
 \begin{figure}
\centering
\begin{subfigure}{0.51\textwidth}
\centering
\includegraphics[height=6.5cm]{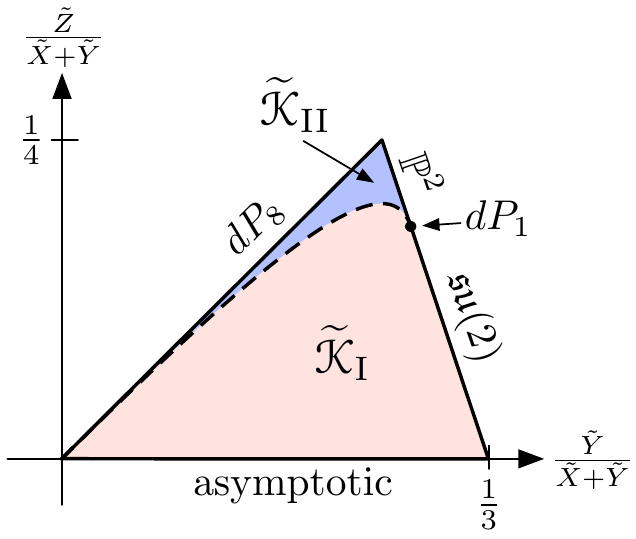}
\caption{The cone of dual coordinates $\Kv$.} \label{sfig:kmvKvPhases}
\end{subfigure}
\begin{subfigure}{0.48\textwidth}
\centering
\includegraphics[height=6.5cm]{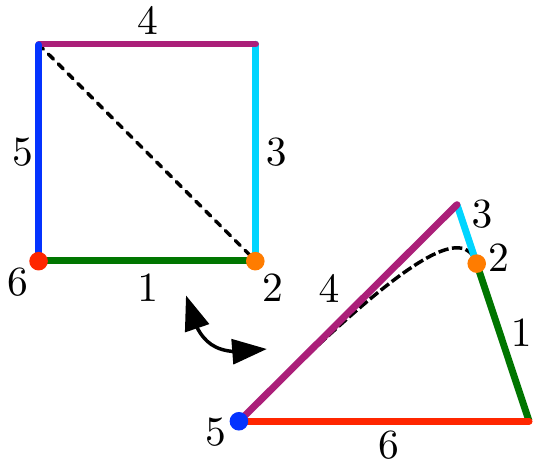}
\caption{Map between the boundaries of $\Kah$ and $\Kv$.} \label{sfig:KMVboundaries}
\end{subfigure}
\caption{\subref{sfig:kmvKvPhases} The cone of dual coordinates $\Kv$ for the KMV geometry sliced through the $\tilde{X}+\tilde{Y}=1$ plane, with labeled boundaries. \subref{sfig:KMVboundaries} The map between the boundaries of $\Kah$ and $\Kv$ is non-trivial. Facets and rays are exchanged on asymptotic boundaries, whereas two different finite-distance $\Kah$ facets map to different parts of the same $\Kv$ facet.}
\label{fig:KMVcones2}
\end{figure}
 
After examining the boundaries of phase II in an analogous fashion, we obtain the explicit picture of $\Kv$ shown in Figure~\ref{sfig:kmvKvPhases}, with the complicated map between the boundaries of $\Kah$ and $\Kv$ illustrated in Figure~\ref{sfig:KMVboundaries}. Unlike this boundary map, the map between the interiors of $\Kah$ and $\Kv$ is necessarily bijective; its exact form (explicitly determined by the prepotential) will not be too important. However, it is notable that the dual-coordinate image of the $Z=0$ flop is the quadric
\begin{equation}
 (\tilde{X} - 2 \tilde{Y} - \tilde{Z}) (\tilde{Y} - \tilde{Z}) = \biggl[\frac{3}{2} \tilde{Z} - \tilde{Y}\biggr]^2 , \qquad \frac{3}{2} \tilde{Z} - \tilde{Y} \ge 0 \,.
\end{equation}
Since this is a curved surface, neither $\Kv_{\text{I}}$ nor $\Kv_{\text{II}}$ is a polyhedral cone, and the latter is not even convex. This illustrates the non-trivial nature of the~\S\ref{sec:boundaries} results implying that $\Kv = \Kv_{\text{I}} \cup \Kv_{\text{II}}$ is convex.

As final comment, we observe that just like at its finite-distance boundaries, a divisor also shrinks at the $\tilde{Z} = 0$ asymptotic boundary of $\Kv$, specifically the charge $(0,0,1)$ toric divisor $x_1 = 0$ corresponding to the fiber of the K3 fibration over $\mathbb{P}^1$. The effective cone $\Eff$ is therefore generated by the $x_1 = 0$, $x_4=0$, and $x_7=0$ toric divisors, with $\Kv = \Eff^\ast$ in agreement with the general results of~\S\ref{sec:boundaries}.

\subsubsection{GV invariants and the T/sLWGC}

The genus 0 GV invariants for phase I of this geometry are shown in Table~\ref{tab:KMVinvs}. As in both previous examples, there is a holomorphic curve in every non-trivial class within $\Kv$ to the degree calculated, in agreement with the $k=1$ T/sLWGC for these directions.

\afterpage{ 
\begin{table}[H] 
{\scriptsize \renewcommand{\arraystretch}{0.9} 
\begin{align*}
\rotatebox[origin=c]{90}{\underline{$q_2=0$}} &\noeq
\begin{array}{c|ccccccc}
 \mathdiagbox[width=0.7cm,height=0.5cm,innerleftsep=0.1cm,innerrightsep=0cm]{q_3}{q_1} & 0 & 1 & 2 & 3 & 4 & 5 & 6 \\ \hline
 0& \cellcolor{KvColor}- & \cellcolor{KvColor}480 & \cellcolor{KvColor}480 & \cellcolor{KvColor}480 & \cellcolor{KvColor}480 & \cellcolor{KvColor}480 & \cellcolor{KvColor}480 \\
 \hhline{~|-|------}
 1& \multicolumn{1}{c|}{1} & \cellcolor{KapColor}252 & \cellcolor{KapColor}5130 & \cellcolor{KapColor}54760 & \cellcolor{KapColor}419895 & \cellcolor{KapColor}2587788 & \cellcolor{KapColor}13630694 \\
 \hhline{~|~|-|>{\arrayrulecolor{KapColor}}----->{\arrayrulecolor{black}}}
 2& 0 & \multicolumn{1}{c|}{0} & \cellcolor{KapColor}-9252 & \cellcolor{KapColor}-673760 & \cellcolor{KapColor}-20534040 & \cellcolor{KapColor}-389320128 & \cellcolor{KapColor}-5398936120 \\
  \hhline{~|~~|-|>{\arrayrulecolor{KapColor}}---->{\arrayrulecolor{black}}}
  3&0 & 0 & \multicolumn{1}{c|}{0} & \cellcolor{KapColor}848628 & \cellcolor{KapColor}115243155 & \cellcolor{KapColor}6499779552 & \cellcolor{KapColor}219488049810 \\
   \hhline{~|~~~|-|>{\arrayrulecolor{KapColor}}--->{\arrayrulecolor{black}}}
 4& 0 & 0 & 0 & \multicolumn{1}{c|}{0} & \cellcolor{KapColor}-114265008 & \cellcolor{KapColor}-23064530112 & \cellcolor{KapColor}-1972983690880 \\
  \hhline{~|~~~~|-|>{\arrayrulecolor{KapColor}}-->{\arrayrulecolor{black}}}
 5& 0 & 0 & 0 & 0 & \multicolumn{1}{c|}{0} & \cellcolor{KapColor}18958064400 & \cellcolor{KapColor}5105167984850 \\
  \hhline{~|~~~~~|-|>{\arrayrulecolor{KapColor}}->{\arrayrulecolor{black}}}
 6& 0 & 0 & 0 & 0 & 0 & \multicolumn{1}{c|}{0} & \cellcolor{KapColor}-3589587111852 \\
\end{array} \\[5pt]
\rotatebox[origin=c]{90}{\underline{$q_2=1$}} &\noeq
\begin{array}{c|ccccccc}
 \mathdiagbox[width=0.7cm,height=0.5cm,innerleftsep=0.1cm,innerrightsep=0cm]{q_3}{q_1} & 0 & 1 & 2 & 3 & 4 & 5 & 6 \\ \hline
0& \multicolumn{1}{c|}{\cellcolor{KaColor}-2} & \multicolumn{1}{c|}{\cellcolor{KvpColor}480} & \cellcolor{KvColor}282888 & \cellcolor{KvColor}17058560 & \cellcolor{KvColor}477516780 & \cellcolor{KvColor}8606976768 & \cellcolor{KvColor}115311621680 \\
\hhline{~|>{\arrayrulecolor{KaColor}}->{\arrayrulecolor{black}}|-|-|>{\arrayrulecolor{KvColor}}---->{\arrayrulecolor{black}}}
1& \cellcolor{KaColor}3 & \multicolumn{1}{c|}{\cellcolor{KaColor}-960} & \multicolumn{1}{c|}{\cellcolor{KapColor}118170} & \cellcolor{KvColor}186280704 & \cellcolor{KvColor}20229416355 & \cellcolor{KvColor}1010446016832 & \cellcolor{KvColor}31547177925210 \\
\hhline{~|->{\arrayrulecolor{KaColor}}->{\arrayrulecolor{black}}|-|----}
2& \multicolumn{1}{c|}{0} & \cellcolor{KaColor}0 & \multicolumn{1}{c|}{\cellcolor{KaColor}-10260} & \cellcolor{KapColor}1569600 & \cellcolor{KapColor}1634529240 & \cellcolor{KapColor}220572616320 & \cellcolor{KapColor}14589435693600 \\
\hhline{~|~|-|>{\arrayrulecolor{KaColor}}->{\arrayrulecolor{black}}|-|>{\arrayrulecolor{KapColor}}--->{\arrayrulecolor{black}}}
3& 0 & \multicolumn{1}{c|}{0} & \cellcolor{KaColor}0 & \multicolumn{1}{c|}{\cellcolor{KaColor}1347520} & \cellcolor{KapColor}-112869600 & \cellcolor{KapColor}-227921708160 & \cellcolor{KapColor}-43955742846400 \\
\hhline{~|~~|-|>{\arrayrulecolor{KaColor}}->{\arrayrulecolor{black}}|-|>{\arrayrulecolor{KapColor}}-->{\arrayrulecolor{black}}}
4& 0 & 0 & \multicolumn{1}{c|}{0} & \cellcolor{KaColor}0 & \multicolumn{1}{c|}{\cellcolor{KaColor}-230486310} & \cellcolor{KapColor}2796677280 & \cellcolor{KapColor}40270571823240 \\
\hhline{~|~~~|-|>{\arrayrulecolor{KaColor}}->{\arrayrulecolor{black}}|-|>{\arrayrulecolor{KapColor}}->{\arrayrulecolor{black}}}
5& 0 & 0 & 0 & \multicolumn{1}{c|}{0} & \cellcolor{KaColor}0 & \multicolumn{1}{c|}{\cellcolor{KaColor}46129060224} & \cellcolor{KapColor}2744302285440 \\
\hhline{~|~~~~|-|>{\arrayrulecolor{KaColor}}->{\arrayrulecolor{black}}|-|>{\arrayrulecolor{KapColor}}>{\arrayrulecolor{black}}}
6& 0 & 0 & 0 & 0 & \multicolumn{1}{c|}{0} & \cellcolor{KaColor}0 & \cellcolor{KaColor}\cdots \\
\end{array} \\[5pt]
\rotatebox[origin=c]{90}{\underline{$q_2=2$}} &\noeq
\begin{array}{c|cccccc}
\mathdiagbox[width=0.7cm,height=0.5cm,innerleftsep=0.1cm,innerrightsep=0cm]{q_3}{q_1} & 0 & 1 & 2 & 3 & 4 & 5  \\ \hline
 0&\cellcolor{KaColor}0 & \multicolumn{1}{c|}{\cellcolor{KaColor}0} & \cellcolor{KvpColor}480 & \multicolumn{1}{c|}{\cellcolor{KvpColor}17058560} & \cellcolor{KvColor}8606976768 & \cellcolor{KvColor}1242058447872 \\
 \hhline{~|>{\arrayrulecolor{KaColor}}-->{\arrayrulecolor{black}}|-|-|-|>{\arrayrulecolor{KvColor}}->{\arrayrulecolor{black}}}
 1&\cellcolor{KaColor}5 & \cellcolor{KaColor}-1920 & \multicolumn{1}{c|}{\cellcolor{KaColor}339120} & \multicolumn{1}{c|}{\cellcolor{KapColor}-68861720} & \multicolumn{1}{c|}{\cellcolor{KvpColor}28474940475} & \cellcolor{KvColor}58862632312080  \\
  \hhline{~|>{\arrayrulecolor{KaColor}}--->{\arrayrulecolor{black}}|-|-|-}
2& \cellcolor{KaColor}-6 & \cellcolor{KaColor}2400 & \cellcolor{KaColor}-473640 & \multicolumn{1}{c|}{\cellcolor{KaColor}58929120} & \cellcolor{KapColor}-45370251660 & \cellcolor{KapColor}6216836921280  \\
 \hhline{~|-|>{\arrayrulecolor{KaColor}}--->{\arrayrulecolor{black}}|-|>{\arrayrulecolor{KapColor}}->{\arrayrulecolor{black}}}
3& \multicolumn{1}{c|}{0} & \cellcolor{KaColor}0 & \cellcolor{KaColor}25650 & \cellcolor{KaColor}-6387920 & \multicolumn{1}{c|}{\cellcolor{KaColor}751266900} & \cellcolor{KapColor}-496866067920 \\
 \hhline{~|~|-|>{\arrayrulecolor{KaColor}}--->{\arrayrulecolor{black}}|-}
4& 0 & \multicolumn{1}{c|}{0} & \cellcolor{KaColor}0 & \cellcolor{KaColor}-3368800 & \cellcolor{KaColor}492546480 & \cellcolor{KaColor}-27563993280 \\
 \hhline{~|~~|-|>{\arrayrulecolor{KaColor}}--->{\arrayrulecolor{black}}}
5& 0 & 0 & \multicolumn{1}{c|}{0} & \cellcolor{KaColor}0 & \cellcolor{KaColor}576215775 & \cellcolor{KaColor}-24186268224 \\
\end{array} \\[5pt]
\rotatebox[origin=c]{90}{\underline{$q_2=3$}} &\noeq
\begin{array}{c|cccccc}
 \mathdiagbox[width=0.7cm,height=0.5cm,innerleftsep=0.1cm,innerrightsep=0cm]{q_3}{q_1} & 0 & 1 & 2 & 3 & 4 & 5  \\ \hline
 0&\cellcolor{KaColor}0 & \cellcolor{KaColor}0 & \multicolumn{1}{c|}{\cellcolor{KaColor}0} & \cellcolor{KvpColor}480 & \cellcolor{KvpColor}477516780 & \cellcolor{KvpColor}1242058447872 \\
  \hhline{~|>{\arrayrulecolor{KaColor}}--->{\arrayrulecolor{black}}|-|-|>{\arrayrulecolor{KvpColor}}->{\arrayrulecolor{black}}}
 1&\cellcolor{KaColor}7 & \cellcolor{KaColor}-2880 & \cellcolor{KaColor}565200 & \multicolumn{1}{c|}{\cellcolor{KaColor}-137832960} & \multicolumn{1}{c|}{\cellcolor{KapColor}24736992255} & \cellcolor{KvpColor}-3999822251328 \\
   \hhline{~|>{\arrayrulecolor{KaColor}}---->{\arrayrulecolor{black}}|-|-}
 2&\cellcolor{KaColor}-32 & \cellcolor{KaColor}16800 & \cellcolor{KaColor}-4126140 & \cellcolor{KaColor}715799200 & \multicolumn{1}{c|}{\cellcolor{KaColor}-173786110560} & \cellcolor{KapColor}25655678309952 \\
    \hhline{~|>{\arrayrulecolor{KaColor}}----->{\arrayrulecolor{black}}|-}
 3&\cellcolor{KaColor}27 & \cellcolor{KaColor}-15360 & \cellcolor{KaColor}4166190 & \cellcolor{KaColor}-718097280 & \cellcolor{KaColor}189132227700 & \cellcolor{KaColor}-33989873269248 \\
     \hhline{~|-|>{\arrayrulecolor{KaColor}}----->{\arrayrulecolor{black}}}
4& \multicolumn{1}{c|}{0} & \cellcolor{KaColor}0 & \cellcolor{KaColor}-164160 & \cellcolor{KaColor}58002560 & \cellcolor{KaColor}-10048515390 & \cellcolor{KaColor}2249322638400 \\
     \hhline{~|~|-|>{\arrayrulecolor{KaColor}}---->{\arrayrulecolor{black}}}
5& 0 & \multicolumn{1}{c|}{0} & \cellcolor{KaColor}0 & \cellcolor{KaColor}21560320 & \cellcolor{KaColor}-5100306330 & \cellcolor{KaColor}\cdots \\
\end{array} \\[5pt]
\rotatebox[origin=c]{90}{\underline{$q_2=4$}} &\noeq
\begin{array}{c|cccc}
 \mathdiagbox[width=0.7cm,height=0.5cm,innerleftsep=0.1cm,innerrightsep=0cm]{q_3}{q_1} & 0 & 1 & 2 & 3 \\ \hline
 0&\cellcolor{KaColor}0 & \cellcolor{KaColor}0 & \cellcolor{KaColor}0 & \cellcolor{KaColor}0 \\
 1&\cellcolor{KaColor}9 & \cellcolor{KaColor}-3840 & \cellcolor{KaColor}791280 & \cellcolor{KaColor}-206749440 \\
2& \cellcolor{KaColor}-110 & \cellcolor{KaColor}64800 & \cellcolor{KaColor}-18267600 & \cellcolor{KaColor}3851600480 \\
 3&\cellcolor{KaColor}286 & \cellcolor{KaColor}-192000 & \cellcolor{KaColor}61299990 & \cellcolor{KaColor}-12935977120 \\
4& \cellcolor{KaColor}-192 & \cellcolor{KaColor}137280 & \cellcolor{KaColor}-47740800 & \cellcolor{KaColor}10676880000 \\
     \hhline{~|-|>{\arrayrulecolor{KaColor}}--->{\arrayrulecolor{black}}}
5& \multicolumn{1}{c|}{0} &\cellcolor{KaColor} 0 & \cellcolor{KaColor}1467180 & \cellcolor{KaColor}-675755000 \\
\hhline{~|~|-|>{\arrayrulecolor{KaColor}}-->{\arrayrulecolor{black}}}
 6& 0 & \multicolumn{1}{c|}{0} & \cellcolor{KaColor}0 & \cellcolor{KaColor}\cdots \\
\end{array}
\end{align*}}
\caption{Genus 0 GV invariants of degree $(q_1, q_2, q_3)$ for the KMV geometry. The blue region $q_1\ge 2 q_2 + q_3$, $q_2 \ge q_3 \ge 0$ is the cone $\Kv$ where the BPS and black hole extremality bounds agree unconditionally and the magenta region is the remainder of its Weyl orbit $\cW(\Kv)$. The yellow regions form the rest of $\Kah^\ast$, with the dark yellow region within $\cW(\Kah)^\ast$. 
See~\S\ref{sssec:GVrevisit} for further discussion of the significance of $\cW(\Kv)$ and $\cW(\Kah)^\ast$.
Invariants recorded as ``$\cdots$'' were not calculated.}  
\label{tab:KMVinvs}
\end{table}\clearpage}

However, unlike before there are vanishing GV invariants within $\Kah^\ast$, although only on its boundary (and always outside $\Kv$). In particular, there are directions along the boundary of $\Kah^\ast$ in which only finitely many GV invariants are nonvanishing. If the BPS and black hole extremality bounds agreed in these directions (due to indeterminate flows resolving to good flows), then we would be in danger of violating the T/sLWGC.\footnote{A violation could still be avoided if there were an infinite tower of holomorphic curves that happened to cancel from the GV invariants.} 

For the time being, we limit our discussion of this phenomenon to a single example.
Consider the direction parallel to the $q_I = (0,1,0)$ charge of the $\mathfrak{su}(2)$ $W^\pm$ bosons, whose presence is reflected in the $-2$ GV invariant at this degree. BPS states with charges in this direction become massless at the $Y=0$ boundary of $\Kah$ where the $\mathfrak{su}(2)$ enhancement occurs. Since the lower-energy effective theory remains weakly coupled at this boundary, we do not expect infinite towers of BPS states to become massless, and indeed there are no further nonvanishing GV invariants in this direction.

Likewise, if there were BPS black holes in this charge direction they would become massless at the $Y=0$ boundary, but massless black holes are inconsistent with positive energy theorems (e.g.,~\cite{Witten:1981mf, Gibbons:1982jg}), so the black hole extremality bound must differ from the BPS bound in this direction, and there is no contradiction with the T/sLWGC.

We will have more to say about the other boundaries of $\Kah^\ast$ that lack infinite towers of nonvanishing GV invariants in~\S\ref{sssec:GVrevisit}.

\subsubsection{BPS versus extremal} \label{sssec:KMVbpsExt}

As in the previous examples, the first step in determining the black hole extremality bound is to locate the attractor points, i.e., the local minima of $\mathcal{Q}(\phi)$ in the interior of the moduli space. Given such a minimum,~\eqref{eqn:WPDE} can be integrated to obtain a (local) fake superpotential $W(\phi)$ with the same minimum, with the associated (quasi)extremal black holes given by solutions to the gradient flow equations~\eqref{eq:fogf}.\footnote{The $W(\phi)$ obtained in this way cannot necessarily be extended across the entire moduli space, i.e., there may be multiple \emph{local} fake superpotentials associated to different attractor points. So long as at least one local fake superpotential is defined at each point in the moduli space (and ignoring the issue of indeterminate flows) the extremality bound is determined by the minimum of all the local fake superpotentials defined at the point in question~\cite{BenBH}.}

For certain charges $q_I$, the attractor points will migrate to the boundaries of the moduli space. This is technically different from the above case, but because in the previous examples this only happened for non-generic choices of $q_I$, we have glossed over this difference so far. In particular, the behavior of these boundary attractor points can be understood by taking a limit of an interior attractor point as the charge $q_I$ approaches a special value.

In the present case, however, a puzzle arises: it is no longer the case that $\cQ(\phi)$ has a local minimum within $\Kah$ for a generic charge $q_I$. In particular, although $\cQ(\phi)$ can only have a minimum along the asymptotic and CFT boundaries for non-generic $q_I$, this is not true of the $\mathfrak{su}(2)$ boundary. For example, letting $q_I$ approach the $\mathfrak{su}(2)$ boundary of $\Kv$, the corresponding BPS attractor point $\Yd_I \propto q_I$ likewise approaches the $\mathfrak{su}(2)$ boundary of $\Kah$, reaching it when $q_I$ reaches the boundary of $\Kv$. However, continuing $q_I$ outside $\Kv$, the minimum of $\mathcal{Q}(\phi)$ does not return to the interior of $\Kah$, but rather remains on the $\mathfrak{su}(2)$ boundary.

\begin{figure}
\centering
\includegraphics[height=3.75cm]{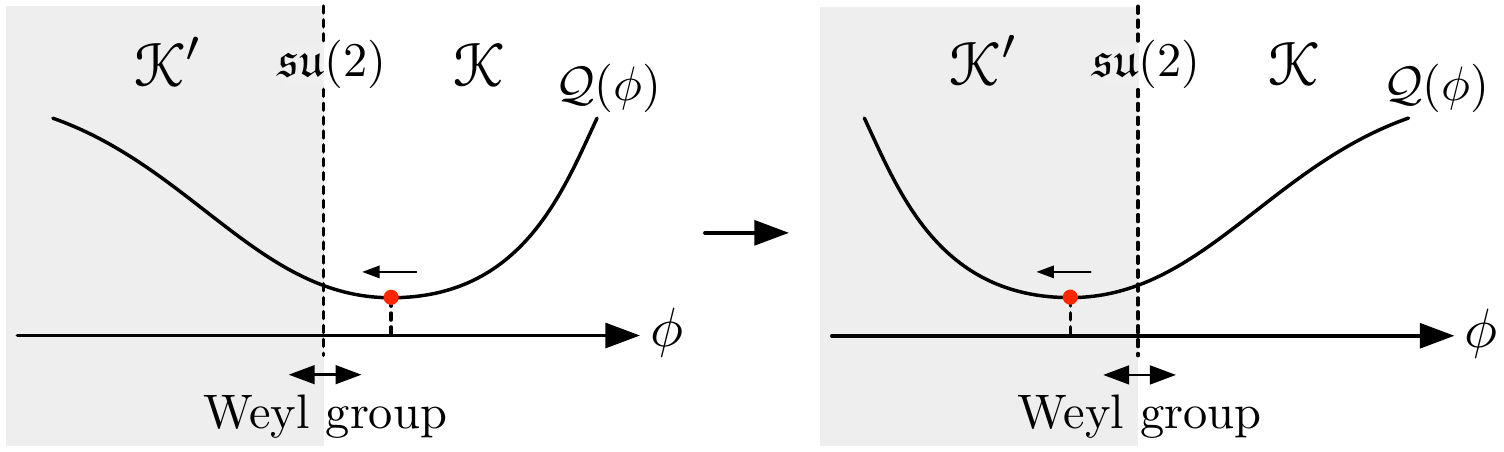}
\caption{An attractor point moves from $\Kah$ into its Weyl reflection $\Kah'$ by passing through the $\mathfrak{su}(2)$ locus.}
\label{fig:su2vanish}
\end{figure}

The explanation for this peculiar behavior is simple.
As discussed in~\S\ref{subsec:PTs} and~\S\ref{subsec:finitedistance}, we can view the moduli space $\Kah$ as the quotient of a larger space $\cW(\Kah)$ by the $\mathbb{Z}_2$ Weyl group of $\mathfrak{su}(2)$. This occurs naturally as the field space of the fundamental scalar fields in the Lagrangian, after truncating the scalars that are charged under the $\mathfrak{u}(1)$ Cartan subalgebra of $\mathfrak{su}(2)$. Because the Weyl group also acts nontrivially on the charge $q_I$, the charge function $\mathcal{Q}(\phi) = \sqrt{a^{I J}(\phi) q_I q_J}$ is not Weyl-invariant, and is better understood as a function on $\cW(\Kah)$ rather than on $\Kah$. In particular, an apparent minimum of $\mathcal{Q}(\phi)$ along the $\mathfrak{su}(2)$ boundary of $\Kah$ need not actually be a minimum when viewed within the larger space $\cW(\Kah)$. In the above example, the attractor point does not actually remain on the $\mathfrak{su}(2)$ boundary as $q_I$ moves outside of $\Kv$. Rather, it continues into the Weyl image of the extended K\"ahler cone, $\Kah'$, as illustrated in Figure~\ref{fig:su2vanish}.

\begin{figure}
\centering
\begin{subfigure}{0.45\textwidth}
\centering
\includegraphics[height=5.75cm]{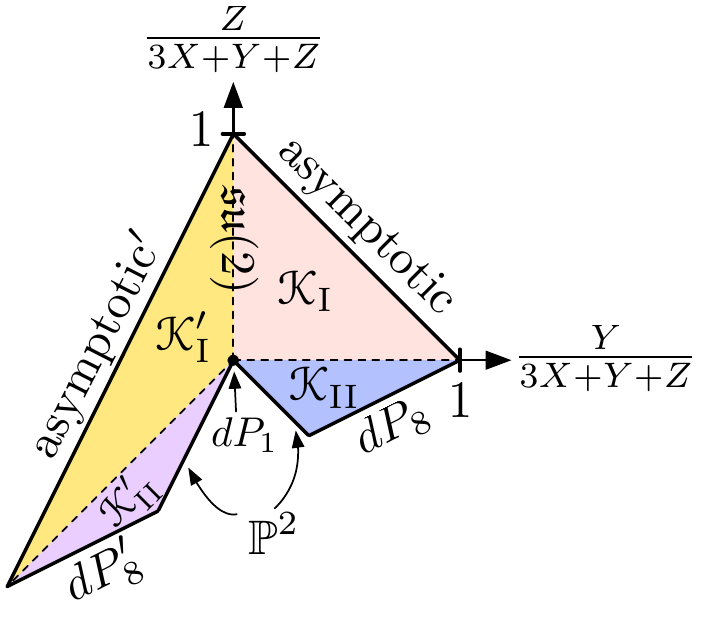}
\caption{$\cW(\Kah)$} \label{sfig:kmvKWeyl}
\end{subfigure}
\begin{subfigure}{0.54\textwidth}
\centering
\includegraphics[height=5.75cm]{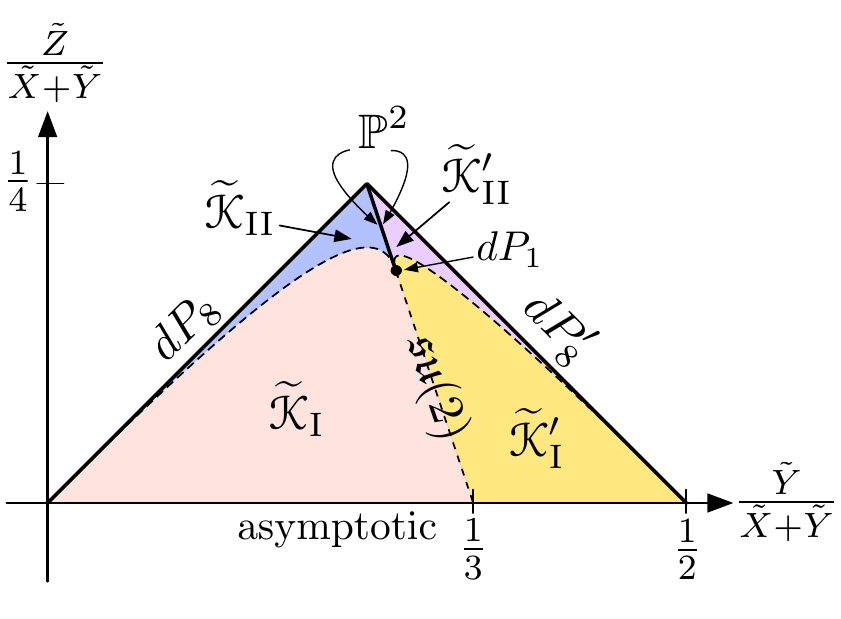}
\caption{$\cW(\Kv)$} \label{sfig:kmvKvWeyl}
\end{subfigure}
\caption{\subref{sfig:kmvKWeyl} The Weyl orbit of the extended K\"ahler cone $\cW(\Kah)$ for the KMV geometry sliced through the $3X+Y+Z=1$ plane, with labeled boundaries. \subref{sfig:kmvKvWeyl} The Weyl orbit of the cone of dual coordinates $\cW(\Kv)$ sliced through the $\tilde{X}+\tilde{Y}=1$ plane.}
\label{fig:KMVconesWeyl}
\end{figure}

Thus, to understand the attractor mechanism in geometries with $\mathfrak{su}(2)$ boundaries, we must consider not just $\Kah$, but the entire Weyl orbit:
\begin{equation}
\cW(\Kah) = \Kah \cup \Kah'\,.
\end{equation}
As a first step, we describe $\cW(\Kah)$ in the present example. The Weyl reflection associated to an $\mathfrak{su}(2)$ boundary can be written in a basis-independent fashion as
\begin{equation}
Y^I \to Y^I - 2 \frac{q_J^{\rm W} Y^J}{q_K^{\rm W}  \tilde{q}^K_{\rm mon}} \tilde{q}^I_{\rm mon} \,, \label{eqn:WeylReflect}
\end{equation}
where $q_I^{\rm W}$ is the charge of the associated W bosons and $\tilde{q}^I_{\rm mon}$ is the charge of the associated 't Hooft-Polyakov monopole strings. In the present instance, $q_I^{\rm W} = (0,1,0)$ and $\tilde{q}^I_{\rm mon} = (1,-2,-1)$, hence we obtain
\begin{equation}
\text{Weyl:} \qquad X \to X+Y \,, \qquad Y\to -Y \,, \qquad Z \to Z - Y \,.
\end{equation}
Applying this to $\Kah$, we obtain the Weyl-reflection $\Kah' = \{X+Y\ge 0, Y \le 0, X+Z \ge 0, Z \ge 2 Y \}$ and the Weyl-orbit $\cW(\Kah) = \Kah \cup \Kah'$, as shown in Figure~\ref{sfig:kmvKWeyl}. Note that $\cW(\Kah)$ is not convex.

The prepotential within $\Kah'$ is fixed by Weyl invariance. For example, in the Weyl image I${}'$ of phase I, we obtain
\begin{multline}
\cF^{\text{(I${}'$)}} = \frac{4}{3} X'^3 + \frac{3}{2} X'^2 Y' + \frac{1}{2} X' Y'^2+ X'^2 Z' + X' Y' Z' \\
= \frac{4}{3} X^3 + \frac{3}{2} X^2 Y + \frac{1}{2} X Y^2 + X^2 Z + X Y Z + \frac{1}{3} Y^3 = \cF^{\text{(I)}} + \frac{1}{3} Y^3\,,
\end{multline}
in agreement with~\eqref{eqn:shrinkingcurve} (since two $W$ boson become massless at the $\mathfrak{su}(2)$ locus). Since the prepotential $\mathcal{F}$ is Weyl-invariant, the dual coordinates $\Yd_I = \frac{1}{\cF} \cF_I$ transform like partial derivatives $\frac{\partial}{\partial Y^I}$, and~\eqref{eqn:WeylReflect} becomes\footnote{Invariance of the central charges further implies that the electric and magnetic charges $q_I$ and $\tilde{q}^I$ transform like $\Yd_I$ and $Y^I$, respectively.}
\begin{equation}
\Yd_I \to \Yd_I - 2 \frac{\tilde{q}^J_{\rm mon} \Yd^J}{q_K^{\rm W}  \tilde{q}^K_{\rm mon}} q_I^{\rm W} \,.
\end{equation}
In the present instance,
\begin{equation} 
\text{Weyl:} \qquad \tilde{X} \to \tilde{X} \,, \qquad \tilde{Y} \to \tilde{X} - \tilde{Y} - \tilde{Z} \,, \qquad \tilde{Z} \to \tilde{Z} \,. 
\end{equation}
Applying this to $\Kv$, we obtain 
\begin{align}
\Kv' &= \{\tilde{Y} + 2\tilde{Z} \le \tilde{X}  \le 2\tilde{Y}+ \tilde{Z}; \tilde{Z} \ge 0\} \,, \\
\cW(\Kv) &= \Kv \cup \Kv' = \{\tilde{X} \ge \tilde{Y} + 2\tilde{Z}; \tilde{Y} \ge \tilde{Z} \ge 0\}\,.
\end{align} 
These regions are shown in Figure~\ref{sfig:kmvKvWeyl}.
Notice that $\cW(\Kv)$ \emph{is} convex, but with an interesting twist: the $\mathbb{P}^2$ and $dP_1$ CFT boundaries impinge into the interior of $\cW(\Kv)$; this will have interesting consequences later on.

\begin{figure}
\centering
\includegraphics[height=8cm]{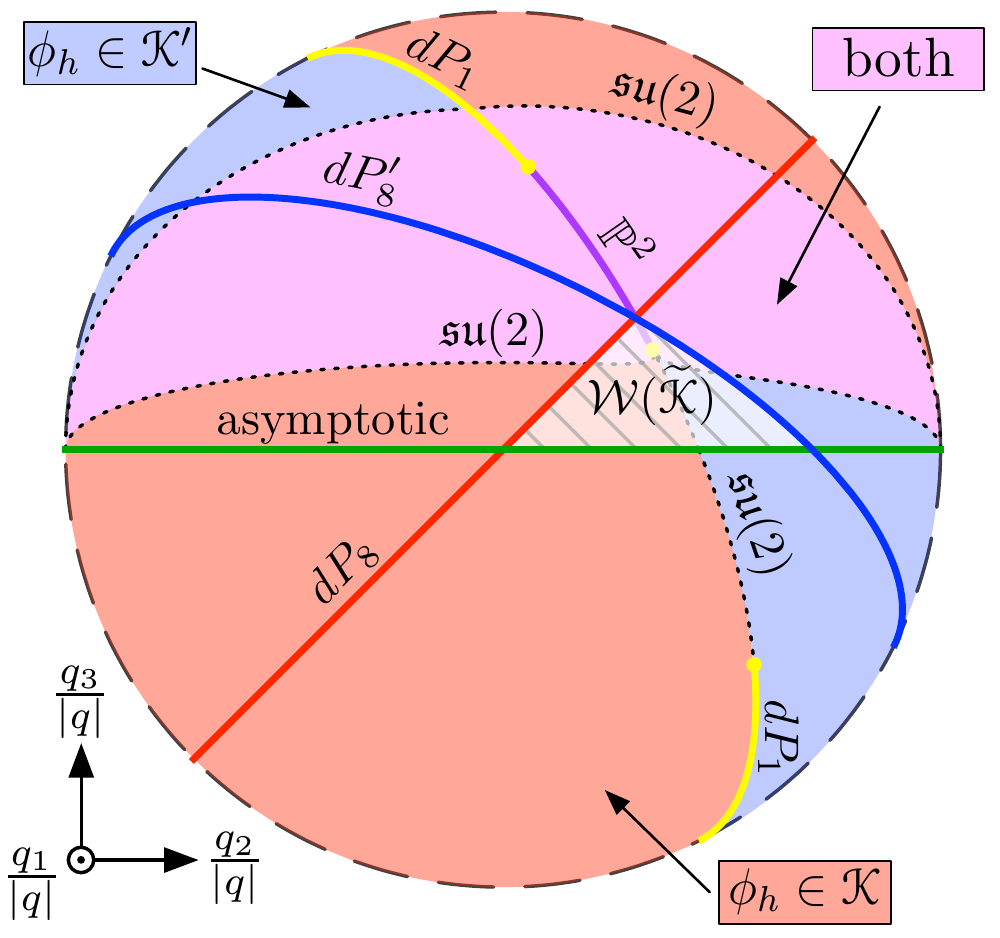}
\caption{The location(s) of attractor point(s) for the KMV geometry as a function of $(q_1,q_2,q_3)$, projected onto the unit sphere and viewed from along the positive $q_1$ axis. In the red (blue) region the attractor(s) lie within the extended K\"ahler cone $\Kah$ (within its Weyl image $\Kah'$), whereas in the magenta region there are attractor points within both $\Kah$ and $\Kah'$. Attractor points reach the indicated boundaries of $\cW(\Kah)$ along the thick colored lines and the $\mathfrak{su}(2)$ locus along the dotted lines. A BPS attractor exists within the crosshatched region $\cW(\Kv)$.}
\label{fig:KMVAttractor}
\end{figure}

Having understood the structure of $\cW(\Kah)$ and $\cW(\Kv)$, we return to the question of attractor points. As illustrated in Figure~\ref{fig:KMVAttractor}, $\cQ(\phi)$ has a minimum in the interior of $\cW(\Kah)$ for generic $q_I$, with the minimum approaching a boundary of $\cW(\Kah)$ when $q_I$ approaches the plane $\tilde{q}^I q_I = 0$ for $\tilde{q}^I$ one of the three generators $(0,0,1)$, $(0,1,-1)$, $(1,-1,-2)$ of $\cW(\Kv)^\ast$, as well as when $q_I$ approaches a certain portion of the plane $\tilde{q}^I q_I = 0$ for $\tilde{q}^I = (1,-2,-1)$ (in the remainder of this plane the minimum lands on the $\mathfrak{su}(2)$ locus).

\begin{figure}
\centering
\includegraphics[height=3.75cm]{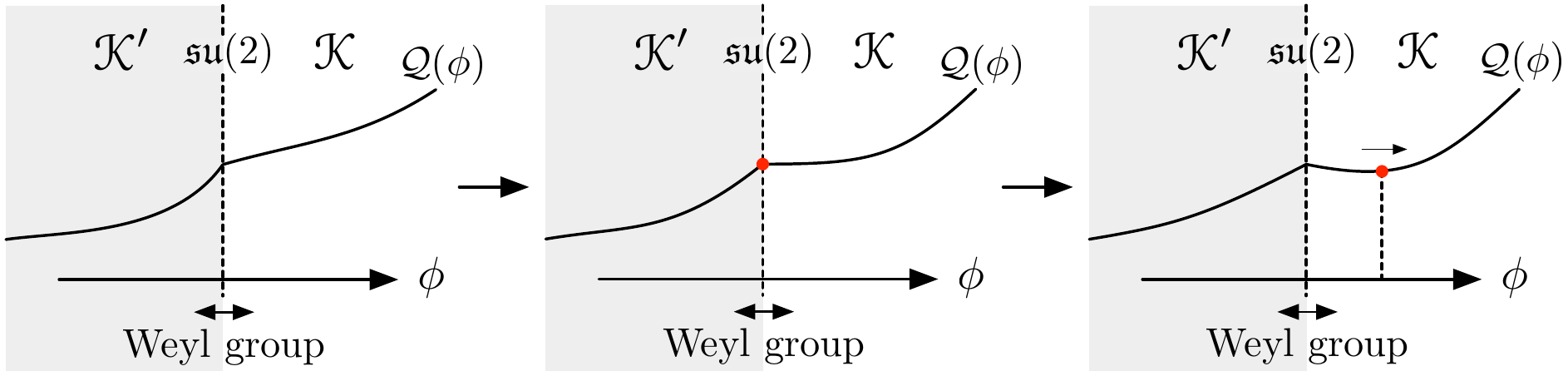}
\caption{An attractor point can be created at the $\mathfrak{su}(2)$ locus because of the upward-directed kink that occurs there.}
\label{fig:su2create}
\end{figure}

In fact, $\cQ(\phi)$ can have more than one minimum within $\cW(\Kah)$. This occurs for two reasons. First, since there is an upwards kink in $\cQ(\phi)$ at the $\mathfrak{su}(2)$ locus (analogous to the downwards kink at a flop as in, e.g., Figure~\ref{fig:ConifoldAttractor}), a local minimum can be created on the ``uphill'' side of the kink, see Figure~\ref{fig:su2create}, resulting in two attractor points, one in $\Kah$ and one in $\Kah'$. This occurs along the dotted lines adjoining the magenta region in Figure~\ref{fig:KMVAttractor}. Second, a single minimum within $\Kah$ can split into two nearby minima separated by a saddle point, and likewise within $\Kah'$. This occurs in small slivers adjoining the $dP_8$ and $dP_8'$ lines in the lower half of Figure~\ref{fig:KMVAttractor} (not pictured).

Since $\cQ(\phi)$ has a minimum within $\cW(\Kah)$ for general $q_I$, we can in principle numerically integrate~\eqref{eqn:WPDE} to obtain a fake superpotential (or multiple local fake superpotentials, in the case of multiple minima), and thereby determine the extremality bound. However, there are some additional complications. To illustrate them, let us focus on the BPS attractors (i.e., those satisfying $\sqrt{3} \cQ(\phi) = \zeta_q(\phi)$ at the attractor point) for which the fake superpotential $W(\phi) = \zeta_q(\phi)$ is known analytically.

\begin{figure}
\centering
\begin{subfigure}{0.48\textwidth}
\centering
\includegraphics[height=3.75cm]{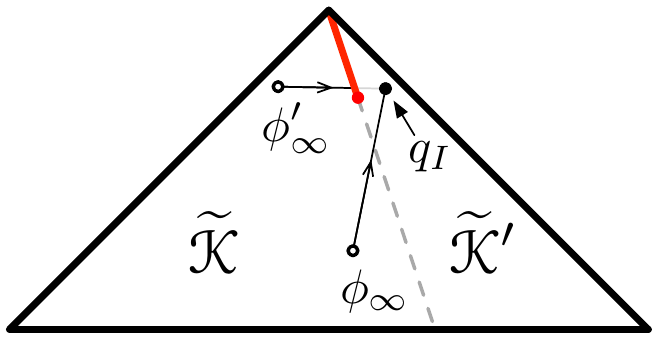}
\caption{Good vs. indeterminate flows for $q_I \in \Kv'$.} \label{sfig:P2indetflow}
\end{subfigure}
\begin{subfigure}{0.48\textwidth}
\centering
\includegraphics[height=3.75cm]{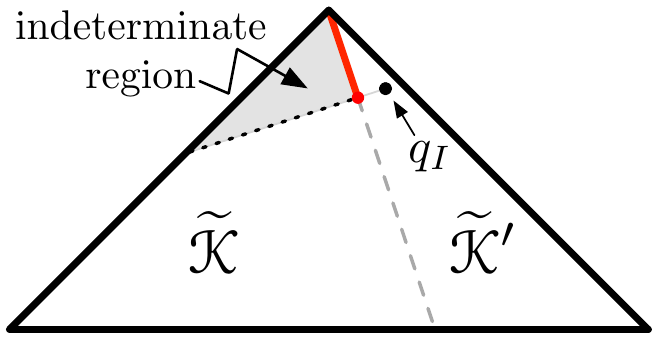}
\caption{The indeterminate region within $\Kv$.} \label{sfig:P2indetshadow}
\end{subfigure}
\caption{\subref{sfig:P2indetflow} When $q_I \in \Kv'$ there is a BPS attractor point within $\Kv'$ but some $\zeta_q$ flows intersect the $\mathbb{P}^2$ CFT boundary and remain indeterminate. \subref{sfig:P2indetshadow} The ``shadowed'' region where indeterminate flows occur, caused by the encroachment of the $\mathbb{P}^2$ boundary within the convex hull of $\cW(\Kv)$. If these indeterminate flows are bad then wall crossing must occur along the dotted line, in conflict with the general arguments of~\S\ref{subsec:wallcrossing}.}
\label{fig:P2Indet}
\end{figure}

There is a unique BPS attractor within $\cW(\Kah)$ for all $q_I \in \cW(\Kv)$, where the attractor point lies within $\Kah$ ($\Kah'$) when $q_I \in \Kv$ ($q_I \in \Kv'$). Trivially extending the results of~\S\ref{subsec:straightline}, $\zeta_q$ flows are straight lines in $\cW(\Kv)$. Since the latter is convex, it appears naively that all $\zeta_q$ flows are good flows within $\cW(\Kah)$ so long as $q_I \in \cW(\Kv)$, implying that the BPS and black hole extremality bounds agree throughout $\cW(\Kv)$. However, the encroachment of the $\mathbb{P}^2$ CFT boundary into the interior of $\cW(\Kv)$ now becomes important. Due to this encroachment, some $\zeta_q$ flows are actually indeterminate when $q_I \in \Kv'$ (fixing $\Yd_I^\infty \in \Kv$ by a choice of gauge), see Figure~\ref{sfig:P2indetflow}.

Thus, while BPS black holes of charge $q_I \in \Kv'$ exist (and can be explicitly constructed) in the unshaded portion of Figure~\ref{sfig:P2indetshadow}, their existence is uncertain in the shaded region, which is ``shadowed'' by the $\mathbb{P}^2$ CFT boundary. If BPS black holes \emph{do not} exist in the shadowed region then they must disappear by wall crossing along the dotted line in Figure~\ref{sfig:P2indetshadow}. However, this putative wall of marginal stability does not lie at a phase boundary, in conflict with the general arguments laid out in~\S\ref{subsec:wallcrossing}. Therefore, by contradiction we reach the conclusion that BPS black holes should exist throughout the moduli space whenever $q_I \in \cW(\Kv)$. Verifying (or refuting) this conclusion through a more careful study of the $dP_1$ and $\mathbb{P}^2$ CFTs coupled to gravity is a very interesting open question that we leave for the future.

Similar issues will affect the determination of the extremality bound outside $\mathcal{W}(\Kah)$, where the attractor points are not BPS. We leave further analysis of such cases to future work.

\subsubsection{GV invariants (revisited)}  \label{sssec:GVrevisit}

Since the BPS and black hole extremality bounds agree across $\cW(\Kv)$---bigger than the region $\Kv$ that we previously considered---we revisit the GV invariants and their consequences for the T/sLWGC. Referring back to Table~\ref{tab:KMVinvs}, we see that there is a holomorphic curve in every non-trivial class within $\cW(\Kv) = \{q_1 \ge q_2 + 2q_3; q_2 \ge q_3 \ge 0\}$ to the degree calculated, once again in agreement with the $k=1$ T/sLWGC. In particular, $\cW(\Kv)$ lies within the subcone $\Kah^\ast = \Mori_\cap$ of the Mori cone $\Mori$ where infinite towers of BPS states are possible, and does not intersect any of the boundaries of $\Kah^\ast$ that lack such towers. This is a satisfying test of both the T/sLWGC and the above reasoning.

Notice that the GV invariants in Table~\ref{tab:KMVinvs} are not invariant under the Weyl reflection, in contrast to, e.g., the GV invariants shown in Table~\ref{tab:SymFlopGV}, which exhibit the same $\mathbb{Z}_2$ reflection symmetry through the flop as the underlying symmetric flop geometry. This implies that wall crossing occurs at the $\mathfrak{su}(2)$ locus. The presence of wall crossing at this locus (where the 't Hooft-Polyakov monopole string of charge $\tilde{q}^I = (1,-2,-1)$ becomes tensionless) and the absence of wall crossing at the flop in the previous two examples (where no BPS strings become tensionless) are both consistent with the general arguments of~\S\ref{subsec:wallcrossing}: 5d wall crossing for BPS particles only occurs at phase transitions where there are tensionless BPS strings.

However, note that the $q_3 = 0$ GV invariants in Table~\ref{tab:KMVinvs} \emph{are} Weyl invariant. Unlike those with $q_3 \ne 0$, BPS particles with $q_3 = 0$ can be made parametrically lighter than the dynamical scale of the $\mathbb{P}^2$ CFT (i.e., the scale of the $q_I \propto +(0,1,1)$ tower of BPS particles associated to it) while remaining near the $\mathfrak{su}(2)$ boundary. Thus, such particles should be describable as the excitations of fundamental fields in a perturbative low-energy effective action valid across the $\mathfrak{su}(2)$ locus, implying that they fall into complete $\mathfrak{su}(2)$ multiplets, and are therefore Weyl invariant as observed.

The latter observation explains why there are infinite towers of vanishing GV invariants within $\Kah^\ast$---at least those that appear along the $q_3 =0$ boundary of $\Kah^\ast$---but it does not address whether BPS black holes are possible in these directions. This is a difficult question in general, because the $\zeta_q$ flows in question are indeterminate. However, some of these indeterminate flows end on the $\mathfrak{su}(2)$ boundary of $\Kah$, and can potentially be resolved by passing through it into $\Kah'$.

In particular, infinite towers of vanishing GV invariants only occur outside $\cW(\Kv)$, so the flows passing through the $\mathfrak{su}(2)$ locus will either: (1) be resolved into bad flows, if they reach a region of $\Kah'$ where $\zeta_q < 0$; or (2) remain indeterminate, if they reach a CFT boundary of $\Kah'$ where $\zeta_q > 0$. The former is only possible if $\zeta_q$ is negative somewhere in $\Kah'$, i.e., $q_I \not\in (\Kah')^\ast$, or equivalently (since $q_I \in \Kah^\ast$ by assumption), $q_I$ lies outside the region $\cW(\Kah)^\ast = \Kah^\ast \cap (\Kah')^\ast$ in which $\zeta_q$ is positive across the entire Weyl-orbit of the moduli space.

Fortunately, all the vanishing GV invariants in Table~\ref{tab:KMVinvs} lie outside $\cW(\Kah)^\ast$ and thus outside $(\Kah')^\ast$. In particular, this implies that the BPS and black hole extremality bounds disagree in these directions for any choice of vacuum $Y^I_\infty$ within $\Kah'$, assuming that wall crossing cannot occur in the interior of $\Kah'$ per~\S\ref{subsec:wallcrossing}. However, wall crossing \emph{may} occur at the $\mathfrak{su}(2)$ locus, so this does not necessarily imply the same result for a choice of vacuum $Y^I_\infty$ within $\Kah$. 

\begin{figure}
\centering
\begin{subfigure}{0.51\textwidth}
\centering
\includegraphics[height=4cm]{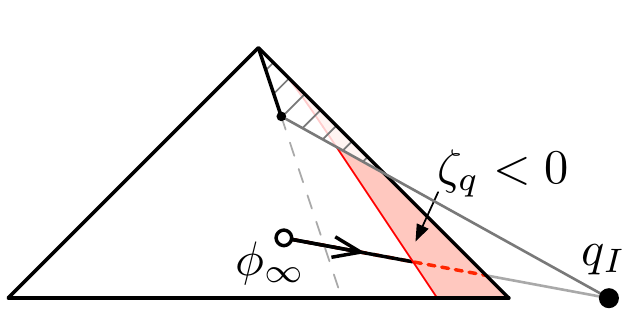}
\caption{$q_I = (2,3,0)$.} \label{sfig:BadFlow1}
\end{subfigure}
\begin{subfigure}{0.47\textwidth}
\centering
\includegraphics[height=4cm]{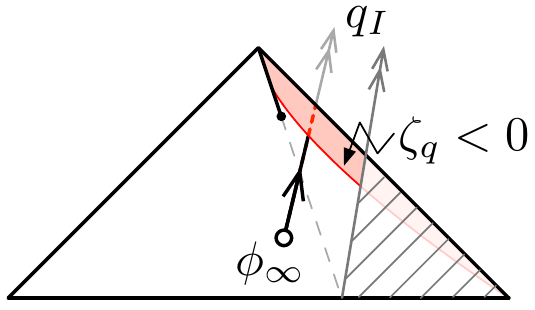}
\caption{$q_I = (1,1,2)$.} \label{sfig:BadFlow2}
\end{subfigure}
\caption{Indeterminate flows ending on the $\mathfrak{su}(2)$ boundary may resolve to bad flows after crossing into $\Kah'$, provided that $q_I$ lies outside $\cW(\Kah)^\ast$, as in the pictured examples. Note that the crosshatched region of $\Kah'$ is inaccessible to (weakly coupled) flows beginning within $\Kah$; for certain $q_I \in \Kah^\ast \setminus \cW(\Kah)^\ast$ (not pictured) the $\zeta_q < 0$ region lies entirely within this inaccessible region, and all flows originating within $\Kah$ are indeterminate.}
\label{fig:BadFlows}
\end{figure}

To reach a definitive answer, we consider all $\zeta_q$ flows beginning within $\Kah$ and check which of them, if any, are bad. Several examples are shown in Figure~\ref{fig:BadFlows}. For the sake of brevity, we omit a complete discussion of the results, which depend on the choice of $q_I$. Most importantly, one can show that for all the directions within $\Kah^\ast$ in which there are infinite towers of vanishing GV invariants (Figures~\ref{sfig:BadFlow1} and~\ref{sfig:BadFlow2} being representative examples) there are bad flows originating within $\Kah$, and therefore in the absence of wall crossing in the interior of $\Kah$, the BPS and black hole extremality bounds disagree in these directions and there is no potential for these vanishing GV invariants to contradict the T/sLWGC. Similar reasoning implies that the BPS and extremality bounds disagree in most other directions outside $\cW(\Kah)^\ast$, though there are some exceptions where we obtain only indeterminate flows upon passing into $\Kah'$, as explained in the caption of Figure~\ref{fig:BadFlows}.

\section{Conclusions and Future Directions}\label{sec:CONC}

The Weak Gravity Conjecture has much richer consequences for the spectrum of BPS particles than one might naively expect. The key to understanding these consequences and to testing whether they hold in specific quantum gravities is to determine in which charge directions the BPS and black hole extremality bounds agree. In these directions, the T/sLWGC require an infinite tower of BPS particles, whereas in other directions the WGC and its tower/sublattice variants place no requirements on the spectrum of BPS particles, as they can be satisfied in these directions by non-BPS particles (which is indeed what happens in perturbative string theory examples where the non-BPS spectrum is calculable). For theories in dimension $D \ge 5$, even the original, ``mild'' form of the WGC has these consequences, since a circle compactification of a $D=d+1$ dimensional theory lacking such infinite towers yields a $d$-dimensional theory violating the mild WGC~\cite{Heidenreich:2015nta} (itself a sensible conjecture for $d \ge 4$).\footnote{Although~\cite{Heidenreich:2015nta} gives a tree-level argument that is potentially sensitive to loop corrections, BPS states are sufficiently protected that the conclusion should remain unchanged.}

In this paper, we focused on 5d $\mathcal{N}=1$ theories arising from M-theory compactified on a Calabi-Yau threefold. We showed that the BPS and black hole extremality bounds agree unconditionally for charges $q_I$ within the cone of dual coordinates $\Kv$, i.e., the dual-coordinate image $\Yd_I = \Yd_I[Y]$ of the extended K\"ahler cone $\Kah$. Moreover, we argued that $\Kv$ is the dual $\Eff^\ast$ of the effective cone $\Eff$, otherwise known as the cone of movable curves~\cite{Boucksom13}. With this information in hand, we were able to formulate a purely geometric conjecture implied by the (tower) WGC: for every non-trivial integral class $q_I \in H_2(X,\mathbb{Z})$ lying within the dual of the effective cone $\Eff^\ast = \Kv$ of a Calabi-Yau three-fold $X$, there is a holomorphic curve in the class $k q_I$ for some integer $k \ge 1$.

This conjecture can be tested by examining the Gopakumar-Vafa invariants for various geometries. We did this for three different $h^{1,1} = 2$ or $h^{1,1}=3$ geometries in the present work, whereas a far more comprehensive search will be undertaken in~\cite{CornellGV}. In all examples considered so far the geometrized conjecture appears to be satisfied, invariably with $k=1$---i.e., up to the degree computed, there is a non-vanishing GV invariant in every non-trivial $H_2(X,\mathbb{Z})$ class within $\Eff^\ast = \Kv$.

Not only is this a highly non-trivial test of the T/sLWGC, it suggests that the lattice WGC (LWGC) formulated in~\cite{Heidenreich:2015nta} deserves renewed attention. Although the LWGC is known to be false in general~\cite{Heidenreich:2016aqi}, these results suggest that it may hold for BPS particles in a wide variety Calabi-Yau compactifications of M-theory. Since $X$ is simply connected in all the examples considered here and in~\cite{CornellGV}, whereas the counterexamples in~\cite{Heidenreich:2016aqi} typically involve freely acting orbifolds with a non-trivial fundamental group, we speculate that violations of the LWGC are somehow linked to orbifolding.\footnote{We thank M.~Demirtas, N.~Gendler, M.~Kim, M.~Lotito, L.~McAllister, J.~Moritz, E.~Palti, and A.~Rios Tascon for discussions on this point.} We defer further consideration of this very interesting topic to future work.

In the absence of finite-distance boundaries within the vector multiplet moduli space, $\Kv$ is simply the dual $\Kah^\ast$ of the extended Kähler cone $\Kah$, where $\Kah^\ast$---equal to the intersection $\Mori_\cap$ of the Mori cones of all the phases within $\Kah$---is also the largest possible region where infinite towers of BPS particles can occur in principle (assuming no wall crossing occurs in the interior of $\Kah$). Thus, in the absence of finite-distance boundaries, the infinite towers of BPS particles that are present are precisely those required by the T/sLWGC.

When there are finite distance boundaries, $\Kv$ becomes a proper subset of $\Kah^\ast$. For charges $q_I$ within $\Kah^\ast$ but outside $\Kv$, the cores of prospective BPS black hole solutions invariably probe the physics at these finite distance boundaries. In the case of CFT boundaries, we are not able to determine in general whether (and for which charges) these prospective solutions are valid due to the strongly coupled physics involved. Resolving this issue is another very interesting question that we defer to future research.

On the other hand, at finite-distance boundaries associated to non-abelian gauge symmetry enhancements (such as $\mathfrak{su}(2)$ boundaries) the physics is perturbative. Such boundaries can be removed by lifting to the Weyl-group orbit $\cW(\Kah)$ of the extended K\"ahler cone $\Kah$, equivalent to the field space of the scalars fields in the low-energy effective action that are neutral under the Cartan subalgebra of the gauge group. We considered one such example in~\S\ref{ssec:KMV}, concluding that the BPS and black hole extremality bounds agree in the larger region $\cW(\Kv) \supset \Kv$. This leads to additional T/sLWGC constraints on the GV invariants that are fortunately also satisfied. We defer a more systematic treatment of non-abelian gauge enhancement boundaries and their WGC consequences to~\cite{CornellGV}.

Another uncertainty affecting our analysis stems from our assumption (common in the literature) that the black hole extremality bound is saturated by spherically symmetric solutions, i.e., there are no black hole solutions with strictly less mass than all spherically symmetric black hole solutions of the same charge. This assumption is necessary to be able to compute the extremality bound using currently available methods, but is not easy to justify a priori. Should it turn out to be incorrect, the black hole extremality bound might end up agreeing with the BPS bound in additional directions outside $\Kv$, leading to a more stringent test of the T/sLWGC. We leave further exploration of this once-again fascinating topic for the future.

Note that the caveats discussed above regarding CFT boundaries, possible failures of spherical symmetry, etc., cannot affect the agreement between the BPS and black hole extremality bounds within $\Kv$, so the predictions we make in this paper are necessary predictions of the T/sLWGC regardless of these subtleties, and their verification remains a highly non-trivial test of these conjectures.

Likewise, none of these subtleties affect the fact that the hypermultiplets becoming massless at flop phase transitions are strictly superextremal. Indeed, this follows from the impossibility of having massless black holes.\footnote{To be precise, this implies that the flopping hypermultiplets are strictly superextremal \emph{at the flop}, which extends to the entire moduli space assuming no wall crossing within $\Kah$ per~\S\ref{subsec:wallcrossing}. Testing and reinforcing the arguments of~\S\ref{subsec:wallcrossing} is another important direction for future research.} More generally, the absence of infinite towers of BPS particles outside $\Kah^\ast$ is perfectly consistent with the T/sLWGC, since the BPS and black hole extremality bounds disagree in these directions, resolving the conifold conundrum of~\S\ref{sec:CC}.

\subsection{Geometric implications}

Our work has interesting geometric consequences. Besides our primary geometric conjecture mandating the existence of infinite towers of holomorphic curves in every direction within the dual of the effective cone, our results imply that the pseudoeffective cone $\overline{\Eff} = \Kv^\ast$ of a Calabi-Yau manifold $X$ can be determined by constructing the extended K\"ahler cone $\Kah$ out of the K\"ahler cones of all the Calabi-Yau manifolds $X'$ birationally equivalent to $X$ (see, e.g.,~\cite{Morrison94}), taking its image under the dual coordinate map $\Yd_I = \Yd_I[Y]$ to obtain $\Kv$, and then taking the dual $\overline\Eff = \Kv^\ast$. Alternately, dualizing~\eqref{eqn:KvFormula} we conclude that the psuedoeffective cone is generated by $\Kah$ together with the divisors that shrink at each of the finite distance boundaries of $\Kah$.

Provided that we can construct $\Kah$, this is a novel and (to our knowledge) previously unknown method by which the psuedoeffective cone of a given Calabi-Yau manifold can be determined. For a related construction of the effective cone itself as well as other geometric consequences, we refer to~\cite{StringWGC}.

\subsection{Other dimensions} \label{sec:4d}

Given the success of our five-dimensional analysis, it is natural to attempt to generalize to other dimensions. Two possibilities with the same amount of supersymmetry are 6d $\mathcal{N}=1$ theories and 4d $\mathcal{N}=2$ theories, see, e.g., \cite{Lee:2018spm, Lee:2019xtm, Gendler:2020dfp} for related previous work. In the 6d case, realized geometrically by compactifying F-theory on an elliptically fibered Calabi-Yau threefold, there are no BPS charged particles but only BPS charged strings (with both electric and magnetic two-form charge). While there is no analog of the tower or sublattice conjectures for strings, the mild WGC for strings should have interesting, testable consequences in such theories. We defer further consideration of these consequences and their geometric interpretation to future work.

The 4d case---realized geometrically by compactifying either type IIA or type IIB string theory on a Calabi-Yau threefold (the two possibilities being related by mirror symmetry)---involves a number of subtleties. As discussed in~\S\ref{subsec:summary}, one issue is that the tower and sublattice conjectures are necessarily modified in four dimensions~\cite{Heidenreich:2016aqi,Heidenreich:2017sim} due to the infrared divergences associated to light charged particles. However, even without knowing the precise form of these modifications, the statement that an infinite tower of BPS particles exists in every direction in which the BPS and black hole extremality bounds agree remains reasonably well motivated, and we can still test it and study its consequences.

A second issue is the prevalence of wall crossing phenomena, making it difficult to determine the spectrum of BPS particles at a given point in the moduli space. Another issue is the presence of infinitely many worldsheet instanton corrections to the prepotential, making it difficult to determine the black hole extremality bound explicitly. Yet another is the presence of non-geometric (e.g., Landau-Ginzburg) phases in the moduli space, further complicating the picture.

Despite (or perhaps because of) these issues, we expect that pursuing these questions in 4d will lead to interesting results, and this is another important direction for future research. One approach would be to consider certain limits of the moduli space where the problem simplifies. Indeed, one such limit is the M-theory limit where D0 branes become light and the M-theory circle becomes large; this limit must reproduce the results of our present work. There may be other interesting limits where the problem simplifies.

\section*{Acknowledgements}

We thank Callum Brodie, Clay C\'ordova, Mehmet Demirtas, I\~naki Garc\'ia Etxebarria, Naomi Gendler, Patrick Jefferson, Manki Kim, Liam McAllister, Jakob Moritz, David Morrison, Eran Palti, Matthew Reece, Andres Rios Tascon, Houri-Christina Tarazi, and Edward Witten for useful discussions, and Naomi Gendler, Matteo Lotito, Jakob Moritz, Cumrun Vafa, and Irene Valenzuela for comments on the manuscript. MA is supported through the DFG Emmy Noether grant AL 1407/2-1. BH was supported by NSF grant PHY-1914934 during the later stages of this work, and by Perimeter Institute for Theoretical Physics during its inception. Research at Perimeter Institute is supported
by the Government of Canada through the Department of Innovation, Science and Economic Development, and by the Province of Ontario through the Ministry of Research,
Innovation and Science. 
The work of TR at the Institute for Advanced Study was supported by the Roger Dashen Membership and by
NSF grant PHY-191129. The work of TR at the University of California, Berkeley, was supported by NSF grant  PHY-1820912, the Simons Foundation, and the Berkeley Center for Theoretical Physics.

\appendix

\addtocontents{toc}{\protect\setcounter{tocdepth}{1}}

\section{Computation of Gopakumar-Vafa invariants}\label{sec:GVinvariants}

In this section we discuss how mirror symmetry can be used to obtain exact results on the BPS degeneracies of the example geometries which we have considered so far. For an overview and background on the techniques of mirror constructions we refer to~\cite{Cox:2000vi,Hori:2003ic}. In the following we adapt the discussion of Hosono, Klemm, Theisen and Yau~\cite{Hosono:1993qy,Hosono:1994ax} to our needs. 

\subsection{Topological data of complete intersection Calabi-Yau threefolds}
The examples we consider in this paper are families of Calabi-Yau (CY) threefolds given as complete intersections of $l$ hypersurfaces in products of $k$ projective spaces. We will denote the family by 
\begin{equation}
 \pi: \mathcal{X}\rightarrow B\,,
\end{equation}
with fibers $\pi^{-1}(b)=X_b$ smooth CY threefolds and $B$ denoting the moduli space of complexified K\"ahler forms of dimension $\textrm{dim}_{\mathbb{C}} B= h^{1,1}(X_b)$.

Before discussing the construction of the mirror families we recall the notation and techniques to specify the topological data of a complete intersection configuration. We denote by $d_j^{(i)}$ the degree of the coordinates of the weighted projective space $\mathbb{P}^{n_i}[\vec{w}^{(i)}]$ in the $j$-th polynomial $p_j$. Where $(i=1,\dots,k)$ and $(j=1,\dots,l)$. The configuration is specified by 
\begin{equation}
\left(   \begin{array}{c|| ccc} \mathbb{P}^{n_1} [w_1^{(1)}\,, \dots, w_{n_1+1}^{(1)}] & d_1^{(1)} &\dots & d_l^{(1)} \\
\vdots & \vdots & & \vdots \\
\mathbb{P}^{n_k} [w_1^{(k)}\,, \dots, w_{n_1+1}^{(k)}] & d_1^{(k)} &\dots & d_l^{(k)}
 \end{array}\right)^{h^{1,1}}_{\chi}\,.
\end{equation}
Associating to the K\"ahler form induced from the $i$-th projective space the formal variable $J_i$, we define a map
$\Pi: \mathbb{Z}[J_1,\dots,J_k] \rightarrow \mathbb{Z}$
given by:
\begin{equation}
 \Pi[P(J)] \df \Biggl(  \prod_{r=1}^k \frac{1}{n_r!} \frac{\partial^{n_r}}{\partial J_r^{n_r}}\Biggr)\! \Biggl( \frac{\prod_{i=1}^k \prod_{j=1}^{n_i+1} (1+ w_j^{(i)} J_i)}{ \prod_{j=1}^l (1+\sum_{i=1}^k d_j^{(i)}J_i)} \Biggr)\! \Biggl(  \frac{\prod_{j=1}^l \sum_{i=1}^k d_j^{(i)} J_i}{ \prod_{i=1}^k \prod_{j=1}^{n_i+1} w_j^{(i)}}\Biggr) P(J)  \Biggr\vert_{J_i = 0}\,,
\end{equation}
where $P(J)$ is some polynomial in the $J_i$.
Applying $\Pi$ to $P(J)$ is equivalent to integrating the wedge product of the corresponding $(1,1)$ forms wedged with the Chern class of the appropriate degree 
over the CY threefold $X$. Therefore, \cite{Hosono:1994ax}:
\begin{equation}
\chi =\int_{X} c_3 = \Pi[1]\,,\quad \int_{X} c_2 \wedge J_m = \Pi[J_m]\,,\quad K_{ijk}^{0}=\int_X J_i \wedge J_j \wedge J_k = \Pi[J_i J_j J_k]\,,
\end{equation}
where $\chi$ denotes the Euler character of the threefold, $c_i$ the $i$-th Chern class, and $K_{ijk}^{0}$ the classical triple intersection numbers. The latter receive quantum corrections from worldsheet instantons in four dimensions, as computed below using mirror symmetry.

\subsubsection{Examples} \label{topexamples}
We focus on the examples discussed in (a)~\S\ref{ssec:GMSV} and (b)~\S\ref{ssec:symFlop}, i.e.,
\begin{equation}
\text{(a): } \left(   \begin{array}{c|| cc} \mathbb{P}^{4} & 4 &1 \\
\mathbb{P}^{1} & 1 & 1
 \end{array}\right)^{2}_{-168}\,, \quad  
\text{(b): } \left(   \begin{array}{c|| cc} \mathbb{P}^{4}[4,1,1,1,1] & 4 &4 \\
\mathbb{P}^{1} & 1 & 1
 \end{array}\right)^{2}_{-168}\,.
\end{equation}
Using the map $\Pi$, we obtain the following topological data:
\begin{enumerate}[(a)]
\item $\chi= -168,\quad \int_{X} c_2 \wedge J_1= 50\,, \quad  \int_{X} c_2 \wedge J_2= 24\,,\quad  K^0= 5 J_1^3 + 4 J_1^2 J_2\,. $
\item $\chi= -168,\quad \int_{X} c_2 \wedge J_1= 44\,, \quad  \int_{X} c_2 \wedge J_2= 24\,,\quad  K^0= 2 J_1^3 + 4 J_1^2 J_2\,, $
\end{enumerate}
where $K^0=\sum_{i,j,k} K^0_{ijk} J_i J_j J_k$.
\subsection{Mirror Calabi-Yau families}\label{mirrorfamilies}
To compute the BPS degeneracies for these geometries we need the exact quantum corrected triple intersections $C_{abc}$ associated to the geometries. These are encoded in the data of the flat connection describing the quantum cohomology of the CY family, which in turn is identified with the flat Gauss-Manin connection describing the variation of Hodge structure of the middle cohomology $H^3(\check{X}_b,\mathbb{C})$ of the mirror family of CY manifolds given by:
\begin{equation}
\check{\pi}: \check{\mathcal{X}} \rightarrow B\, , \quad \check{\pi}^{-1}(b)= \check{X}_b\,.
\end{equation}
In distinguished, special coordinates, the quantum-corrected triple intersections are encoded in the prepotential $F_0(t)$ of the geometry as 
\begin{equation}\label{yukprep}
 C_{abc}= \partial_a \partial_b\partial_c F_0(t)\,,\quad \partial_a:=\frac{\partial}{\partial t^a}\,,
 \end{equation}
where $t^a,\, a=1,\dots,h^{2,1}(\check{X})=h^{1,1}(X)$ denotes a set of special coordinates, which are then identified as the mirror maps giving the identification of the moduli space of complex structures of the mirror family with the moduli space of complexified K\"ahler forms of the original CY family.

In the following we will use the mirror construction of Batyrev and Borisov \cite{Batyrev:1994hm,Batyrev:1994pg}, reviewed in \cite{Hosono:1994ax}. The idea is to first determine the mirror of the ambient space, where defining equations of the complete intersection configuration are set to zero,  as a toric variety and then set appropriate equations to zero in the mirror ambient space. The ambient space in question is a product of weighted projective spaces. Each projective space $\mathbb{P}^{n_i}[\vec{w}^{(i)}]$ is a toric variety which can be encoded in the data of a reflexive simplicial polyhedron $\Delta_i$ with vertices in $\mathbb{R}^{n_i}$.\footnote{See~\cite{Hosono:1994ax} for definitions and background.} The ambient space as a product of weighted projective spaces is also a toric variety associated to the reflexive polyhedron:
\begin{equation}
\Delta = \Delta_1\times \dots \times \Delta_k \, \quad \textrm{in} \,\quad  \mathbb{R}^{n_1} \times \dots \times \mathbb{R}^{n_k}\,.
\end{equation}
The mirror ambient space is constructed out of the data of the dual polyhedron $\Delta^*$. The corners of $\Delta^*_i$ are the integral points:
\begin{equation}
\begin{array}{ccccccc}
\nu_{i,1}^* & =& (1,&\dots,&0)\,,\\
\vdots &&&&&\\
\nu_{i,n_i}^* & =& (0,&\dots,&1)\,,\\
\nu_{i,n_i+1}^* & =& (-w_1^{(i)},&\dots,&-w_{n_i}^{(i)})\,,
\end{array}
\end{equation}
in $\mathbb{R}^{n_i}$ of $\mathbb{R}^{n_1}\times \dots \times \mathbb{R}^{n_k}$. These vertices satisfy the relation
\begin{equation} \label{vertexrel}
\sum_{j=1}^{n_i+1} w_j^{(i)} \nu_{i,j}^* =0\,,\quad i=1,\dots,k\,.
\end{equation}

Next the set of vertices 
\begin{equation}
E= \bigcup_{i=1}^k  \bigcup_{j=1}^{n_i+1} \left\{  \nu_{i,j}^*\right\}\,,
\end{equation}
 is grouped into $l$ sets $E_m (m=1,\dots,l)\,$ such that the set $E_m$ contains $d_m^{(i)}$ vertices from $\bigcup_{j=1}^{n_i+1} \left\{  \nu_{i,j}^*\right\}$ for each $i=1,\dots,k$. Each vertex $\nu_{i,j}^*$ of $E_m$ is then extended to a vertex $\overline{\nu}_{i,j}^*=(\vec{e}^{(m)}, \nu_{i,j})$ in $\mathbb{R}^l \times \mathbb{R}^{n_1}\times\dots\times\mathbb{R}^{n_k}$ with $\vec{e}^{(m)}$ being the $m$-th unit vector in $\mathbb{R}^l$. Since we are considering CY complete intersections, the first Chern class vanishes which implies the following relation:
\begin{equation}
c_1 = c_1^a J_a= \left( \sum_{i=1}^{n_a+1} w_i^{(a)} - \sum_{i=1}^l d_i^{(a)}\right) J_a =0\,.
\end{equation} 
Together with the relation~\eqref{vertexrel} this implies that the $\sum_{i=1}^k n_i+2$ extended vertices satisfy $k$ independent linear relations:
\begin{equation}
\sum_{\alpha=1}^{l+\sum_{i=1}^k n_i} (l^{(s)})_{\alpha} (\overline{\nu}_{i,j}^*)_{\alpha}=0\,.
\end{equation}
The $l^{(s)}$ are given by:
\begin{equation}
l^{(s)}=(-d_1^{(s)}\,,\dots, -d_l^{(s)}; \dots,w_1^{(s)},\dots,w^{(s)}_{n_s+1},0,\dots) =: ( \{l_{0j}^{(s)}\};\{ l_i^{(s)}\})\,.
\end{equation}
The mirror manifold $\check{X}$ is conjecturally obtained as the complete intersection of the following Laurent polynomials in the $\mathbb{C}^*$ variables $X_{m,n}, m=1,\dots,k, n=1,\dots,n_m$:
\begin{equation}\label{mirroreq}
P_r = a_r -\sum_{\nu_{i,j}^* \in E_r} a_{i,j} X^{\nu_{i,j}^*}\,, \quad (r=1,\dots,l)\,.
\end{equation}
The vanishing loci of the $l$ equations~\eqref{mirroreq} are considered in a toric variety $\mathbb{P}_{\Delta^*}$, where $\Delta^*=\sum_{i=1}^l \Delta^*_{(i)}$ and $\Delta^*_{(i)}$ being the convex hull of $\{0\}$ and the set $E_i$.
The holomorphic three-form on $\check{X}$ can be obtained from a residue construction from the natural holomorphic form in $(\mathbb{C}^*)^{\sum_{i=1}^k n_i}$ as:
\begin{equation}
\Omega = \textrm{Res}_{P_1=\dots=P_l=0} \frac{a_1\dots a_l}{P_1 \dots P_l} \prod_{m=1}^k \prod_{n=1}^{n_m} \frac{dX_{m,n}}{X_{m,n}}\,.
\end{equation}

We are interested in the explicit expressions for the period integrals of this holomorphic form over the various three-cycles of $\check{X}$. One particular cycle $\Gamma_0$ is determined by $|X_{m,n}|=1$ for all $m,n$; the period of the holomorphic three-form over this cycle is:
\begin{equation}
\pi^0(a) = \int_{\Gamma_0} \Omega\,.
\end{equation}
The parameters $\{ a_i\} , \{ a_{m,n}\}, i=1,\dots,l \,, m=1,\dots,k , n=1,\dots,n_m\,,$ are homogeneous coordinates overparameterizing the complex structure deformations of $\check{X}$. The various rescalings of the $\mathbb{C}^*$ coordinates $X_{m,n}$ can be used to obtain $k$ independent coordinates:
\begin{equation}
z_s:= \frac{a_{s,1}^{w_1^{(s)}} \dots a_{s,n_s+1}^{w_{n_s+1}^{(s)}} }{ a_1^{d_1^{(s)}} \dots a_l^{d_l^{(s)}}}=: a^{l^{(s)}}\, \quad s=1,\dots,k\,.
\end{equation}

The period $\pi^0(a)$ can be given as:
\begin{equation}\label{holperiod}
\pi^0(z) = \sum_{n_s\ge0} c(n) z^n\,, \quad z^n:=\prod_{s=1}^k z_{s}^{n_s}\,,
\end{equation}
with 
\begin{equation}
c(n) = \frac{\prod_{j=1}^l \left(  \sum_{i=1}^k n_i d_j^{(i)}\right)!}{ \prod_{i=1}^k \prod_{j=1}^{n_i+1} \left( w_j^{(i)} n_i\right) !} \,.
\end{equation}
It satisfies the $k$ linear GKZ differential operators:
\begin{multline}\label{GKZ}
\mathcal{L}^s = \prod_{j=1}^{n_s+1} \left( w_j^{(s)} \theta_s \right) \left( w_j^{(s)} \theta_s -1 \right) \dots \left( w_j^{(s)} \theta_s -w_j^{(s)} +1\right)  \\
- \prod_{j=1}^l \left( \sum_{i=1}^k d_j^{(i)} \theta_i \right) \dots   \left( \sum_{i=1}^k d_j^{(i)} \theta_i -d_j^{(s)} +1 \right) z_s \,, \quad \theta_i:=z_i \frac{\partial}{\partial z_i}
\end{multline}

\subsubsection{Examples}
The mirror geometries of the examples discussed in \S\ref{topexamples} are specified by the vertices of $\Delta^*$ given by the vertices of the two reflexive polyhedra $\Delta_1^*,\Delta_2^*$ in $\mathbb{R}^4\times \mathbb{R}$ shown in Table~\ref{tab:DeltaVertices}. 
\begin{table}
\centering
\begin{subtable}{0.48\textwidth}
\centering
$\begin{array}{cccccccccc}
\nu_{1,1}^* & =& (&1,&0,&0,&0;&0&)\,,\\
\nu_{1,2}^* & =& (&0,&1,&0,&0;&0&)\,,\\
\nu_{1,3}^* & =& (&0,&0,&1,&0;&0&)\,,\\
\nu_{1,4}^* & =& (&0,&0,&0,&1;&0&)\,,\\
\nu_{1,5}^* & =& (&-1,&-1,&-1,&-1;&0&)\,,\\
\nu_{2,1}^* & =& (&0,&0,&0,&0;&1&)\,,\\
\nu_{2,2}^* & =& (&0,&0,&0,&0;&-1&)\,.
\end{array}$
\caption{$\Delta_1^*$ vertices} \label{stab:Delta1}
\end{subtable}
\begin{subtable}{0.48\textwidth}
\centering
$\begin{array}{cccccccc}
\nu_{1,1}^* & =& (1,&0,&0,&0;&0)\,,\\
\nu_{1,2}^* & =& (0,&1,&0,&0;&0)\,,\\
\nu_{1,3}^* & =& (0,&0,&1,&0;&0)\,,\\
\nu_{1,4}^* & =& (0,&0,&0,&1;&0)\,,\\
\nu_{1,5}^* & =& (-4,&-1,&-1,&-1;&0)\,,\\
\nu_{2,1}^* & =& (0,&0,&0,&0;&1)\,,\\
\nu_{2,2}^* & =& (0,&0,&0,&0;&-1)\,.
\end{array}$
\caption{$\Delta_2^*$ vertices} \label{stab:Delta2}
\end{subtable}
\caption{The vertices of the reflexive polyhedra~\subref{stab:Delta1} $\Delta_1^*$ and~\subref{stab:Delta2} $\Delta_2^*$ within $\mathbb{R}^4\times \mathbb{R}$.}
\label{tab:DeltaVertices}
\end{table}
We now discuss the mirror geometry, complex structure moduli and GKZ operator for the first example, the second one being analogous.
We group the vertices 
into two sets:
\begin{equation}
E_1= \{\nu^*_{1,1},\nu^*_{1,2},\nu^*_{1,3},\nu^*_{1,4},\nu^*_{2,2} \}\,,\quad E_2=\{\nu^*_{1,5},\nu^*_{2,1} \}\,,
\end{equation}
and define the extended vertices $\overline{\nu}^*=(\vec{e}_{1,2},\nu^*)$ in $\mathbb{R}^7$, where we choose $\vec{e}_1=(1,0)$ and $\vec{e}_2=(0,1)$ for the vertices in the first and in the second set respectively. We obtain the following two Laurent polynomials:
\begin{align}
P_1&= a_1 - a_{1,1} X_{1,1} - a_{1,2} X_{1,2} - a_{1,3} X_{1,3} - a_{1,4} X_{1,4} - \frac{a_{2,2}}{X_{2,1}}\, ,\\
P_2&=a_2- a_{2,1}X_{2,1}-\frac{a_{1,5}}{X_{1,1} X_{1,2} X_{1,3}X_{1,4}}\,.
\end{align}
The two vectors $l^{(a)}$, $a=1,2$, giving the relations between the extended vertices are:
\begin{equation}
\begin{array}{cccccccccccc}
l^1& =& (&-4,&-1;& 1,&1,&1,&1,&0,&0&) \,,\\
l^2& =& (& -1,&-1;&0,&0,&0,&0,&1,&1&) \,.
\end{array}
\end{equation}
These vectors also specify the $U(1)$ charges of the fields in the GLSM construction of the geometry \cite{Witten:1993yc}. We obtain the local coordinates on the two dimensional moduli space of complex structures:\footnote{The negative sign for $z_1$ is following a convention of \cite{Hosono:1994ax}}
\begin{equation}
z_1=-\frac{a_{1,1} a_{1,2} a_{1,3} a_{a,4}}{a_1^4 a_2}\,, \quad z_{2}= \frac{a_{2,1}a_{2,2}}{a_1 a_2}
\end{equation}

From (\ref{GKZ}) we can now obtain the two GKZ operators:
\begin{align}
\mathcal{L}^1 &= \theta_1^5-z_1 (\theta_1+\theta_2+1) \prod_{i=1}^4 (4\theta_1+\theta_2+i)\,, \\
\mathcal{L}^2 &= \theta_2^2 -z_2 (4\theta_1+\theta_2 +1)(\theta_1 +\theta_2+1)\,, \quad \theta_i:=z_i \frac{\partial}{\partial z_i}\,.
\end{align} 
We reduce the operators to obtain a system of Picard-Fuchs equations in the following way:
\begin{equation}
\mathcal{L}^1- \theta_1^3 \mathcal{L}^2= (\theta_1+\theta_2) \tilde{\mathcal{L}^1}\,, \quad 4 \tilde{\mathcal{L}}^1 -\frac{5}{4} \theta_1^2 \mathcal{L}^2= (4\theta_1+\theta_2) L_1\,.
\end{equation}
The two Picard-Fuchs operators read:
\begin{align}
L_1 &= \left( 1+ 4 z_2\right) \theta_1^3 - 4z_1 \prod_{i=1}^3 (4 \theta_1 + \theta_2+i) -\frac{5}{4} \theta_1^2 \theta_2 + \frac{5}{4} z_2 \theta_1^2 (\theta_1+ \theta_2+1)\,, \\
L_2 &= \theta_2^2 -z_2 (4\theta_1+\theta_2 +1)(\theta_1 +\theta_2+1)\,.
\end{align}
The discriminants of this system of operators can be computed to be:
\begin{align}
\Delta_1&= (z_2-1)^5-65536 z_1^2+z_1 \left(-27 z_2^4+144
   z_2^3-320 z_2^2+2816 z_2+512\right)\,,\\
\Delta_2 &= 384 z_1-5 (z_2-1)^2\,.
\end{align}

To obtain the prepotential we use the fact that its triple derivative in the special coordinates gives the Yukawa couplings. We therefore introduce the Yukawa couplings in algebraic coordinates:
\begin{equation}
\tilde{C}_{ijk}:= - \int_{\check{X}} \Omega \wedge \theta_i \theta_j \theta_k \Omega\,, \quad i,j,k=1,2\,,
\end{equation}
which relate to the $C_{abc}$ as:
\begin{equation}
C_{abc} = \frac{1}{\pi_0^2} \sum_{i,j,k} \frac{\partial z_i}{\partial t^a} \frac{\partial z_j}{\partial t^b} \frac{\partial z_k}{\partial t^c} \frac{1}{z_i z_j z_k} \tilde{C}_{ijk}\,.
\end{equation}

The Picard Fuchs equations generate an ideal $\mathcal{I}$ in the ring $\mathcal{R}=\mathbb{C}[z_1,z_2,\theta_1,\theta_2]$, the relations in $\mathcal{R}/\mathcal{I}$ together with Griffiths' transversality of the Gauss-Manin connection can be used to deduce relations among the $\tilde{C}_{ijk}$ as well as differential equations satisfied by these.\footnote{See \cite{Cox:2000vi} for more details.} We obtain:
\begin{align}\label{Yukawa}
\tilde{C}_{111} &= \frac{c \left(5 (z_2-1)^3-16 z_1 \left(9 z_2^2-32
   z_2+48\right)\right)}{\Delta_1}\,, \\
 \tilde{C}_{112} &=\frac{c \left(64 z_1 \left(9 z_2^2-20
   z_2+16\right)-(z_2-1)^2 (21
   z_2+4)\right)}{\Delta_1}\,, \\
   \tilde{C}_{122} &= \frac{c z_2 \left(5 \left(17 z_2^2-9
   z_2-8\right)-256 z_1 (9
   z_2-8)\right)}{\Delta_1}\,, \\
  \tilde{C}_{222} &= -\frac{c z_2 \left(341 z_2^2+268 z_2-1024 z_1 (9
   z_2+4)+16\right)}{\Delta_1}\,.
\end{align}
Fixing the integration constant $c$ to $1$ leads to a matching of the leading expansion of $C_{abc}$ with the triple intersection numbers of the geometry $X$.

\subsection{Mirror maps and genus 0 GV invariants}
The periods of the holomorphic three-form $\Omega$ of $\check{X}$ can be found as solutions to the Picard Fuchs system. Using the toric data, the holomorphic period near $z_1,z_2=0$ can be found using the expression given in (\ref{holperiod}). The periods which are used to construct the special coordinates on the moduli space, which become the mirror maps to the complexified K\"ahler moduli space of $X$ are obtained using the Frobenius method.\footnote{See, e.g., \cite{Hosono:1993qy,Hosono:1994ax} for details.} These have the form:
\begin{align}
   \pi^1&= \frac{1}{2\pi i} \pi^0 \log z_1 + S_1(z_1,z_2)\,, \\
   \pi^2&= \frac{1}{2\pi i}\pi^0 \log z_2 + S_2 (z_1,z_2)\,.  
   \end{align}
The special coordinates are given by:
\begin{equation} t_1= \frac{\pi^1}{\pi^0}\,, \quad t_2=\frac{\pi^2}{\pi^0}\,. \end{equation}
To obtain the inverse of these, i.e., $z_1(t_1,t_2)$ and $z_2(t_1,t_2)$ we introduce:
\begin{equation}
    \mathfrak{q}_a := \exp (2\pi i t_a),\quad a=1,2,
\end{equation}
which gives:
\begin{align}
z_1 &= \mathfrak{q}_1+\left(-104 \mathfrak{q}_1^2-5 \mathfrak{q}_2 \mathfrak{q}_1\right)+
   \left(6444 \mathfrak{q}_1^3+282 \mathfrak{q}_2 \mathfrak{q}_1^2+10 \mathfrak{q}_2^2
   \mathfrak{q}_1\right)+O\left(\mathfrak{q}^4\right)\,, \\
   z_2 &= \mathfrak{q}_2-74  (\mathfrak{q}_1 \mathfrak{q}_2)+ \left(1581 \mathfrak{q}_2 \mathfrak{q}_1^2+16
   \mathfrak{q}_2^2 \mathfrak{q}_1\right)+O\left(\mathfrak{q}^4\right) \,.
\end{align}

The genus 0 Gopakumar-Vafa (GV) BPS invariants can be extracted from the prepotential $F_0(\mathfrak{q}_1,\mathfrak{q}_2)$, the latter can be obtained by integrating the relations\footnote{An alternative way to obtain the prepotential using the Frobenius method was given in \cite{Hosono:1994ax} and is implemented in a mathematica program (INSTANTON) by Albrecht Klemm. We have used this program to verify our results.} 
\begin{equation}
C_{abc}= \frac{\partial^3}{\partial t^a \partial t^b \partial t^c} F_0(t)\,,
\label{eq:Yukawa}
\qquad \text{where} \quad
F_0(t)=\sum_{\beta>0} n_{\beta}^0 \textrm{Li}_3(\mathfrak{q}^{\beta})\,, \quad \mathfrak{q}^{\beta}= \exp(2\pi i t^{\beta})\,.
\end{equation}
Here, Li$_3$ is the polylogarithm of order 3, and $n_{\beta}^0$ is the genus 0 GV invariant of homology class $\beta$. In this way, we find the genus 0 invariants shown in Tables~\ref{tab:GMSVinvs0} and~\ref{tab:GMSVinvs}. 

The second geometry is specified by the toric geometry charge vectors:
\begin{equation}
\begin{array}{ccccccccccccc}
l^1& =& (&-4,&-4;& 4,&1,&1,&1,&1,&0,&0&) \,,\\
l^2& =& (& -1,&-1;&0,&0,&0,&0,&0,&1,&1&) \,,
\end{array}
\end{equation}
and the genus 0 invariants can be obtained similarly. The results are shown in Table \ref{tab:SymFlopGV}.

\subsection{Genus 1 GV invariants}
The generating function for the genus 1 GV invariants can be computed by integrating the holomorphic anomaly equation of Bershadsky, Cecotti, Ooguri and Vafa (BCOV) \cite{Bershadsky:1993ta} and imposing some boundary conditions discussed in~\cite{Bershadsky:1993ta,Vafa:1995ta}. To state the genus 1 BCOV anomaly equation we need to introduce further ingredients of the special geometry of the moduli space $\mathcal{M}$ of complex structures of $\check{X}$. $\mathcal{M}$ is a projective special K\"ahler manifold with K\"ahler potential:
\begin{equation}
e^{-K} := i\int_{\check{X}} \Omega \wedge \overline{\Omega}\,,
\end{equation}
and metric  $G_{i\bar{j}} := \partial_i \partial_{\bar{j}} K$ in local coordinates $z^i,i=1,\dots,h^{2,1}(\check{X})$.
 The $C_{ijk}$ are defined as before and we further introduce
$\overline{C}_{\overline{\imath}}^{jk}= e^{2K} G^{j \bar{m}} G^{k \bar{n}} \overline{C_{imn}}$.
 
 The holomorphic anomaly equation at genus 1 takes the form \cite{Bershadsky:1993ta}
\begin{equation}
\bar{\partial}_{\bar{i}} \partial_j \mathcal{F}^{(1)}= \frac{1}{2} \overline{C}_{\bar{i}}^{kl} C_{jkl}+\left( 1-\frac{\chi}{24}\right) G_{\bar{i}j} \, .
\end{equation}
This can be integrated to give:
\begin{equation}
\mathcal{F}^{(1)}= \frac{1}{2} \left( 3+h^{2,1}(\check{X}) -\frac{\chi(\check{X})}{12}\right) K +\frac{1}{2} \log \det G^{-1} +\sum_i s_i \log z_i + \sum_a r_a \log  \Delta_a \,,
\end{equation}
where $i=1,\dots,h^{2,1}$ and $a$ runs over the number of discriminant components. The coefficients $s_i$ and $r_a$ are fixed by the leading singular behavior of $\mathcal{F}^{(1)}$, given by
\begin{equation}
 \mathcal{F}^{(1)} \sim -\frac{1}{24} \sum_i \log z_i \int_X c_2 J_i \, ,
 \end{equation}
for the algebraic coordinates $z_i$, for a discriminant $\Delta_{con}$ corresponding to a conifold singularity the leading behavior is given by 
\begin{equation}
  \mathcal{F}^{(1)} \sim -\frac{1}{12} \log \Delta_{con} \,.
\end{equation}
Using the holomorphic limit of $G$,
$G_{\bar{i}j} \rightarrow  A_{\bar{i} a} \theta_j t^{a}$
for some constant $A$, where the convention of multiplying the indices with coordinates is adopted. We find for the ambiguities of $\mathcal{F}^{(1)}$, $s_1=-\frac{50}{24}, s_2=-1$ and $b_1=-\frac{1}{12}, b_2=0$, since only the discriminant $\Delta_1$ appears in the denominator of the Yukawa couplings of~\eqref{Yukawa} and corresponds to a conifold type singularity in the moduli space. We obtain the genus 1 GV invariants shown in Table~\ref{stab:GMSVgenus1phase1}.

\begin{table}
\centering
\begin{subtable}{0.55\textwidth}
\centering
$\arraycolsep=5pt
\begin{array}{c|ccccc}
 \mathdiagbox[width=1cm,height=0.75cm,innerleftsep=0.1cm,innerrightsep=0cm]{q_2}{q_1} &0&1&2&3&4\\
 \hline
0& -& 0 & 0 & -1280 & -317864 \\
 1&0 & 0 & 0 & 10240 & 3922880 \\
 2&0 & 0 & 0 & 356368 & 484562136 \\
 3&0 & 0 & 0 & 243328 & 2025329024 \\
 4&0 & 0 & 0 & -15708 & 1183070468 \\
 5&0 & 0 & 0 & 34320 & 15848448 \\
 6&0 & 0 & 0 & -32032 & 21037432 \\
 7&0 & 0 & 0 & 21840 & -22437984 \\
 8&0 & 0 & 0 & -10920 & 17668602 \\
 9&0 & 0 & 0 & 3920 & -11445504 \\
 10&0 & 0 & 0 & -960 & 6174504 \\
 11&0 & 0 & 0 & 144 & -2713248 \\
 12&0 & 0 & 0 & -10 & 933780 \\
 13&0 & 0 & 0 & 0 & -27492 \\
\end{array}
$
\caption{Phase I} \label{stab:GMSVgenus1phase1}
\end{subtable}
\begin{subtable}{0.42\textwidth}
\centering
$\arraycolsep=5pt
\begin{array}{c|cccc}
\mathdiagbox[width=1cm,height=0.8cm,innerleftsep=0.08cm,innerrightsep=0cm]{q_2'}{q_1'} &0&1&2&3\\
 \hline
 0&- & 0 & 0 & -10 \\
 1&0 & 0 & 0 & 144 \\
 2&0 & 0 & 0 & -960 \\
 3&0 & 0 & 0 & 3920 \\
 4&0 & 0 & 0 & -10920 \\
 5&0 & 0 & 0 & 21840 \\
 6&0 & 0 & 0 & -32032 \\
 7&0 & 0 & 0 & 34320 \\
 8&0 & 0 & 0 & -15708 \\
 9&0 & 0 & 0 & 243328 \\
 10&0 & 0 & 0 & 356368 \\
 11&0 & 0 & 0 & 10240 \\
 12&0 & 0 & 0 & -1280 \\
 13&0 & 0 & 0 & 0 \\
 \end{array}
$
\caption{Phase II} \label{stab:GMSVgenus1phase2}
\end{subtable}
\caption{Genus 1 GV invariants of degree $(q_1,q_2)$ for \subref{stab:GMSVgenus1phase1} phase I and \subref{stab:GMSVgenus1phase2} phase II of the GMSV geometry.} 
\label{tab:GMSVgenus1}
\end{table}

\subsection{Conifold flop transition of the geometry}
The conifold flop in the geometries $X$ specified in Sec.~\ref{topexamples} is the analytic continuation of the imaginary part of the complexified K\"ahler moduli $t_2$ to negative values. On the mirror geometries $\check{X}$ this is reflected by a different choice of local special coordinates. The flop transition in both cases is given by the special coordinate transformations,
$t_1 \rightarrow \tilde{t}_1= t_1 + 4 t_2$ and 
$t_2 \rightarrow \tilde{t}_2 = -t_2$. 
which have the effect
\begin{equation}\label{flop}
\mathfrak{q}_1 \rightarrow \tilde{\mathfrak{q}}_1= \mathfrak{q}_1 \mathfrak{q}_2^4\,, \qquad
\mathfrak{q}_2 \rightarrow \tilde{\mathfrak{q}}_2 = \frac{1}{\mathfrak{q}_2} \,.
\end{equation}

We first determine the fate of the 16 particles becoming massless at the conifold singularity given by $z_2=0$ or $\mathfrak{q}_2=1$, these correspond to the GV invariant at degree $(d_1,d_2)=(0,1)$. Noting that this GV invariant can also be extracted directly from the Yukawa coupling $C_{222}$ in the limit $\mathfrak{q}_1\rightarrow 0$, we find:
\begin{equation}
C_{222}|_{\mathfrak{q}_1=0}= 16 \frac{\mathfrak{q}_2}{1-\mathfrak{q}_2} 
\end{equation}
after the flop we compute the new Yukawa couplings by:
\begin{equation}
\tilde{C}_{abc}= \sum_{l,m,n} \frac{\partial \tilde{t}^l}{\partial t^a} \frac{\partial \tilde{t}^m}{\partial t^b} \frac{\partial \tilde{t}^n}{\partial t^c} C_{lmn},
\end{equation}
and obtain
\begin{equation}
\tilde{C}_{222}|_{\tilde{\mathfrak{q}}_1=0}= 16 \frac{\tilde{\mathfrak{q}}_2}{1-\tilde{\mathfrak{q}}_2} .
\end{equation}
The other invariants can be obtained directly from the prepotential by making the replacement of~(\ref{flop}).
At genus 0, we obtain the invariants shown in Table~\ref{tab:GMSVflopped}, whereas at genus 1 we obtain those shown in Table~\ref{stab:GMSVgenus1phase2}.
For the second geometry, since the flopped phase is isomorphic to the original phase, we once again obtain precisely the genus 0 invariants in Table~\ref{tab:SymFlopGV}.

\subsection{Hypersurfaces and three moduli flop transition}
The mirror constructions reviewed in~\S\ref{mirrorfamilies} also apply to hypersurfaces in toric varieties, as discussed in, e.g., \cite{Hosono:1993qy}. The motivation for the first example of a complete intersection discussed previously comes from \cite{Greene:1996dh}, where an equivalent geometry to the $(2,86)$ complete intersection was given in terms of the quintic hypersurface in $\mathbb{P}^4$ where the ambient space is blown up along a $\mathbb{P}^2$. We find that the genus 0 GV invariants of the $(2,86)$ geometry agree with the GMSV geometry in the phase specified by the toric charge vectors.
We consider the geometry given in \cite{Greene:1996dh}, described by the toric charge vectors:
\begin{equation}
\begin{array}{cccccccc}
l^1= &(-4; &1,&0,&0,&1,&1,&1) \,, \\
l^2=& (-1;&-1&1&1&0&0&0) \,.
\end{array} 
\end{equation}
Since the geometric data is slightly different in this case, we introduce the local coordinates
\begin{equation}
z_1=\frac{a_1 a_2 a_3 a_4}{a_0^4}\,,\quad z_2 = -\frac{a_3 a_4}{a_1 a_0}\,.
\end{equation}
The corresponding Picard-Fuchs operators are: 
\begin{align}
\mathcal{L}^1 &= \theta_1^3 - 4 z_1 \prod_{i=1}^3 (4\theta_1+\theta_2+i) -\frac{5}{4} \theta_1^2(\theta_2 +z_2 (\theta_1-\theta_2)) \,,\\
\mathcal{L}^2 &= \theta_2^2 - z_2 (\theta_1-\theta_2) (4\theta_1 + \theta_2 +1)\,, \quad \theta_i:=z_i \frac{\partial}{\partial z^i}\,,
\end{align}
and the discriminant is:
 \begin{equation}
 \Delta= (z_2+1)^3+65536 z_1^2-z_1
   \left(3125 z_2^4+10000 z_2^3+11200 z_2^2+4864
   z_2+512\right) \,.
 \end{equation}
We find the following expressions for the Yukawa couplings:
\begin{align}
\tilde{C}_{111} &=  \frac{5 (z_2+1)^3+16 z_1 \left(25 z_2^2+64
   z_2+48\right)}{\Delta}\,, \\
 \tilde{C}_{112} &= \frac{64 z_1 \left(25 z_2^2+44 z_2+16\right)-(z_2+1)^2 (5
   z_2+4)}{\Delta}\,, \\
   \tilde{C}_{122} &= \frac{-z_2 \left(5 z_2^2+13 z_2+256 z_1 (25
   z_2+24)+8\right)}{\Delta}\,, \\
  \tilde{C}_{222} &= \frac{-z_2 \left(5 z_2^2+12 z_2-1024 z_1 (25
   z_2+4)+16\right)}{\Delta}\,.
\end{align}
We have computed the genus $0$ and $1$ GV invariants for this geometry and found agreement with the $(2,86)$ complete intersection.

As a last example, we study a geometry used by Klemm, Mayr and Vafa in \cite{Klemm:1996hh}. The geometry is obtained by resolution of singularities of a degree 18 hypersurface in $\mathbb{P}_{1,1,1,6,9}$ and subsequent blow up of a point in the base $\mathbb{P}^2$ of the resulting elliptic fibration. This geometry has a phase where it becomes an elliptic fibration over a Hirzebruch surface $\mathbb{F}_1$. The toric data of this geometry is given explicitly in \cite{Klemm:1996hh}; to avoid repetition we will only give the toric charge vectors:
\begin{equation}
\begin{array}{ccccccccc}
l^1= l^{(E)}=&(-6; &3,&2,&1,&0,&0,&0,&0,)\,,\\
l^2=l^{(F)}=& (0;&0,&0,&-2,&1,&1,&0,&0,)\,,\\
l^3=l^{(D)}=& (0;&0,&0,&-1,&0,&-1,&1,&1,)\,.
\end{array}
\end{equation}
The resulting genus 0 GV invariants are shown in Table~\ref{tab:KMVinvs}.\footnote{The invariants for $q_2=d_F=0$ were already given in \cite{Klemm:1996hh}.}

\section{Analytic fake superpotential for the symmetric flop} \label{app:analytic}

In this appendix, we briefly describe how the fake superpotential can be found analytically (as a parametric equation) for
the example described in {\textsection}\ref{ssec:symFlop}. While the
numerical method described in the main text is wholly sufficient, this
analytic approach provides a useful crosscheck.

Rather than attacking (\ref{eqn:symWPDE}) directly, it turns out to be easier
to solve (\ref{eqn:BHeqns}) first, which become
\begin{subequations}
\begin{align}
  \ddot{\psi} &= \frac{1}{3} P_+^2 e^{2 (\psi + 2 \phi)} + \frac{2}{3} Q_+^2
  e^{2 (\psi - \phi)}, &
  \ddot{\phi} &= \frac{1}{3} P_+^2 e^{2 (\psi + 2 \phi)} - \frac{1}{3} Q_+^2
  e^{2 (\psi - \phi)}, \\
  \dot{\psi}^2 + 2 \dot{\phi}^2 &= \frac{1}{3} P_+^2 e^{2 (\psi + 2 \phi)} +
  \frac{2}{3} Q_+^2 e^{2 (\psi - \phi)},  \label{eqn:psiphicons}
\end{align}
\end{subequations}
in the region $\phi > 0$. Taking linear combinations of the first two
equations, we obtain
\begin{equation}
  \ddot{\psi} + 2 \ddot{\phi} = P_+^2 e^{2 (\psi + 2 \phi)}, \qquad
  \ddot{\psi} - \ddot{\phi} = Q_+^2 e^{2 (\psi - \phi)} .
\end{equation}
The general solution to $\ddot{\psi} = Q^2 e^{2 \psi}$
is $e^{- \psi} = \frac{Q}{\kappa} \sinh (\kappa (z - z_0))$ for
constants $\kappa$, $z_0$, hence
\begin{equation}
  e^{- \psi - 2 \phi} = \frac{| P_+ |}{k_+} \sin (k_+  (z + z_P^+)), \qquad
  e^{- \psi + \phi} = \frac{\sqrt{2} | Q_+ |}{k_+} \sinh \biggl(
  \frac{k_+}{\sqrt{2}}  (z + z_Q^+) \biggr), \qquad (\phi \geqslant
  0). \label{eqn:genPosSoln}
\end{equation}
after constraining the integration constants using~\eqref{eqn:psiphicons}.
Similarly,
\begin{equation}
  e^{- \psi + 2 \phi} = \frac{| P_- |}{k_-} \sin (k_-  (z + z_P^-)), \qquad
  e^{- \psi - \phi} = \frac{\sqrt{2} | Q_- |}{k_-} \sinh \biggl(
  \frac{k_-}{\sqrt{2}}  (z + z_Q^-) \biggr), \qquad (\phi \leqslant
  0).
\end{equation}
A priori $k_{\pm}$ could be either real or imaginary, but only the real case will be needed.

In the special case $k_\pm \to 0$, the solution becomes
\begin{equation}
  e^{- \psi \mp 2 \phi} = | P_{\pm} |  (z + z_P^{\pm}), \qquad e^{- \psi \pm
  \phi} = | Q_{\pm} |  (z + z_Q^{\pm}), \qquad (\pm \phi \geqslant
  0). \label{eqn:zerok}
\end{equation}
In particular, only (\ref{eqn:zerok}) gives a regular solution as $z
\rightarrow \infty$ (near the horizon). If there are no flops for $0 \leqslant
z < \infty$ then we obtain (for $\phi_h > 0$):\footnote{Note that the attractor is located at $
  \phi_h = \begin{cases}
    \frac{1}{3} \log \frac{| Q_+ |}{| P_+ |}, & | Q_+ | \geqslant | P_+ |,\\
    - \frac{1}{3} \log \frac{| Q_- |}{| P_- |}, & | Q_- | \geqslant | P_- |,\\
    0, & | Q_{\pm} | < | P_{\pm} |.
  \end{cases}
$}
\begin{equation}
  e^{- \psi - 2 \phi} = e^{- 2 \phi_{\infty}} + | P_+ | z, \qquad e^{- \psi +
  \phi} = e^{\phi_{\infty}} + | Q_+ | z,
\end{equation}
upon setting $\psi (z = 0) = 0$, $\phi (z = 0) = \phi_{\infty}$. Thus,
\begin{equation}
  W (\phi_{\infty}) = - 3 \dot{\psi}_{\infty} = | P_+ | e^{2 \phi_{\infty}} +
  2 | Q_+ | e^{- \phi_{\infty}} .
\end{equation}
This is true independent of $\phi_{\infty} > 0$, so we conclude that
\begin{equation}
  \text{if $\phi_h > 0$ then} \qquad W (\phi) = | P_+ | e^{2 \phi} + 2 | Q_+ |
  e^{- \phi} \qquad \text{for $\phi \geqslant 0$} .
\end{equation}
Likewise, if $\phi_h < 0$ then $W (\phi) = | P_- | e^{- 2 \phi} + 2 | Q_- |
e^{\phi}$ for $\phi \leqslant 0$.

When $\phi_{\infty}$ and $\phi_h$ have opposite signs, a flop occurs at some
$0 < z_1 < \infty$. If $\phi_h > 0$,
\begin{equation}
  e^{- \psi - 2 \phi} = e^{- \psi_1} + | P_+ |  (z - z_1), \qquad e^{- \psi +
  \phi} = e^{- \psi_1} + | Q_+ |  (z - z_1), \qquad \text{for $z \geqslant
  z_1$,}
\end{equation}
where $\psi_1 = \psi (z_1)$. For $z < z_1$, the solution takes the general
form (\ref{eqn:genPosSoln}), which can be rewritten as
\begin{align}
  e^{- \psi + 2 \phi} &= \frac{A}{k} \sin (k (z - z_1)) + e^{- \psi_1} \cos
  (k (z - z_1)), \\
  e^{- \psi - \phi} &= \frac{\sqrt{2} B}{k} \sinh \biggl( \frac{k}{\sqrt{2}} 
  (z - z_1) \biggr) + e^{- \psi_1} \cosh \biggl( \frac{k}{\sqrt{2}}  (z - z_1)
  \biggr) ,
\end{align}
where $k^2 = 2 e^{2 \psi_1}  (B^2 - Q_-^2) = e^{2 \psi_1}  (P_-^2 - A^2)$.
Matching $\phi, \psi$ and $\dot{\phi}, \dot{\psi}$ at the flop, we obtain $A =
\frac{4}{3}  | Q_+ | - \frac{1}{3}  | P_+ |$ and $B = \frac{2}{3}  | P_+ | +
\frac{1}{3}  | Q_+ |$, hence
\begin{equation}
  k^2 = \frac{8}{9} e^{2 \psi_1}  (| P_+ Q_+ | - P_+ Q_+) .
\end{equation}
When $P_+ Q_+ \geqslant 0$, $k = 0$ and we obtain
\begin{equation}
  e^{- \psi + 2 \phi} = e^{- \psi_1} + | P_- |  (z - z_1), \qquad e^{- \psi -
  \phi} = e^{- \psi_1} + | Q_+ |  (z - z_1), \qquad \text{for $z \leqslant
  z_1$,}
\end{equation}
which leads to $W (\phi) = | P_- | e^{- 2 \phi} + 2 | Q_- | e^{\phi}$. The
result is simple because $W (\phi) = | \zeta_q (\phi) |$ ($q_I$ lies with
$\Kv$ or $-\Kv$).

By contrast, when $P_+ Q_+ < 0$, $k = \frac{4}{3} e^{\psi_1}  \sqrt{|
P_+ Q_+ |} \ne 0$. Defining $u \df k (z - z_1)$ and
\begin{align}
  F_{\pm} (u) &= \frac{4}{3}  \sqrt{| P_{\mp} Q_{\mp} |} \cosh \biggl(
  \frac{u}{\sqrt{2}} \biggr) + \sqrt{2}  \biggl( \frac{2}{3}  | P_{\mp} | +
  \frac{1}{3} | Q_{\mp} | \biggr) \sinh \biggl( \frac{u}{\sqrt{2}} \biggr),\\
  G_{\pm} (u) &= \frac{4}{3}  \sqrt{| P_{\mp} Q_{\mp} |} \cos (u) + \biggl(
  \frac{4}{3}  | Q_{\mp} | - \frac{1}{3}  | P_{\mp} | \biggr) \sin (u),
\end{align}
we find $e^{- \psi - \phi} = \frac{1}{k} F_- (u)$ and $e^{- \psi + 2 \phi} =
\frac{1}{k} G_- (u)$ for $z \leqslant z_1$, or
\begin{equation}
  W (u) = - 3 \dot{\psi} e^{- \psi} = 2 F'_- (u) e^{\phi} + G'_- (u) e^{- 2
  \phi}, \qquad e^{- 3 \phi (u)} = \frac{F_- (u)}{G_- (u)} .
\end{equation}
This gives a parametric equation for $W (\phi)$ in the region $\phi < 0$ in
terms of the parameter $u \in \Bigl[- \tan^{- 1} \Bigl( \frac{\sqrt{| P_+ Q_+ |}}{| Q_+ |
- \frac{1}{4}  | P_+ |} \Bigr) , 0\Bigr]$. Similarly, when $\phi_h < 0$,
\begin{equation}
  W (u) = 2 F'_+ (u) e^{- \phi} + G'_+ (u) e^{2 \phi}, \qquad e^{3 \phi (u)} =
  \frac{F_+ (u)}{G_+ (u)},
\end{equation}
gives a parametric equation for $W (\phi)$ in the region $\phi > 0$ in terms
of the parameter $u \in \Bigl[- \tan^{- 1} \Bigl( \frac{\sqrt{| P_- Q_- |}}{| Q_- | -
\frac{1}{4}  | P_- |} \Bigr), 0\Bigr]$.

The solution is rather different when $\phi_h = 0$ (apart from the special
cases $| Q_+ | = | P_+ |$ and $| Q_- | = | P_- |$ arising from limits of
solutions where $\phi_h > 0$ or $\phi_h < 0$), because then $\mathcal{Q}^2
(\phi)$ has a vee-shaped minimum at $\phi = 0$ (see, e.g., Figure
\ref{fig:ConifoldAttractor}). In this case, $\phi$ stays at the attractor
point $\phi_h = 0$ throughout a finite region $z \geqslant z_1$ outside the
horizon. The solution in this region is simply
\begin{equation}
  \phi = 0, \qquad e^{- \psi} = e^{- \psi_1} + \sqrt{\frac{1}{3} P_+^2 +
  \frac{2}{3} Q_+^2}  (z - z_1) .
\end{equation}
Assuming $\phi_{\infty} > 0$, we match onto the general solution
(\ref{eqn:genPosSoln}) at $z = z_1$ as before. Now we obtain
\begin{equation}
  W (u) = 2 F'_{0 +} (u) e^{- \phi} + G'_{0 +} (u) e^{2 \phi}, \qquad e^{3
  \phi (u)} = \frac{F_{0 +} (u)}{G_{0 +} (u)},
\end{equation}
where
\begin{align}
  F_{0 \pm} (u) &= \sqrt{\frac{2}{3} P_{\pm}^2 - \frac{2}{3} Q_{\pm}^2} \cosh
  \biggl( \frac{u}{\sqrt{2}} \biggr) + \sqrt{2} \sqrt{\frac{1}{3} P_{\pm}^2 +
  \frac{2}{3} Q_{\pm}^2} \sinh \biggl( \frac{u}{\sqrt{2}} \biggr), \\
  G_{0 \pm} (u) &= \sqrt{\frac{2}{3} P_{\pm}^2 - \frac{2}{3} Q_{\pm}^2} \cos
  (u) + \sqrt{\frac{1}{3} P_{\pm}^2 + \frac{2}{3} Q_{\pm}^2} \sin (u) .
\end{align}
This gives a parametric equation for $W (\phi)$ in the region $\phi \geqslant
0$ in terms of the parameter $u \in \Bigl[ - \tan^{- 1} \sqrt{\frac{P_+^2 - Q_+^2}{P_+^2 /
2 + Q_+^2}}, 0\Bigr]$. Likewise, in the region $\phi \leqslant 0$, we have
\begin{equation}
W (u) = 2 F'_{0
-} (u) e^{\phi} + G'_{0 -} (u) e^{- 2 \phi} ,~~~~~ e^{- 3 \phi (u)} =
\frac{F_{0 -} (u)}{G_{0 -} (u)}, 
\end{equation}
for $u \in \Bigl[- \tan^{- 1} \sqrt{\frac{P_-^2 -
Q_-^2}{P_-^2 / 2 + Q_-^2}}, 0\Bigr]$.

\bibliographystyle{JHEP}
\bibliography{refs}
\end{document}